\documentclass{article}

\usepackage[a4paper, total={6in, 10in}]{geometry}

\usepackage[utf8]{inputenc} 
\usepackage[T1]{fontenc}    
\usepackage{lmodern}    
\usepackage{url}            
\usepackage{booktabs}       
\usepackage{nicefrac}       
\usepackage{microtype}      

\usepackage{natbib}
\usepackage{doi}
\usepackage{xcolor, comment}
\usepackage{cancel}
\usepackage{amsfonts, amsthm, amssymb,latexsym,amsmath,mathrsfs}
\usepackage{array}

\usepackage{caption}

\usepackage{alphalph}  
\usepackage{bm}
\usepackage{graphicx}

\usepackage{colortbl}
\usepackage{makecell}
\usepackage{array}

\usepackage{ulem}

\newcommand{\R}{\mathbb{R}}
\newcommand{\Y}{\mathcal{Y}}
\newcommand{\dy}{\;{\rm d}y}

\newcommand{\la}{\langle}
\newcommand{\ra}{\rangle}
\newcommand{\w}{w}
\newcommand{\vv}{v}

\definecolor{ggreen}{cmyk}{1,     0,      1,      0}

\newcommand{\rr}{}

\newcommand{\rrsout}[1]{}

\definecolor{Mercury}{rgb}{0.85,0.85,0.85}
\definecolor{Mercury2}{rgb}{0.95,0.95,0.95}

\usepackage{tikz}

\definecolor{pythonAzul}{HTML}{1F77B4}
\definecolor{pythonLaranja}{HTML}{FF7F0E}

\newcommand{\circlemarker}{%
	\tikz[baseline=-0.5ex] \draw[pythonLaranja, line width=1.5pt] (0,0) circle (2.5pt);
}

\newcommand{\squaremarker}{%
	\tikz[baseline=-0.5ex] \draw[pythonAzul, line width=1.5pt]
	(-2.5pt,-2.5pt) rectangle (2.5pt,2.5pt);
}

\makeatletter
\newtheoremstyle{boldplain}
  {}{}{\normalfont}{}{\bfseries}{.}{ }%
  {\thmname{#1}~\thmnumber{#2}\thmnote{#3}}
\makeatother

\theoremstyle{boldplain}

\newcounter{assumptioncounter}
\newcounter{subassumption}[assumptioncounter]

\renewcommand{\thesubassumption}{\arabic{assumptioncounter}\AlphAlph{\value{subassumption}}}

\newtheorem{remark}{Remark}

\newcommand{\brho}{\widetilde{\rho}}

\author{Davide Cusseddu\thanks{Department of Mathematical Sciences ``G. L. Lagrange'', Politecnico di Torino, 10129 Torino, Italy (davide.cusseddu@gmail.com)}, Gaetana Gambino\thanks{Department of Mathematics and Computer Science, University of Palermo, Via Archirafi 34, Palermo, 90123, Italy (gaetana.gambino@unipa.it)}, Tommaso Lorenzi\thanks{Department of Mathematical Sciences ``G. L. Lagrange'', Politecnico di Torino, 10129 Torino, Italy (tommaso.lorenzi@polito.it)}}

\title{On a phenotype-structured Shigesada–Kawasaki–Teramoto model: Turing instability and pattern selection under fast phenotype switching}
\begin{document}
\maketitle

\begin{abstract}
The Shigesada--Kawasaki--Teramoto (SKT) model has become a classical modelling framework for studying spatial segregation and cross-diffusion-driven pattern formation in competing populations. This model assumes phenotypic homogeneity, but phenotypic variability persists within any population and can strongly influence both ecological and evolutionary dynamics. In this paper, we present a generalised phenotype-structured formulation of the SKT model that accounts for phenotypic variability. In this formulation, the competing populations are continuously structured across some phenotype state spaces. Population members move and compete in phenotype-dependent ways, and can also switch between different phenotype states. First we show how a form of the classical SKT model, wherein parameters are written in terms of continuous weighted averages of the phenotype-dependent functions of the generalised structured model, with weights given by the phenotype distributions of the two populations, can be obtained in the quasi-invariant regime of fast phenotype switching. Then, still assuming fast phenotype switching and extending classical Turing-like linear and weakly nonlinear analyses, we explore the conditions for the emergence of spatial patterns, identify a Turing-type bifurcation threshold leading to pattern formation, and investigate the nature of such a bifurcation (super- or sub-critical) as well as the stability of the patterned state. The results obtained make it possible to draw connections between phenotype-dependent model functions and the emergence of population-scale aggregate spatial dynamics, showing in particular how phenotype distributions can act as effective control parameters for Turing instability and pattern selection. These findings are complemented by numerical simulations, which validate the formal asymptotics and confirm the predictions of the pattern formation analyses.
\end{abstract}

\section{Introduction}
Population dynamics strongly depend on the ability of population members to self-organise, move, and interact with the members of other neighbouring populations. These aspects are often closely intertwined and may be equally important in shaping ecological processes. 

In particular, when two or more populations compete for resources, the success of one of them may rely on its ability to sense the competitors and move away from them, thereby establishing its niche, rather than engaging in purely local competition. More generally, species movement is not only determined by the intrinsic tendency of the individuals to explore the spatial environment, but it may depend on several external factors, including the presence and distribution of other coexisting species. This is for instance the case, for opposite reasons, for both prey and predators: prey tend to avoid areas with high predator density, whereas predators may preferentially move towards regions where prey are abundant. Similarly, when two species compete for the same resources, their survival may depend on their ability to redistribute their presence in space so as to reduce local competition. 

From an ecological perspective, these movement responses highlight the fact that \rr{spatial} dispersal of species may be impacted by the local presence of other species. This is naturally captured by cross-diffusion models, which have then been successfully used to describe a variety of biological and ecological phenomena -- see e.g. \citep{BisiCussedduSoaresTravaglini2026, conforto2014rigorous, consolo2025extended, daus2020cross, desvillettes2024triangular, desvillettes2019non, EigentlerEtAl2022, FarivarGambinoEtAl2025, giannino2025beyond, LiDing2026, ruiz2013mathematical, soresina2022hopf, YongLiHendrix2012} and references therein.

Cross-diffusion is a key element of the well-known Shigesada--Kawasaki--Teramoto (SKT) model \citep{Shigesada1979Spatial}, which has become a classical modelling framework for studying spatial segregation and cross-diffusion-driven pattern formation in competing populations. The SKT model comprises the following coupled system of reaction-cross-diffusion equations for the densities of two competing populations labelled by an index $i=1,2$, $\rho_1(t,x)$ and $\rho_2(t,x)$, at time $t > 0$ and position $x\in\Omega$:
\begin{equation}\label{eqn:SKT}
	\begin{cases}
		\partial_t \rho_1 - \Delta \left[ \left( d_1 + a_{11} \rho_1 + a_{12} \rho_2\right)\rho_1 \right]
		= 
		r_1 \rho_1
		-
		\left( b_{11} \rho_1 + b_{12} \rho_2\right)\rho_1, 
		\\ 
		\partial_t \rho_2 - \Delta \left[ \left( d_2 + a_{22} \rho_2 + a_{21} \rho_1\right)\rho_2 \right]
		= 
		r_2 \rho_2
		- 
		\left( b_{22} \rho_2 + b_{21} \rho_1\right)\rho_2.
	\end{cases}
\end{equation}
In the SKT model~\eqref{eqn:SKT}, the reaction terms correspond to competitive Lotka--Volterra dynamics~\citep{EdelsteinKeshet2005,Murray_I}, with intrinsic growth rate $r_i\geq 0$ and intraspecific $(i=j)$ / interspecific $(i\neq j)$ competition coefficient $b_{ij}\geq 0$. Moreover, the diffusion terms contain both linear and density-dependent components. Specifically, the parameters $d_i\geq 0$ and $a_{ij}\geq 0$ are, respectively, the linear diffusion and the self-diffusion $(i=j)$ / cross-diffusion $(i\neq j)$ coefficients of the members of population $i$. Here, the non-negativity of $a_{ij}$ reflects the tendency of members of population $i$ to avoid regions where there is a high density of members of population $j$.

Cross-diffusion plays a fundamental role in the formation of spatial patterns within the framework of the SKT model~\eqref{eqn:SKT}. In fact, linear diffusion alone typically stabilises the spatially homogeneous coexistence equilibrium of the underlying Lotka--Volterra dynamics and does not lead to diffusion-driven instabilities. On the contrary, the inclusion of nonlinear density-dependent diffusion can be a catalyst for spatial segregation of the competing populations, thus providing the substrate for the emergence of spatial patterns, which has driven the popularity of the SKT model.
\rr{Spatially structured microbial communities provide, for instance, a possible ecological setting in which motility, competition, and spatial segregation are strongly intertwined~\citep{GudeEtAl2020}, while competition avoidance has been proposed as a mechanism driving spatial segregation in microbial communities~\citep{MatteiArenas2024}. In this work, we focus on avoidance scenarios, corresponding to non-negative cross-diffusion coefficients, without implying that other movement responses, including attraction, are not biologically relevant.}

Phenotypic variability amongst individuals persists within any population and can strongly influence both ecological and evolutionary dynamics -- see e.g. \citep{Hughes2008GeneticDiversity}. Despite this, in classical ecological models of spatiotemporal dynamics, such as the diffusive Lotka--Volterra system and the SKT system, populations are represented as homogeneous entities, implicitly averaging out differences amongst population members through mean representative features, although extensions accounting for phenotypic variability have begun to emerge -- see, for instance, \citep{Maynard2019Phenotypic} and~\citep{DuongSpillVanRensburg2026}.

Based on these observations, in this work we present a phenotype-structured formulation of the SKT model~\eqref{eqn:SKT}, wherein the competing populations are continuously structured across some phenotype state spaces. In this formulation, population members move and compete in phenotype-dependent ways, and can also switch between different phenotype states. In contrast to most of the works on phenotype-structured models of biological movement (see the recent review~\citep{LorenziPainterVilla2025}), where the effects of phenotype changes are incorporated into the model through differential terms, here we represent phenotype switching through integral terms, as similarly done, for instance, in~\citep{arnold2012existence,barles2009concentration,calsina2004small,bernardi2025heterogeneously,lorenzi2026phenotype,lorenzi2023modelling,lorz2013populational}. Considering a quasi-invariant regime of fast phenotype switching, in the vein of previous work on phenotype-structured integro-differential models of evolutionary dynamics~\citep{bernardi2025heterogeneously,lorenzi2026phenotype}, and building upon previous works on weakly nonlinear analysis of pattern formation in cross-diffusion systems~\citep{gambino2012turing,gambino2013pattern}, we carry out, on the formal level, Turing-type linear and weakly nonlinear analyses to investigate the onset and characteristics of spatial patterns produced by this model. In so doing, we complement the existing literature on the formation of spatial patterns in continuously structured population models, which is still at an early stage and comprises, for instance, \citep{LorenziPainter2025} and~\citep{jewell2026phenotype}, where Turing-type linear analysis for a phenotypically structured model of chemotaxis and a phenotypically structured nonlocal aggregation model was performed, and~\citep{ridgway2025weakly}, where Turing-type weakly nonlinear analysis for a chemically structured model of signalling bacteria was carried out. 

 \rr{We remark that phenotypic variation does not arise only from slow evolutionary changes in heritable traits driven by natural selection; instead, it may reflect the occurrence of fast transitions between reversible physiological or behavioural states. For instance, in microbial populations, competition-associated reversible switching between phenotypes characterised by different motilities has been observed experimentally~\citep{BeerEtAl2011}, while transitions between motile and sessile states are a well-established feature of bacterial physiology~\citep{RomlingEtAl2013}. In this respect, the regime of fast phenotype switching considered here is intended as a stylised way of representing transitions between such reversible phenotypic states occurring on a shorter timescale than population-level spatial and competitive dynamics.} 

The remainder of the paper is organised as follows. We first present a phenotype-structured extension of the SKT model~\eqref{eqn:SKT} (see Section~\ref{Sec:MODEL}). We follow by carrying out pattern formation analyses of such a phenotype-structured SKT model (see Section~\ref{Sec:PatternFormation}). We then integrate the analytical results obtained with the results of numerical simulations (see Section~\ref{sec:numerics}). We conclude with a summary of key findings and a discussion of future directions (see Section~\ref{sec:discussion}).

\section{A phenotype-structured SKT model} \label{Sec:MODEL}
In this section, we first present a possible phenotype-structured generalisation of the SKT model~\eqref{eqn:SKT} accounting for phenotypic heterogeneity, where the variability extends across movement and competition of the population members. This model relies on a form of continuous structuring, whereby population 1 is structured according to phenotype $y_1 \in \Y_1$ and population 2 is structured according to phenotype $y_2 \in \Y_2$, where $\Y_1 \subset \mathbb{R}$ and $\Y_2 \subset \mathbb{R}$ are closed and bounded intervals. Note that $\Y_1$ and $\Y_2$ may possibly be different as the two populations may not share the same phenotype state space (see Section~\ref{sec:mod0}). We then consider a quasi-invariant regime of fast phenotype switching and, introducing an appropriately scaled version of such a phenotype-structured model, we show that a model of the form of the SKT model~\eqref{eqn:SKT} can formally be derived  (see Section~\ref{sec:mod1}). In this model, the parameters are written in terms of continuous weighted averages of the phenotype-dependent functions of the structured model, with weights given by the phenotype distributions of the two populations. 

\subsection{The model}
\label{sec:mod0}

\subsubsection{Model equations}\label{sec:model_equations}
We represent the \textit{phenotype densities} of the two populations through the functions $n_1(t,x,y_1)$ and $n_2(t,x,y_2)$, from which the corresponding \textit{(total) densities}, $\rho_1(t,x)$ and $\rho_2(t,x)$, are obtained by integrating over the respective phenotype state spaces, that is,
\begin{equation*}
    \rho_i(t,x) := \int_{\Y_i} n_i(t,x,y_i) \dy_i, \qquad i=1,2.
\end{equation*}
We let the spatial domain $\Omega \subset \mathbb{R}^d$, with $d \geq 1$, be a bounded and connected set with smooth boundary $\partial \Omega$, and postulate that the dynamics of $n_1$ and $n_2$ are governed by the following coupled system of non-local reaction-cross-diffusion equations:
\begin{equation}
\label{eqn:PDE_n1_n2_mutational_kernel}
\begin{cases}
\displaystyle{\partial_t n_1 - \Delta_x \left[ (D_1(y_1) + \alpha_{11}(y_1)\rho_1 + \alpha_{12}(y_1)\rho_2 ) n_1 \right]  =
\left(\rr{r_1(y_1)} - \mathcal{K}_1[n_1,n_2]\right)n_1 + \theta_1 \, \mathcal{M}_1 \left[ n_1 \right],}
\\\\
\displaystyle{\partial_t n_2 - \Delta_x \left[ (D_2(y_2) + \alpha_{21}(y_2)\rho_1 + \alpha_{22}(y_2)\rho_2 ) n_2 \right] = 
\left(\rr{r_2(y_2)} - \mathcal{K}_2[n_1,n_2]\right)n_2 + \theta_2 \, \mathcal{M}_2 \left[ n_2 \right],}
\\\\
\displaystyle{\rho_1(t,x) := \int_{\Y_1} n_1(t,x,y_1) \dy_1, \quad \rho_2(t,x) := \int_{\Y_2} n_2(t,x,y_2) \dy_2.}
\end{cases}
\end{equation}

In the diffusion terms of the phenotype-structured SKT model~\eqref{eqn:PDE_n1_n2_mutational_kernel}, the non-negative functions $D_i \in C(\Y_i)$ and $\alpha_{ij}\in C(\Y_i)$ represent, respectively, the linear diffusion and the self-diffusion $(i=j)$ / cross-diffusion $(i\neq j)$ coefficients of the members of population $i$ with phenotype $y_i$. Consistently with the unstructured SKT model~\eqref{eqn:SKT}, here we assume the function $\alpha_{ij}$ to be non-negative, in order to take into account the fact that the members of population $i$ tend to avoid regions where the (total) density of population $j$ is high. \rr{Moreover, we assume that the members of population $i$ with phenotype $y_i$ proliferate at a non-negative rate $r_i \in C(\Y_i)$ and} we use the following definitions of the competition terms
\begin{align}
\label{eqn:K1}
    \mathcal{K}_1[n_1,n_2](y_1) := \beta_{11} \int_{\Y_1} K_{11}(y_1,y_1')n_1(t,x,y_1')\dy_1'
    + \beta_{12}
    \int_{\Y_2} K_{12}(y_1,y_2')n_2(t,x,y_2')\dy_2', 
    \\
    \label{eqn:K2}
    \mathcal{K}_2[n_1,n_2](y_2) := \beta_{21} \int_{\Y_1} K_{21}(y_2,y_1')n_1(t,x,y_1')\dy_1'
    + \beta_{22}
    \int_{\Y_2} K_{22}(y_2,y_2')n_2(t,x,y_2')\dy_2',
\end{align}
where $K_{ij} \in C(\Y_i\times\Y_j)$ is a non-negative interaction kernel, which models the effects, on the dynamics of population $i$, of competitive intra-population $(i=j)$ / inter-population $(i\neq j)$ interactions, which occur at rate $\beta_{ij}\geq 0$ -- i.e. $K_{ij}$ is the \textit{competition kernel} and $\beta_{ij}$ is the \textit{competition rate}. 

\rrsout{Note also that here we do not consider trait variation in proliferation. Hence, similarly to the unstructured SKT model~\eqref{eqn:SKT}, the phenotype-structured model~\eqref{eqn:PDE_n1_n2_mutational_kernel} implicitly relies on the assumption that all members of population $i$ proliferate at the same rate $r_i \geq 0$ and, therefore, $r_i$ is also the intrinsic growth rate of the population.}

Finally, the last terms on the right-hand side of the non-local reaction-cross-diffusion equations~\eqref{eqn:PDE_n1_n2_mutational_kernel}, where
\begin{align}\label{eq:mathcal_M_definition}
   \mathcal{M}_i[n_i](y_i) := \int_{\Y_i} M_i(y_i|y_i') n_i(t,x,y_i')  \dy_i' - n_i(t,x, y_i), \qquad i=1,2,
\end{align}
model the effects of phenotype switching on the dynamics of population $i$. These terms rely on the assumption that, as a result of phenotype changes, which occur at rate $\theta_i > 0$, members of population $i$ may switch from phenotype $y_i'$ to phenotype $y_i$ with probability $M_i(y_i|y_i')$. We let 
\begin{align}\label{assMi}
M_i \in \mathcal{P}_{>0}(\Y_i;C(\Y_i)), \qquad i=1,2,
\end{align}
where $\mathcal{P}_{>0}(\Y_i)$ denotes the set of positive probability distributions defined on the measurable space $(\Y_i, \mathcal{B}(\Y_i))$, with $\mathcal{B}(\Y_i)$ being the Borel $\sigma-$algebra of $\Y_i$. Note that, under assumption~\eqref{assMi}, the following property holds
\begin{align}\label{eqn:int mathcal M = 0}
 \int_{\Y_i} M_i(y_i|y_i') \dy_i = 1 \;\; \text{for all } y_i'\in\Y_i, \qquad i=1,2.
\end{align}
In the remainder of the paper, we will refer to $M_i$ as the \textit{phenotype switching kernel} and $\theta_i$ as the \textit{phenotype switching rate}. 

\subsubsection{Initial data and boundary conditions}
We complement the coupled system of non-local reaction-cross-diffusion equations~\eqref{eqn:PDE_n1_n2_mutational_kernel} with initial data such that the following conditions hold
\begin{align*}
    n_i(0,x, y_i) \in C^{2}(\Omega)\times L^1 \cap L^\infty(\Y_i), \quad n_i(0,x, y_i) \geq 0, \qquad i=1,2.
\end{align*}
Moreover, we impose the following homogeneous Neumann (i.e. zero-flux) boundary conditions 
\begin{align}\label{eq:bcs}
    \nabla_x n_i(t, x, y_i) \cdot \nu = 0
    \;\; \text{for all }
    (t,x,y_i) \in (0,\infty) \times \partial\Omega \times \Y_i, \qquad i=1,2,
\end{align}
where $\nu$ denotes the outward unit normal to $\partial\Omega$. 

\subsection{Quasi-invariant regime of fast phenotype switching}
\label{sec:mod1}
\subsubsection{Rescaled model}
To consider a quasi-invariant regime of fast phenotype switching, we introduce a small parameter $0< \varepsilon \ll 1$ and set the phenotype switching rates $\theta_1 = \theta_2 = \varepsilon^{-1}$ whilst considering the other terms in the phenotype-structured SKT model~\eqref{eqn:PDE_n1_n2_mutational_kernel} to be of order $O(1)$. This leads to the following rescaled version of the coupled system of non-local reaction-cross-diffusion equations~\eqref{eqn:PDE_n1_n2_mutational_kernel}
\begin{equation}
\label{eqn:PDE_n1_n2_mutational_kernel_rescaled}
\begin{cases}
\displaystyle{\partial_t n^\varepsilon_1 - \Delta_x \left[ (D_1(y_1) + \alpha_{11}(y_1)\rho^\varepsilon_1 + \alpha_{12}(y_1)\rho^\varepsilon_2 ) n^\varepsilon_1 \right]  =
\left(r_1\rr{(y_1)} - \mathcal{K}_1[n^\varepsilon_1, n^\varepsilon_2]\right) n^\varepsilon_1 + \dfrac{1}{\varepsilon} \, \mathcal{M}_1 \left[ n^\varepsilon_1 \right],}
\\\\
\displaystyle{\partial_t n^\varepsilon_2 - \Delta_x \left[ (D_2(y_2) + \alpha_{21}(y_2)\rho^\varepsilon_1 + \alpha_{22}(y_2)\rho^\varepsilon_2 ) n^\varepsilon_2 \right] = 
\left(r_2\rr{(y_2)} - \mathcal{K}_2[n^\varepsilon_1, n^\varepsilon_2]\right) n^\varepsilon_2 + \dfrac{1}{\varepsilon} \, \mathcal{M}_2 \left[ n^\varepsilon_2 \right],}
\\\\
\displaystyle{\rho^\varepsilon_1(t,x) := \int_{\Y_1} n^\varepsilon_1(t,x,y_1) \dy_1, \quad \rho^\varepsilon_2(t,x) := \int_{\Y_2} n^\varepsilon_2(t,x,y_2) \dy_2,}
\end{cases}
\end{equation}
complemented with definitions~\eqref{eqn:K1}-\eqref{eq:mathcal_M_definition} and subject to the following analogue of boundary conditions~\eqref{eq:bcs}
\begin{align}\label{eq:bcs_rescaled}
    \nabla_x n^\varepsilon_i(t, x, y_i) \cdot \nu = 0
    \;\; \text{for all }
    (t,x,y_i) \in (0,\infty) \times \partial\Omega \times \Y_i, \qquad i=1,2.
\end{align}

Integrating the first equation in~\eqref{eqn:PDE_n1_n2_mutational_kernel_rescaled} over $\Y_1$ and the second equation over $\Y_2$, and then using property~\eqref{eqn:int mathcal M = 0}, yields
\begin{equation}
\label{eqn:PDE_n1n2_mutational_kernel_rescaled_integrated}
\begin{cases}
\displaystyle{\partial_t \rho^\varepsilon_1 -  \int_{\Y_1} \Delta_x \left[ (D_1 + \alpha_{11}\rho^\varepsilon_1 + \alpha_{12}\rho^\varepsilon_2 ) n^\varepsilon_1 \right] \dy_1 
=  \rr{\int_{\Y_1}  \left( r_1  -\mathcal{K}_1[n^\varepsilon_1,n^\varepsilon_2] \right)\, n^\varepsilon_1  \dy_1}}, 
\\\\
\displaystyle{\partial_t \rho^\varepsilon_2  - \int_{\Y_2} \Delta_x \left[ (D_2 + \alpha_{21}\rho^\varepsilon_1 + \alpha_{22}\rho^\varepsilon_2 ) n^\varepsilon_2 \right] \dy_2 
= \rr{\int_{\Y_2} \left( r_2 -  \mathcal{K}_2[n^\varepsilon_1,n^\varepsilon_2] \right) \, n^\varepsilon_2  \dy_2}}.
\end{cases}
\end{equation}
Moreover, integrating~\eqref{eq:bcs_rescaled} over $\Y_i$ gives the following boundary conditions for the system~\eqref{eqn:PDE_n1n2_mutational_kernel_rescaled_integrated}
\begin{align}\label{eq:bcs_rescaledrho}
    \nabla_x \rho^\varepsilon_i(t, x) \cdot \nu = 0
    \;\; \text{for all }
    (t,x) \in (0,\infty) \times \partial\Omega, \qquad i=1,2.
\end{align} 
Note that~\eqref{eqn:PDE_n1n2_mutational_kernel_rescaled_integrated} is not a closed system for the densities $\rho^\varepsilon_1$ and $\rho^\varepsilon_2$, as it still requires the knowledge of the phenotype densities $n^\varepsilon_1$ and $n^\varepsilon_2$. However, from~\eqref{eqn:PDE_n1_n2_mutational_kernel_rescaled} and~\eqref{eqn:PDE_n1n2_mutational_kernel_rescaled_integrated}, a self-consistent model describing the spatiotemporal dynamics of the densities of the two populations can be formally obtained in the asymptotic regime $\varepsilon \to 0$, as demonstrated in the next section.

\subsubsection{Formal asymptotics for $\varepsilon \to 0$}
Denoting by $n^0_i$ and $\rho^0_i$ the leading order terms of the asymptotic expansions of $n_i^{\varepsilon}$ and $\rho_i^{\varepsilon}$ as $\varepsilon \to 0$, letting $\varepsilon \to 0$ in~\eqref{eqn:PDE_n1_n2_mutational_kernel_rescaled} and recalling~\eqref{eq:mathcal_M_definition}, we find that $n^0_i$ formally satisfies
\begin{equation}
\label{eq:eqn0}
\int_{\Y_i} M_i(y_i|y_i') n^0_i(t,x,y_i')  \dy_i' = n^0_i(t,x,y_i), \qquad i=1,2.
\end{equation}
Moreover, letting $\varepsilon \to 0$ in~\eqref{eqn:PDE_n1n2_mutational_kernel_rescaled_integrated} and~\eqref{eq:bcs_rescaledrho} formally gives the following system
\begin{equation}
\label{eqn:PDE_n1n2_mutational_kernel_rescaled_integrated0}
\begin{cases}
\displaystyle{\partial_t \rho^0_1 -  \int_{\Y_1} \Delta_x \left[ (D_1 + \alpha_{11}\rho^0_1 + \alpha_{12}\rho^0_2 ) n^0_1 \right] \dy_1 
=  \rr{\int_{\Y_1}  \left( r_1 - \mathcal{K}_1[n^0_1,n^0_2] \right) \, n^0_1  \dy_1}}, 
\\\\
\displaystyle{\partial_t \rho^0_2  - \int_{\Y_2} \Delta_x \left[ (D_2 + \alpha_{21}\rho^0_1 + \alpha_{22}\rho^0_2 ) n^0_2 \right] \dy_2 
= \rr{\int_{\Y_2} \left( r_2  -  \mathcal{K}_2[n^0_1,n^0_2] \right) \, n^0_2  \dy_2}},
\end{cases}
\end{equation}
subject to the following boundary conditions
\begin{align*}
    \nabla_x \rho^0_i(t, x) \cdot \nu = 0
    \;\; \text{for all }
    (t,x) \in (0,\infty) \times \partial\Omega, \qquad i=1,2.
\end{align*} 

From~\eqref{eq:eqn0}, recalling that $\int_{\Y_i} n^0_i(t,x,y_i) \dy_i =: \rho^0_i(t,x)$, we find 
\begin{equation}
\label{eq:ansatz}
n^0_i(t,x,y_i) = \rho^0_i(t,x) \, \psi_i(y_i), \qquad i = 1,2,
\end{equation}
where $\psi_i$ is the solution of the following integral problem
\begin{equation}
\label{eq:steadystatenlimpsi}
\begin{cases}
\displaystyle{\int_{\Y_i} M_i(y_i|y_i') \psi_i(y_i')  \dy_i' = \psi_i(y_i),}
\\\\
\displaystyle{\int_{\Y_i} \psi_i(y_i) \dy_i = 1,}
\end{cases}
\qquad i=1,2,
\end{equation}
that is, $\psi_i$ is a normalised eigenfunction, associated with the eigenvalue $1$, of the integral operator $\psi \to \int_{\Y_i} M_i(y_i|y_i') \psi(y_i')  \dy_i'$ mapping $L^1(\Y_i)$ into itself. Under assumptions~\eqref{assMi}, the Krein-Rutman theorem ensures that there exists a unique positive function $\psi_i$ satisfying the integral problem~\eqref{eq:steadystatenlimpsi}. Such a function can be regarded as the phenotype distribution of population $i$ in the quasi-invariant regime of fast phenotype switching.

\begin{remark}\label{rem:psiM}
In particular, note that if the phenotype switching kernel $M_i$ is such that the following additional assumptions hold
\begin{equation}
\label{eq:assumption_M_indipendent_of_y'}
M_i(y_i|y_i')\equiv M_i(y_i), \qquad i = 1,2,
\end{equation}
that is, if the new phenotype states acquired by the population members upon phenotype switching are independent of the original ones, then
\begin{equation}
\label{eq:psiunderassumption_M_indipendent_of_y'}
\psi_i(y_i) = M_i(y_i), \qquad i = 1,2,
\end{equation}
and thus, substituting into~\eqref{eq:ansatz}, we have
\begin{equation}
\label{eq:ansatzM}
n^0_i(t,x,y_i) = \rho^0_i(t,x) \, M_i(y_i), \qquad i = 1,2.
\end{equation}

\end{remark}

Moreover, substituting~\eqref{eq:ansatz} into~\eqref{eqn:PDE_n1n2_mutational_kernel_rescaled_integrated0}, introducing the following definitions
\begin{align}\label{eqn:def<Di><alphaij>}
    &\la{D}_i\ra := \int_{\Y_i} D_i(y_i) \psi_i(y_i)  \dy_i, 
    \quad
    \la{\alpha}_{ij}\ra := \int_{\Y_i} \alpha_{ij}(y_i) \psi_i(y_i)  \dy_i, \qquad i, j = 1,2,
\end{align}
\begin{align}\label{eqn:def<Kij>}
	&\rr{\la r_i \ra :=  \int_{\Y_i}
		r_{i}(y_i) \psi_i(y_i) \dy_i}, \quad 
	\la K_{ij} \ra := \int_{\Y_i}\int_{\Y_j}
	K_{ij}(y_i, y_j') \psi_i(y_i) \psi_j(y_j') \dy_i \dy_j', \qquad i, j = 1,2,
\end{align}
and dropping the superscripts ``0'' for convenience, we obtain the following closed system for the densities $\rho_1$ and $\rho_2$ of the two competing populations 
\begin{equation}
    \label{eqn:SKT_phenotype}
\begin{cases}
\partial_t \rho_1 - \Delta \left[ \left( \la{D}_1\ra + \la{\alpha}_{11}\ra \rho_1 + \la{\alpha}_{12}\ra \rho_2\right)\rho_1 \right]
= 
\rr{\la r_1\ra} \rho_1
- 
\left( \beta_{11}\la{K}_{11}\ra \rho_1 + \beta_{12}\la{K}_{12}\ra \rho_2\right)\rho_1, 
\\ 
\partial_t \rho_2 - \Delta \left[ \left( \la{D}_2\ra + \la{\alpha}_{22}\ra \rho_2 + \la{\alpha}_{21} \ra\rho_1\right)\rho_2 \right]
= 
\rr{\la r_2 \ra} \rho_2
- 
\left( \beta_{22}\la{K}_{22}\ra \rho_2 + \beta_{21}\la{K}_{21}\ra \rho_1\right)\rho_2, 
\end{cases}
\end{equation}
with boundary conditions
\begin{align}\label{eq:bcs_rescaledrho_limit}
\nabla_x \rho_i(t, x) \cdot \nu = 0
\;\; \text{for all }
(t,x) \in (0,\infty) \times \partial\Omega, \qquad i=1,2.
\end{align} 
The system~\eqref{eqn:SKT_phenotype} is of the form of the unstructured SKT model~\eqref{eqn:SKT}, and the roles of the parameters $d_i$, $a_{ij}$, \rr{$r_i$}, and $b_{ij}$ of~\eqref{eqn:SKT} are played in~\eqref{eqn:SKT_phenotype} by the effective parameters $\la{D}_i\ra$, $\la{\alpha}_{ij}\ra$, \rr{$\la r_i\ra$}, and $\beta_{ij}\la{K}_{ij}\ra$ defined via~\eqref{eqn:def<Di><alphaij>} and~\eqref{eqn:def<Kij>}, respectively, which are obtained as continuous weighted averages of the phenotype-dependent functions $D_i$, $\alpha_{ij}$, $r_i$ and $K_{ij}$, with weights given by the phenotype distributions $\psi_i$ and $\psi_j$. 
\rr{\begin{remark}
In this paper, we let the functions $K_{ij}(y_i,y_j)$ and $\alpha_{ij}(y_i)$ in the system~\eqref{eqn:PDE_n1_n2_mutational_kernel} be all non-negative, in line with the assumptions made by Shigesada, Kawasaki, and Teramoto~\citep{Shigesada1979Spatial}, so as to consider scenarios where all population members undergo competitive interactions and move in space with a tendency to avoid competitors. Hence, each of the effective parameters $\la{K}_{ij}\ra$ and $\la{\alpha}_{ij}\ra$ in the limiting system~\eqref{eqn:SKT_phenotype} is non-negative. Yet, the results of pattern formation analysis presented in the remainder of the paper also apply to cases in which $K_{ij}(y_i,y_j)$ and $\alpha_{ij}(y_i)$ take negative values for some values of $y_i$ and $y_j$, provided that the eﬀective parameters $\la{K}_{ij}\ra$ and $\la{\alpha}_{ij}\ra$ are non-negative.
\end{remark}}

\section{Pattern formation analyses}\label{Sec:PatternFormation}
In this section, we carry out pattern formation analyses for the rescaled phenotype-structured SKT model~\eqref{eqn:PDE_n1_n2_mutational_kernel_rescaled} complemented with definitions~\eqref{eqn:K1}-\eqref{eq:mathcal_M_definition} and subject to the boundary conditions~\eqref{eq:bcs_rescaled}. Specifically, we start by performing Turing-type linear stability analysis, characterising coexistence uniform-in-space steady-state solutions (Section~\ref{Sec:Turing_steady-state}) and analysing first the conditions for stability to uniform-in-space perturbations (Section~\ref{Sec:Turing_stability}) and then the conditions for instability to space-dependent perturbations (Section~\ref{Sec:Turing_instability}). While the full problem is analytically intricate, illumination can be provided by investigating the limiting case $\varepsilon \to 0$, which here we study on the formal level. In so doing, we find a Turing instability threshold and we identify the most unstable mode. We then carry out weakly nonlinear analysis on the limiting problem to assess how the critical spatial mode saturates beyond onset of spatial patterns. In particular, through asymptotic expansions in a small control parameter, we derive a closed equation for the evolution of the pattern amplitude in the vicinity of the Turing-type bifurcation (see Section~\ref{sec:wnla}). 

\rr{We remark that, while relying on classical techniques for SKT systems, Turing-type linear and weakly nonlinear analyses are carried out in this section within the additional scope of building an analytical bridge between the rescaled phenotype-structured SKT model~\eqref{eqn:PDE_n1_n2_mutational_kernel_rescaled} and the corresponding phenotype-averaged limiting model~\eqref{eqn:SKT_phenotype}. In particular, since the effective coefficients of the limiting model depend on the phenotype distributions in the quasi-invariant regime of fast phenotype switching, these analyses enable relating phenotype structuring of the two populations to the onset of Turing instability, the selection of the critical spatial mode, and the super- or sub-critical nature of the corresponding bifurcation.}

\subsection{Coexistence uniform-in-space steady-state solutions}
\label{Sec:Turing_steady-state}
Coexistence uniform-in-space steady states of the system~\eqref{eqn:PDE_n1_n2_mutational_kernel_rescaled}, complemented with definitions~\eqref{eqn:K1}-\eqref{eq:mathcal_M_definition}, are pairs $(\overline{n}^\varepsilon_1(x,y_1), \overline{n}^\varepsilon_2(x,y_2)) \equiv (\overline{n}_1^\varepsilon(y_1), \overline{n}_2^\varepsilon(y_2))$ where $\overline{n}_1^\varepsilon \geq 0$ and $\overline{n}_2^\varepsilon \geq 0$ are the components of the solutions to the following system of integral equations
\begin{equation}
\label{eq:steadystate}
\begin{cases}
\displaystyle{\left(r_1 - \mathcal{K}_1[\overline{n}_1^\varepsilon,\overline{n}_2^\varepsilon]\right)\overline{n}_1^\varepsilon +  \frac1\varepsilon \mathcal{M}_1 \left[\overline{n}^\varepsilon_1 \right] = 0},
\\\\
\displaystyle{\left(r_2 - \mathcal{K}_2[\overline{n}_1^\varepsilon,\overline{n}_2^\varepsilon]\right)\overline{n}_2^\varepsilon +  \frac1\varepsilon \mathcal{M}_2 \left[\overline{n}^\varepsilon_2 \right] = 0},
\end{cases}
\end{equation}
with $\overline{n}_1^\varepsilon$ and $\overline{n}_2^\varepsilon$ such that
$$
\overline{\rho}^\varepsilon_1 := \int_{\Y_1} \overline{n}^\varepsilon_1(y_1) \dy_1 > 0 \quad \text{ and } \quad \overline{\rho}^\varepsilon_2 := \int_{\Y_2} \overline{n}^\varepsilon_2(y_2) \dy_2 > 0.
$$
Note that integrating the first equation in~\eqref{eq:steadystate} over $\Y_1$ and the second equation over $\Y_2$, recalling~\eqref{eqn:K1} and~\eqref{eqn:K2}, and then using property~\eqref{eqn:int mathcal M = 0} gives
\begin{equation}
\label{eq:steadystate_INTEGRATED}
\begin{cases}
\displaystyle{\int_0^1 \left[
	\rr{r_1(y_1)}  
	- \beta_{11} \int_{\Y_1} K_{11}(y_1,y_1')\overline{n}^\varepsilon_1(y_1')\dy_1' 
	-  \beta_{12} \int_{\Y_2} K_{12}(y_1,y_2')\overline{n}^\varepsilon_2(y_2')\dy_2' \right]\overline{n}^\varepsilon_1(y_1) \dy_1  = 0},
\\\\
\displaystyle{\int_0^1 \left[
	\rr{r_2(y_2)}
	- \beta_{21} \int_{\Y_1} K_{21}(y_2,y_1')\overline{n}^\varepsilon_1(y_1')\dy_1' 
	- \beta_{22} \int_{\Y_2} K_{22}(y_2,y_2')\overline{n}^\varepsilon_2(y_2')\dy_2'\right] \overline{n}^\varepsilon_2(y_2)\dy_2 = 0}.
\end{cases}
\end{equation}

\paragraph{Formal asymptotics for $\varepsilon\to 0$.} Denoting by $\overline{n}_i$ and $\overline{\rho}_i$ the leading order terms of the asymptotic expansions of $\overline{n}_i^\varepsilon$ and $\overline{\rho}_i^\varepsilon$ as $\varepsilon \to 0$, letting $\varepsilon \to 0$ in~\eqref{eq:steadystate} and recalling~\eqref{eq:mathcal_M_definition}, we find that $\overline{n}_i$ formally satisfies
\begin{equation}
\label{eq:steadystatenlim}
\int_{\Y_i} M_i(y_i|y_i') \overline{n}_i(y_i')  \dy_i' = \overline{n}_i(y_i), \qquad i=1,2,
\end{equation}
while letting $\varepsilon \to 0$ in~\eqref{eq:steadystate_INTEGRATED} formally gives 
\begin{equation}
\label{eq:steadystate_INTEGRATEDlim}
\begin{cases}
\displaystyle{\int_0^1 \left[
	\rr{r_1(y_1)}
	- \beta_{11} \int_{\Y_1} K_{11}(y_1,y_1')\overline{n}_1(y_1')\dy_1' 
	- \beta_{12} \int_{\Y_2} K_{12}(y_1,y_2')\overline{n}_2(y_2')\dy_2' \right]\overline{n}_1(y_1) \dy_1  = 0},
\\\\
\displaystyle{\int_0^1 \left[
	\rr{r_2(y_2)}
	- \beta_{21} \int_{\Y_1} K_{21}(y_2,y_1')\overline{n}_1(y_1')\dy_1' 
	- \beta_{22} \int_{\Y_2} K_{22}(y_2,y_2')\overline{n}_2(y_2')\dy_2'\right] \overline{n}_2(y_2)\dy_2 = 0}.
\end{cases}
\end{equation}
From~\eqref{eq:steadystatenlim}, recalling that $\int_{\Y_i} \overline{n}_i(t,x,y_i) \dy_i =: \overline{\rho}_i$, we find 
\begin{equation}
\label{eq:steadystateansatz}
\overline{n}_i(y_i) = \overline{\rho}_i  \, \psi_i(y_i), \qquad i = 1,2,
\end{equation}
where, as noted in Section~\ref{sec:mod1}, under assumptions~\eqref{assMi}, $\psi_i$ is the unique positive solution of the integral problem~\eqref{eq:steadystatenlimpsi}. Moreover, substituting~\eqref{eq:steadystateansatz} into~\eqref{eq:steadystate_INTEGRATEDlim}, recalling~\eqref{eqn:def<Kij>}, and rearranging terms, we find that $(\overline{\rho}_1, \overline{\rho}_2)$ are the component-wise positive solutions of the following coupled algebraic system
\begin{equation}
\label{eq:steadystate_rho_simplified}
\begin{cases}
\displaystyle{\beta_{11}\la{K}_{11}\ra \overline{\rho}_1 +  \beta_{12}\la {K}_{12}\ra \overline{\rho}_2 = \rr{\la r_1\ra}},
\\
\displaystyle{\beta_{22}\la{K}_{22} \ra \overline{\rho}_2 + \beta_{21}\la {K}_{21}\ra \overline{\rho}_1 = \rr{\la r_2\ra}}.
\end{cases}
\end{equation}
Provided that the additional conditions below hold
\begin{equation}
\label{ass:coex}
\beta_{11}\beta_{22}\la{K}_{11}\ra \la{K}_{22}\ra - \beta_{12}\beta_{21} \la{K}_{12}\ra \la{K}_{21}\ra \neq 0,
\end{equation}
\begin{equation}
\label{ass:poscoex1}
{\rm sgn} \left(\rr{\la r_1\ra} \beta_{22} \la {K}_{22} \ra - \rr{\la r_2\ra} \beta_{12} \la {K}_{12} \ra\right) = {\rm sgn} \left(\beta_{11}\beta_{22}\la{K}_{11}\ra \la{K}_{22}\ra - \beta_{12}\beta_{21} \la{K}_{12}\ra \la{K}_{21}\ra \right),
\end{equation}
and
\begin{equation}
\label{ass:poscoex2}
{\rm sgn} \left(\rr{\la r_2\ra} \beta_{11} \la{K}_{11}\ra - \rr{\la r_1\ra} \beta_{21}\la{K}_{21}\ra\right) = {\rm sgn} \left(\beta_{11}\beta_{22}\la{K}_{11}\ra \la{K}_{22}\ra - \beta_{12}\beta_{21} \la{K}_{12}\ra \la{K}_{21}\ra \right),
\end{equation}
the algebraic system~\eqref{eq:steadystate_rho_simplified} admits a unique component-wise positive solution, that is, 
\begin{align}\label{eq:steadystate_rho}
    \left(\overline{\rho}_1, \overline{\rho}_2\right) = \left( 
    \frac{\rr{\la r_1\ra} \beta_{22} \la {K}_{22} \ra - \rr{\la r_2\ra}  \beta_{12} \la {K}_{12} \ra}{\beta_{11}\beta_{22}\la{K}_{11}\ra \la{K}_{22}\ra - \beta_{12}\beta_{21} \la{K}_{12}\ra \la{K}_{21}\ra}
    ,
    \frac{\rr{\la r_2\ra}  \beta_{11} \la{K}_{11}\ra - \rr{\la r_1\ra}\beta_{21}\la{K}_{21}\ra}{\beta_{11}\beta_{22}\la{K}_{11}\ra \la{K}_{22}\ra - \beta_{12}\beta_{21} \la{K}_{12}\ra \la{K}_{21}\ra}
    \right).
\end{align}

In summary, we formally found that, under assumptions~\eqref{assMi} and the additional assumptions \eqref{ass:coex}-\eqref{ass:poscoex2}, in the asymptotic regime $\varepsilon\to 0$, there is a unique coexistence uniform-in-space steady-state solution of the system~\eqref{eqn:PDE_n1_n2_mutational_kernel_rescaled}, complemented with definitions~\eqref{eqn:K1}-\eqref{eq:mathcal_M_definition}, the components of which are of the form~\eqref{eq:steadystateansatz} with $\psi_i$ given by the integral problem~\eqref{eq:steadystatenlimpsi} and $\overline{\rho}_i$ given by~\eqref{eq:steadystate_rho}.

\subsection{Stability to uniform-in-space perturbations}\label{Sec:Turing_stability}
To investigate the stability of the coexistence uniform-in-space steady state to small uniform-in-space perturbations, we make the ansatz
\begin{equation}
\label{eq:ansatzls1}
n^\varepsilon_i(t,x,y_i) \equiv n^\varepsilon_i(t,y_i) =
\overline{n}_i^\varepsilon(y_i) + \widetilde{\rho}_i^\varepsilon {\varphi_i}^\varepsilon(y_i) \exp\left(\lambda^\varepsilon t\right), 
\qquad i=1,2,
\end{equation}
where 
\begin{align}\label{eqn:unit_integral_perturbation}
    0<
    |\widetilde{\rho_i}^\varepsilon | \ll 1
       \quad
    \text{ and } 
    \quad  
    \int_{\Y_i} {\varphi_i}^\varepsilon(y_i) \dy_i = 1,
    \qquad 
    i = 1,2.
\end{align} 
Substituting the ansatz~\eqref{eq:ansatzls1} into the rescaled phenotype-structured SKT model~\eqref{eqn:PDE_n1_n2_mutational_kernel_rescaled}, using the fact that $(\overline{n}_1^\varepsilon(y_1), \overline{n}_2^\varepsilon(y_2))$ satisfies~\eqref{eq:steadystate}, and retaining only $O(\widetilde{\rho_i}^\varepsilon)$ terms yields
\begin{equation}
\label{eq:steadystate_perturbed_only_phenotype}
\begin{cases}
\displaystyle{\lambda^\varepsilon \widetilde{\rho}_1^\varepsilon {\varphi}_1^\varepsilon 
=  -
\mathcal{K}_1[ \widetilde{\rho}_1^\varepsilon {\varphi}_1^\varepsilon , \widetilde{\rho}_2^\varepsilon {\varphi}_2^\varepsilon]
\overline{n}_1^\varepsilon 
+
\left(r_1
- \mathcal{K}_1[\overline{n}_1^\varepsilon,\overline{n}_2^\varepsilon] 
\right)
\widetilde{\rho}_1^\varepsilon {\varphi}_1^\varepsilon 
+ 
\frac1\varepsilon \widetilde{\rho}_1^\varepsilon 
\mathcal{M}_1[{\varphi}_1^\varepsilon ]},
\\\\
\displaystyle{\lambda^\varepsilon \widetilde{\rho}_2^\varepsilon {\varphi}_2^\varepsilon 
= -
\mathcal{K}_2[\widetilde{\rho}_1^\varepsilon {\varphi}_1^\varepsilon,  \widetilde{\rho}_2^\varepsilon {\varphi}_2^\varepsilon ]
\overline{n}_2^\varepsilon 
+
\left(r_2
- \mathcal{K}_2[\overline{n}_1^\varepsilon,\overline{n}_2^\varepsilon] 
\right)
\widetilde{\rho}_2^\varepsilon {\varphi}_2^\varepsilon 
+ 
\frac1\varepsilon \widetilde{\rho}_2^\varepsilon 
\mathcal{M}_2[{\varphi}_2^\varepsilon ]}.
\end{cases}
\end{equation}
Moreover, integrating the first equation in~\eqref{eq:steadystate_perturbed_only_phenotype} over $\Y_1$ and the second equation over $\Y_2$, recalling~\eqref{eqn:K1} and~\eqref{eqn:K2}, and then using property~\eqref{eqn:int mathcal M = 0} along with the normalisation conditions~\eqref{eqn:unit_integral_perturbation} on~${\varphi_i}^\varepsilon(y_i)$ for $i=1,2$, we find  
\begin{equation}
\label{eq:steadystate_perturbed_only_phenotype_integrated}
\begin{cases}
\displaystyle{\lambda^\varepsilon \widetilde{\rho}_1^\varepsilon 
= -
\int_{\Y_1}
\mathcal{K}_1[ \widetilde{\rho}_1^\varepsilon {\varphi}_1^\varepsilon , \widetilde{\rho}_2^\varepsilon {\varphi}_2^\varepsilon]
\overline{n}_1^\varepsilon \dy_1
\rr{
+ \int_{\Y_1} \left( r_1 - \mathcal{K}_1[\overline{n}_1^\varepsilon,\overline{n}_2^\varepsilon] \right)
\widetilde{\rho}_1^\varepsilon {\varphi}_1^\varepsilon 
\dy_1}},
\\\\
\displaystyle{\lambda^\varepsilon \widetilde{\rho}_2^\varepsilon 
= -
\int_{\Y_2}
\mathcal{K}_2[\widetilde{\rho}_1^\varepsilon {\varphi}_1^\varepsilon,  \widetilde{\rho}_2^\varepsilon {\varphi}_2^\varepsilon ]
\overline{n}_2^\varepsilon \dy_2
\rr{+
\int_{\Y_2} 
\left( r_2
-
\mathcal{K}_2[\overline{n}_1^\varepsilon,\overline{n}_2^\varepsilon] \right)
\widetilde{\rho}_2^\varepsilon {\varphi}_2^\varepsilon \dy_2}.}
\end{cases}
\end{equation}

\paragraph{Formal asymptotics for $\varepsilon\to 0$.}
Denoting by $\overline{\rho}_i$, $\widetilde{\rho_i}$, $\lambda_i$, and $\varphi_i$ the leading order terms of the asymptotic expansions of $\overline{\rho}_i^\varepsilon$, $\widetilde{\rho_i}^\varepsilon$, ${\lambda_i}^\varepsilon$, and ${\varphi_i}^\varepsilon$ as $\varepsilon \to 0$, letting $\varepsilon \to 0$ in~\eqref{eq:steadystate_perturbed_only_phenotype} and recalling~\eqref{eq:mathcal_M_definition}, we formally find that $\varphi_i \equiv \psi_i$, where $\psi_i$ is the unique positive solution of the integral problem \eqref{eq:steadystatenlimpsi}. Furthermore, letting $\varepsilon \to 0$ in~\eqref{eq:steadystate_perturbed_only_phenotype_integrated}, rearranging terms, using~\eqref{eq:steadystate_rho_simplified}, and recalling~\eqref{eqn:def<Kij>}, we formally obtain the following eigenvalue problem
$$
		J{\brho} = \lambda {\brho}, \quad \text{ with } \quad
		{\brho}=\left(\begin{array}{cc}
			\widetilde{\rho_1}\\
			\widetilde{\rho_2}
		\end{array}\right),
$$
where
\begin{equation}\label{J}
		J:= - \left(\begin{array}{ll}
			\beta_{11}\la K_{11} \ra \overline{\rho}_1 & \beta_{12}\la K_{12} \ra\overline{\rho}_1 \\
			\beta_{21}\la K_{21} \ra\overline{\rho}_2 & \beta_{22}\la K_{22} \ra\overline{\rho}_2
		\end{array}\right),
	\end{equation}
and $\overline{\rho}_1$ and $\overline{\rho}_2$ are defined via~\eqref{eq:steadystate_rho}. The characteristic equation of the above eigenvalue problem is
$$
		\lambda^2-{\rm tr}(J)\lambda+ {\rm det}(J)=0.
$$
Under the non-negativity assumptions on the competition kernels $K_{ij}$,  all the entries of $J$ are negative and thus a necessary and sufficient condition ensuring stability to uniform-in-space perturbations is ${\rm det} (J)>0$. Hence, in the remainder of the article, we assume
\begin{equation}\label{C_cond3}
\beta_{11}\beta_{22}\la{K}_{11}\ra\la{K}_{22}\ra-	\beta_{12}\beta_{21}\la{K}_{12}\ra\la{K}_{21}\ra>0,
\end{equation}
and then, in order for conditions~\eqref{ass:poscoex1} and \eqref{ass:poscoex2} to hold, we also assume
\begin{equation}\label{C_cond12}
	\rr{\la r_1\ra} \beta_{22}\la{K}_{22}\ra - \rr{\la r_2\ra}\beta_{12}\la{K}_{12}\ra >0, \quad \rr{\la r_2\ra}\beta_{11}\la{K}_{11}\ra -\rr{\la r_1\ra}\beta_{21}\la{K}_{21}\ra >0.
\end{equation}

\subsection{Instability to space-dependent perturbations}\label{Sec:Turing_instability}
We then restrict to a one-dimensional spatial scenario, i.e. $\Omega \equiv (-\pi,\pi)$, and investigate instability of the coexistence uniform-in-space steady state to small space-dependent perturbations by making the ansatz
\begin{equation}
\label{eq:ansatzls2}
n^\varepsilon_i(t,x,y_i) =
\overline{n}_i^\varepsilon(y_i) + \widetilde{\rho}_i^\varepsilon {\varphi_i}^\varepsilon(y_i) \phi_k(x) \exp\left(\lambda^\varepsilon t\right), 
\qquad i=1,2,
\end{equation}
where $\widetilde{\rho}_i^\varepsilon$ and ${\varphi_i}^\varepsilon$ satisfy assumptions \eqref{eqn:unit_integral_perturbation} and $\phi_k(x)$ are the eigenfunctions of the Laplace operator, acting on functions defined on $(-\pi,\pi)$ and subject to homogeneous Neumann boundary conditions, indexed by the wavenumber $k>0$.

Substituting the ansatz~\eqref{eq:ansatzls2} into the rescaled phenotype-structured SKT model~\eqref{eqn:PDE_n1_n2_mutational_kernel_rescaled}, using the fact that $(\overline{n}_1^\varepsilon(y_1), \overline{n}_2^\varepsilon(y_2))$ satisfies~\eqref{eq:steadystate}, and retaining only $O(\widetilde{\rho_i}^\varepsilon)$ terms gives
\begin{equation}
\label{eqn:stability_spatial_perturbation_O(epsilon)}
\begin{cases}
\lambda^\varepsilon \widetilde{\rho}_1^\varepsilon  {\varphi}_1^\varepsilon(y_1) 
+ \rr{ \left[ \alpha_{11}(y_1) \widetilde{\rho}_1^\varepsilon + \alpha_{12}(y_1) \widetilde{\rho}_2^\varepsilon \right] \overline{n}_1^\varepsilon(y_1) k^2 }
\\
\quad \phantom{\lambda^\varepsilon \widetilde{\rho}_1^\varepsilon  {\varphi}_1^\varepsilon(y_1) } +
\left[ D_1(y_1) 
+ \alpha_{11}(y_1) \overline{\rho}_1^\varepsilon  + \alpha_{12}(y_1) \overline{\rho}_2^\varepsilon \right] \widetilde{\rho}_1^\varepsilon   {\varphi}_1^\varepsilon(y_1)  
k^2 \\
\qquad \phantom{\lambda^\varepsilon \widetilde{\rho}_1^\varepsilon  {\varphi}_1^\varepsilon(y_1) } 
= - \mathcal{K}_1[ \widetilde{\rho}_1^\varepsilon {\varphi}_1^\varepsilon(y_1) , \widetilde{\rho}_2^\varepsilon {\varphi}_2^\varepsilon(y_2)]
\overline{n}_1^\varepsilon(y_1) 
+
\left(\rr{r_1(y_1)} 
- \mathcal{K}_1[\overline{n}_1^\varepsilon(y_1),\overline{n}_2^\varepsilon(y_2)] 
\right)
\widetilde{\rho}_1^\varepsilon {\varphi}_1^\varepsilon(y_1) 
\\
\qquad \quad \phantom{\lambda^\varepsilon \widetilde{\rho}_1^\varepsilon  {\varphi}_1^\varepsilon(y_1) } 
+ \dfrac1\varepsilon \widetilde{\rho}_1^\varepsilon 
\mathcal{M}_1[{\varphi}_1^\varepsilon ],
\\\\
\lambda^\varepsilon \widetilde{\rho}_2^\varepsilon {\varphi}_2^\varepsilon(y_2)
+ \rr{ \left[\alpha_{21}(y_2) \widetilde{\rho}_1^\varepsilon
+ \alpha_{22}(y_2) \widetilde{\rho}_2^\varepsilon \right]
\overline{n}_2^\varepsilon(y_2) k^2} \\
\quad \phantom{\lambda^\varepsilon \widetilde{\rho}_2^\varepsilon {\varphi}_2^\varepsilon(y_2) } 
+ \left[D_2(y_2)
+ \alpha_{21}(y_2) \overline{\rho}_1^\varepsilon
+ \alpha_{22}(y_2) \overline{\rho}_2^\varepsilon \right]
\widetilde{\rho}_2^\varepsilon {\varphi}_2^\varepsilon(y_2) k^2 \\
\qquad  \phantom{\lambda^\varepsilon \widetilde{\rho}_2^\varepsilon {\varphi}_2^\varepsilon(y_2) } 
= - \mathcal{K}_2[
\widetilde{\rho}_1^\varepsilon {\varphi}_1^\varepsilon(y_1),
\widetilde{\rho}_2^\varepsilon {\varphi}_2^\varepsilon(y_2)]
\overline{n}_2^\varepsilon(y_2)
+ \left(\rr{r_2(y_2)} 
- \mathcal{K}_2[\overline{n}_1^\varepsilon(y_1),\overline{n}_2^\varepsilon(y_2)]\right)
\widetilde{\rho}_2^\varepsilon {\varphi}_2^\varepsilon(y_2) 
\\
\qquad\quad \phantom{\lambda^\varepsilon \widetilde{\rho}_2^\varepsilon {\varphi}_2^\varepsilon(y_2) } 
+ \dfrac{1}{\varepsilon} \widetilde{\rho}_2^\varepsilon 
\mathcal{M}_2[{\varphi}_2^\varepsilon ].
\end{cases}
\end{equation}
Moreover, integrating the first equation in~\eqref{eqn:stability_spatial_perturbation_O(epsilon)} over $\Y_1$ and the second equation over $\Y_2$, and then using property~\eqref{eqn:int mathcal M = 0} along with the normalisation conditions~\eqref{eqn:unit_integral_perturbation} on~${\varphi_i}^\varepsilon(y_i)$ for $i=1,2$, we find 
\begin{equation}
\label{eqn:stability_spatial_perturbation_O(epsilon)_integrated}
\begin{cases}
\displaystyle{\lambda^\varepsilon \widetilde{\rho}_1^\varepsilon  
+ k^2 \int_{\Y_1}
\rr{\left[ (\alpha_{11}(y_1) \widetilde{\rho}_1^\varepsilon
+ \alpha_{12}(y_1) \widetilde{\rho}_2^\varepsilon ) \right]
\overline{n}_1^\varepsilon(y_1) \dy_1} }
\\[0.9em]
\hspace{0.5em}
\displaystyle{+\, k^2 \int_{\Y_1}
\left[ D_1(y_1) + \alpha_{11}(y_1) \overline{\rho}_1^\varepsilon
+ \alpha_{12}(y_1) \overline{\rho}_2^\varepsilon \right]
\widetilde{\rho}_1^\varepsilon {\varphi}_1^\varepsilon(y_1)
\dy_1} \\[0.9em]
\quad
\displaystyle{= - \int_{\Y_1}
\mathcal{K}_1[ \widetilde{\rho}_1^\varepsilon {\varphi}_1^\varepsilon(y_1),
\widetilde{\rho}_2^\varepsilon {\varphi}_2^\varepsilon(y_2)]
\overline{n}_1^\varepsilon (y_1) \dy_1}
\rr{\displaystyle{+
\int_{\Y_1} \left( r_1(y_1) - \mathcal{K}_1[\overline{n}_1^\varepsilon(y_1),\overline{n}_2^\varepsilon(y_2)]\right)
\widetilde{\rho}_1^\varepsilon {\varphi}_1^\varepsilon(y_1) \dy_1}},
\\\\
\displaystyle{\lambda^\varepsilon \widetilde{\rho}_2^\varepsilon  
+ k^2 \int_{\Y_2}
\rr{ \left[ \alpha_{21}(y_2) \widetilde{\rho}_1^\varepsilon 
+ \alpha_{22}(y_2) \widetilde{\rho}_2^\varepsilon \right]
\overline{n}_2^\varepsilon(y_2) \dy_2} }
\\[0.9em]
\hspace{0.5em}
\displaystyle{+\, k^2 \int_{\Y_2}
\left[ D_2(y_2) + \alpha_{21}(y_2) \overline{\rho}_1^\varepsilon
+ \alpha_{22}(y_2) \overline{\rho}_2^\varepsilon \right]
\widetilde{\rho}_2^\varepsilon {\varphi}_2^\varepsilon(y_2)
\dy_2} 
\\[0.9em]
\quad
\displaystyle{= - \int_{\Y_2}
\mathcal{K}_2[ \widetilde{\rho}_1^\varepsilon {\varphi}_1^\varepsilon(y_1),
\widetilde{\rho}_2^\varepsilon {\varphi}_2^\varepsilon(y_2)]
\overline{n}_2^\varepsilon (y_2) \dy_2}
\rr{\displaystyle{+
\int_{\Y_2} \left( r_2(y_2) - \mathcal{K}_2[\overline{n}_1^\varepsilon(y_1),\overline{n}_2^\varepsilon(y_2)]\right)
\widetilde{\rho}_2^\varepsilon {\varphi}_2^\varepsilon(y_2) \dy_2}}.
\end{cases}
\end{equation}

\paragraph{Formal asymptotics for $\varepsilon\to 0$.} Denoting by $\overline{\rho}_i$, $\widetilde{\rho_i}$, $\lambda_i$, and $\varphi_i$ the leading order terms of the asymptotic expansions of $\overline{\rho}_i^\varepsilon$, $\widetilde{\rho_i}^\varepsilon$, ${\lambda_i}^\varepsilon$, and ${\varphi_i}^\varepsilon$ as $\varepsilon \to 0$, letting $\varepsilon \to 0$ in~\eqref{eqn:stability_spatial_perturbation_O(epsilon)} and recalling~\eqref{eq:mathcal_M_definition}, we formally find that $\varphi_i \equiv \psi_i$, where $\psi_i$ is the unique positive solution of the integral problem \eqref{eq:steadystatenlimpsi}. Moreover, letting $\varepsilon \to 0$ in~\eqref{eqn:stability_spatial_perturbation_O(epsilon)_integrated}, rearranging terms, using~\eqref{eq:steadystate_rho_simplified}, and recalling~\eqref{eqn:def<Kij>} and~\eqref{eqn:def<Di><alphaij>}, we formally obtain the following eigenvalue problem
\begin{equation*}
		\mathcal{L}{\brho} = \lambda {\brho},\quad {\rm with}\quad  \mathcal{L}=J-k^2 D,\qquad  
		{\brho}=\left(\begin{array}{cc}
			\widetilde{\rho_1}\\
			\widetilde{\rho_2}
		\end{array}\right),
	\end{equation*}
where $J$ is defined via~\eqref{J},
	\begin{equation}\label{eqn:D}
		D:=\left(\begin{array}{ll}
			\la{D}_1\ra +2\la{\alpha}_{11}\ra \overline{\rho}_1 +\la{\alpha}_{12}\ra \overline{\rho}_2 & \la{\alpha}_{12}\ra \overline{\rho}_1\\
			\la{\alpha}_{21}\ra \overline{\rho}_2 & 	\la{D}_2\ra +2\la{\alpha}_{22}\ra \overline{\rho}_2+ \la{\alpha}_{21}\ra \overline{\rho}_1
		\end{array}\right),
	\end{equation}
and $\overline{\rho}_1$ and $\overline{\rho}_2$ are defined via~\eqref{eq:steadystate_rho}. Note that, under the non-negativity assumptions that we are making here on the functions $D_i$ and $\alpha_{ij}$, the quantities $\la{D}_i\ra$ and $\la{\alpha}_{ij}\ra$ defined via~\eqref{eqn:def<Di><alphaij>} are also non-negative and ${\rm det} (D)>0$.
 The characteristic equation of the above eigenvalue problem is
	\begin{equation}\label{disp_rel}
		\lambda^2+(k^2{\rm tr}(D)-{\rm tr}(J))\lambda+h(k^2)=0,
	\end{equation}
	where
	\begin{equation*}
		h(k^2)=\det(D)k^4+qk^2+\det(J)
	\end{equation*}	
	and $q$ is the following constant 
	\begin{equation}\label{q_comp}
		\begin{split}
			q =&-J_{11}D_{22}-J_{22}D_{11}+J_{12}D_{21}+J_{21}D_{12}\\	=&\la{\alpha}_{21}\ra \overline{\rho}_1(\beta_{11}\la K_{11} \ra \overline{\rho}_1 - \beta_{12}\la K_{12} \ra\overline{\rho}_2)+
			\la{\alpha}_{12}\ra \overline{\rho}_2(\beta_{22}\la K_{22} \ra\overline{\rho}_2 - \beta_{21}\la K_{21} \ra \overline{\rho}_1)\\
			&+\beta_{11}\la K_{11} \ra \overline{\rho}_1 (\la{D}_2\ra +2\la{\alpha}_{22}\ra \overline{\rho}_2)
			+\beta_{22}\la K_{22} \ra\overline{\rho}_2(\la{D}_1\ra +2\la{\alpha}_{11}\ra \overline{\rho}_1),
		\end{split}
	\end{equation}	
   where $J_{ij}$ and $D_{ij}$ denote the $(i,j)$-entries of the matrices $J$ and $D$ defined~via~\eqref{J} and~\eqref{eqn:D}. Since ${\rm tr}(D)>0$ and ${\rm tr}(J)<0$, the linear coefficient $k^2{\rm tr}(D)-{\rm tr}(J)$ of $\lambda$ in \eqref{disp_rel} is positive; hence, for space-dependent perturbations to grow over time (i.e. for Turing instability to occur), it is necessary that $h(k^2)<0$ for some $k^2$. In particular, since $h(k^2)$ is an upward-opening parabola, we observe a bifurcation when
	\begin{equation}\label{Tur_cond_}
		\min(h(k^2))=0 \quad {\rm with}\ k^2=-\frac{q}{2\det(D)},
	\end{equation}
thus, since ${\rm det} (D)>0$, we have that $q<0$ is a necessary condition for Turing instability to occur. Under assumptions~\eqref{C_cond3} and \eqref{C_cond12}, one can easily verify that only one of the following conditions can be met
	\begin{equation*}
		\beta_{11}\la K_{11} \ra\overline{\rho}_1-\beta_{12}\la K_{12} \ra\overline{\rho}_2<0 
        \quad
        {\rm or}\quad 
        \beta_{22}\la K_{22} \ra\overline{\rho}_2 -\beta_{21}\la K_{21} \ra \overline{\rho}_1<0.
	\end{equation*}
If, without loss of generality, we assume
\begin{equation} \label{furth_cond_local}
		\beta_{11}\la K_{11} \ra\overline{\rho}_1-\beta_{12}\la K_{12} \ra\overline{\rho}_2<0 \quad \text{ and } \quad \beta_{22}\la K_{22} \ra\overline{\rho}_2 -\beta_{21}\la K_{21} \ra \overline{\rho}_1>0,
\end{equation}
then the condition $q<0$ holds only if $\la{\alpha}_{21}\ra$ is sufficiently large, that is, increasing the value of the cross-diffusion term $\la{\alpha}_{21}\ra$ will promote the occurrence of Turing instability, whereas increasing the value of the other cross-diffusion term $\la{\alpha}_{12}\ra$ will have a stabilising effect.
Imposing the condition $q<0$ leads to $\la{\alpha}_{21}\ra>Q_2/Q_1$, where 
\begin{equation*}
Q_1:=\overline{\rho}_1(\beta_{11}\la K_{11} \ra \overline{\rho}_1 - \beta_{12}\la K_{12} \ra\overline{\rho}_2)
\end{equation*}
and
\begin{eqnarray*}
&&Q_2:=\la{\alpha}_{12}\ra\overline{\rho}_2
    (\beta_{21}\la K_{21} \ra \overline{\rho}_1 - \beta_{22}\la K_{22} \ra\overline{\rho}_2)
	-\beta_{11}\la K_{11} \ra \overline{\rho}_1 (\la{D}_2\ra +2 \la{\alpha}_{22}\ra \overline{\rho}_2)\\
	&& \qquad \quad -\beta_{22}\la K_{22} \ra\overline{\rho}_2
	(\la{D}_1\ra +2\la{\alpha}_{11}\ra \overline{\rho}_1).
\end{eqnarray*}
Note that $Q_2/Q_1>0$. We thus seek $\xi>0$ such that if $\la{\alpha}_{21}\ra=Q_2/Q_1+\xi$ then condition \eqref{Tur_cond_} is met. This leads to the problem of finding the zeros of the following polynomial 
\begin{equation*}
	\begin{split}
	Q_1^2\xi^2&-4\det(J)(\la{D}_1\ra+2\la{\alpha}_{11}\ra\overline{\rho}_1)\overline{\rho}_1 \xi\\ &-4\det(J)\left((\la{D}_1\ra +2\la{\alpha}_{11}\ra \overline{\rho}_1 +\la{\alpha}_{12}\ra \overline{\rho}_2)(\la{D}_2\ra +2\la{\alpha}_{22}\ra \overline{\rho}_2)+(\la{D}_1\ra +2\la{\alpha}_{11}\ra \overline{\rho}_1 )\overline{\rho}_1 \frac{Q_2}{Q_1}\right).
	\end{split}
\end{equation*}
According to the \rr{Descartes'} rule of signs, this polynomial admits a unique positive root, $\xi^+$. Hence, the critical value of $\la{\alpha}_{21}\ra$ corresponding to a Turing-type bifurcation leading to pattern formation is
\begin{equation}\label{TT_comp_}
	\la{\alpha}_{21}\ra_c :=\frac{Q_2}{Q_1}+\xi^+,
\end{equation}
and from~\eqref{Tur_cond_} we see that the wavenumber of the critical (i.e. the most unstable) spatial mode, $k_c$, is
\begin{equation}\label{kc_comp}
	k_c :=\sqrt{-\frac{q_c}{2\det(D_c)}},
\end{equation}
where $D_c$ and $q_c$ are obtained by setting $\la{\alpha}_{21}\ra=\la{\alpha}_{21}\ra_c$ in the definition~\eqref{eqn:D} of $D$ and the definition~\eqref{q_comp} of $q$, respectively.
\rr{Note that $k_c$ is obtained by regarding the wavenumber as a continuous variable. Since $\Omega\equiv(-\pi,\pi)$ and homogeneous Neumann boundary conditions are imposed, the admissible wavenumbers are $k_m=m/2$, $m\in\mathbb{N}_0$. Therefore, the onset of Turing instability on the bounded domain requires at least one admissible wavenumber to lie within the instability band.}

\subsection{Weakly nonlinear analysis}\label{sec:wnla}
To assess how the critical mode, $k_c$, given by~\eqref{kc_comp}, saturates beyond onset of spatial patterns, in the vein of~\citep{gambino2012turing}, we now carry out weakly nonlinear analysis for the limiting problem \eqref{eqn:SKT_phenotype}-\eqref{eq:bcs_rescaledrho_limit} in the vicinity of the bifurcation threshold, $\la \alpha_{21}\ra_c$, given by~\eqref{TT_comp_}. 

In more detail, to gain a more complete understanding of how the solution $(\rho_1, \rho_2)$ to the limiting system \eqref{eqn:SKT_phenotype} evolves from the uniform steady state $(\overline{\rho}_1, \overline{\rho}_2)$ given by~\eqref{eq:steadystate_rho}, we set
$$
w 
=
\left(
\begin{array}{c}
w_1\\
w_2
\end{array}
\right)
:= \left(
\begin{array}{c}
\rho_1 - \overline{\rho}_1\\
\rho_2 - \overline{\rho}_2
\end{array}
\right)
$$
and, from the system \eqref{eqn:SKT_phenotype} subject to the boundary conditions~\eqref{eq:bcs_rescaledrho_limit}, we derive the following system
\begin{equation}\label{eqn:WNL-system-linear-nonlinear}
\partial_t \w = \mathscr{L}\w + \nabla^2 Q_D(\w,\w) + Q_K(\w,\w)
\end{equation}
subject to homogeneous Neumann boundary conditions. In the system~\eqref{eqn:WNL-system-linear-nonlinear}, the operator $\mathscr{L}$ is defined as $\mathscr{L} := J + D \nabla^2$, with $J$ and $D$ defined via \eqref{J} and \eqref{eqn:D}, respectively, while $Q_D$ and $Q_K$ comprise nonlinear (quadratic) terms and are defined as 
$$
Q_D({z},{z}') := \left(
\begin{array}{c}
\la \alpha_{11} \ra z_1 z_1' + \la \alpha_{12} \ra z_1 z_2' \\
\la \alpha_{21} \ra z_1' z_2 + \la \alpha_{22} \ra z_2 z_2'
\end{array}
\right)
\quad\text{and}\quad
Q_K({z},{z}') := -\left(
\begin{array}{c}
\beta_{11} \la K_{11} \ra z_1 z_1' + \beta_{12} \la K_{12} \ra z_1 z_2' \\
\beta_{21}\la K_{21} \ra z_1' z_2 + \beta_{22}\la K_{22} \ra z_2 z_2'
\end{array}
\right)
$$
for ${z} = \left(z_1, z_2\right)^{\rm T}$ and ${z}' = \left(z_1', z_2'\right)^{\rm T}$. We then introduce a control parameter $\delta$, defined via
\begin{equation}\label{contr_par}
\delta^2:=\frac{\langle \alpha_{21}\rangle-\langle \alpha_{21}\rangle^c}{\langle \alpha_{21}\rangle^c},
\end{equation}
which measures a relative distance of $\la\alpha_{21}\ra$ from the bifurcation threshold $\la\alpha_{21}\ra_c$. Since in the vicinity of the Turing-type bifurcation (i.e. when $\delta\ll 1$) slow modes dominate the dynamics, we focus on the slow temporal scale $\tau=\delta^2 t$ corresponding to the evolution of the pattern amplitude near onset. Rescaling time and recalling definition~\eqref{contr_par}, from system \eqref{eqn:WNL-system-linear-nonlinear} we obtain the following rescaled system
\begin{eqnarray}\label{eqn:WNL-system-linear-nonlinear_rescaled}
\delta^2 \partial_{\tau} \w &=& \mathscr{L}^c\w + \nabla^2 Q_{D_c}(\w,\w) + Q_K(\w,\w) \nonumber
\\
&& + \,
\delta^2
\left(
\begin{array}{cc}
0 & 0 \\
\la\alpha_{21}\ra_c \overline{\rho}_2 & \la\alpha_{21}\ra_c \overline{\rho}_1
\end{array}
\right)
\nabla^2\w
+
\delta^2
\nabla^2
\left(
\begin{array}{c}
0 \\
\la\alpha_{21}\ra_c w_1w_2
\end{array}
\right)
\end{eqnarray}
subject to homogeneous Neumann boundary conditions. In the system~\eqref{eqn:WNL-system-linear-nonlinear_rescaled}, the operator $\mathscr{L}^c$ is defined as $\mathscr{L}^c := J + D_c\nabla^2$ with 
	\begin{equation*}\label{eqn:Dc}
	D_c:=\left(\begin{array}{ll}
	\la{D}_1\ra +2\la{\alpha}_{11}\ra \overline{\rho}_1 +\la{\alpha}_{12}\ra \overline{\rho}_2 & \la{\alpha}_{12}\ra \overline{\rho}_1\\
	\la{\alpha}_{21}\ra_c \overline{\rho}_2 & 	\la{D}_2\ra +2\la{\alpha}_{22}\ra \overline{\rho}_2+ \la{\alpha}_{21}\ra_c \overline{\rho}_1
	\end{array}\right)
	\end{equation*}
and
\begin{equation*}
Q_{D_c}({z},{z}') := \left(
\begin{array}{c}
\la \alpha_{11} \ra z_1 z_1' + \la \alpha_{12} \ra z_1 z_2' \\
\la \alpha_{21} \ra^c z_1' z_2 + \la \alpha_{22} \ra z_2 z_2'
\end{array}
\right) \quad \text{for} \quad {z} = \left(z_1, z_2\right)^{\rm T} \;\; \text{ and } \;\; {z}' = \left(z_1', z_2'\right)^{\rm T}.
\end{equation*}

We seek solutions to the system \eqref{eqn:WNL-system-linear-nonlinear_rescaled} in the form of asymptotic expansions in the small control parameter $\delta$, that is, solutions of the form
\begin{equation}\label{expansion_sol}
w=\delta w_1+\delta^2 w_2+\delta^3 w_3+O(\delta^4).
\end{equation}
Substituting~\eqref{expansion_sol} into \eqref{eqn:WNL-system-linear-nonlinear_rescaled} and equating the coefficients of first-, second-, and third-order terms in $\delta$, we obtain a hierarchy of equations for the functions $w_1$, $w_2$, and $w_3$ in~\eqref{expansion_sol}. 

First, equating the $O(\delta)$-coefficients, we obtain the stationary homogeneous problem given by the system
\begin{equation*}
\mathscr{L}^c w_1=0
\end{equation*}
subject to homogeneous Neumann boundary conditions, the solution of which can be explicitly calculated and is given by
\begin{equation}\label{sol1}
w_1(\tau,x) =A(\tau)\,{v}\cos(k_c x),\qquad {v}:=\left(\begin{array}{c}1\\v_2\end{array}\right) \ {\rm and}\ v_2=\frac{J_{2,1}+k_c^2D_{c 2,1}}{-J_{2,2}+k_c^2D_{c 2,2}},
\end{equation}
where $A(\tau)$ is the pattern amplitude.

Then, equating the $O(\delta^2)$-coefficients, we obtain the stationary problem given by the system
$
\mathscr{L}^c w_2 + \nabla^2 Q_{D_c}(\w_1, \w_1) + Q_K(\w_1, \w_1) = 0
$,
which can be more conveniently rewritten as
\begin{equation}\label{ord_ep2}
\mathscr{L}^c w_2 = {F}, \qquad
{F} := A^2(\tau) \left[
\frac{1}{2}
Q_K(\vv,\vv)
+
\left(
2 k_c^2
Q_{D_c}(\vv, \vv)
+
\frac{1}{2}
Q_K(\vv,\vv)
\right)
\cos(2 k_c x)
\right],
\end{equation}
with $\vv$ defined via \eqref{sol1}, subject to homogeneous Neumann boundary conditions. Since the forcing term ${F}$ in the system \eqref{ord_ep2} contains only modes with wave-numbers $0$ (i.e. constants) and
$2k_c$, no resonance with the critical mode $k_c$ occurs and the solvability condition is thus automatically satisfied. The solution is of the following form
\begin{equation*}
w_2(\tau,x) =
A(\tau)^2
\left(
w_{20}
+
w_{22} \cos(2 k_c x)
\right),
\end{equation*}
where $w_{20} \in\R^2$ and $w_{22} \in\R^2$ are determined by the model parameters.

Finally, equating the $O(\delta^3)$-coefficients, we obtain the problem given by the system
\begin{equation}\label{ord_ep3}
\mathscr{L}^c w_3={G}, \qquad
{G}=\left(\frac{{\rm d}A}{{\rm d}\tau}{{v}}
+A {G}_1^{(1)}+A^3{G}_1^{(3)}\right)\cos(k_c x)+A^3{G}_3\cos(3k_c x)
\end{equation}
subject to homogeneous Neumann boundary conditions. In the system \eqref{ord_ep3},
\begin{equation*}
{G}_1^{(1)} :=\left(\begin{array}{c}
0\\
\langle \alpha_{21}\rangle^ck_c^2(v_2\bar{\rho}_1+\bar{\rho}_2)\end{array}\right),
\end{equation*}
\begin{equation*}
{G}_1^{(3)} := Q_K^s(\vv, \w_{20}) + \frac12 Q_K^s(\vv, \w_{22})
 - k_c^2
 \left( Q_{D_c}^s(\vv, \w_{20}) +\frac12  Q_{D_c}^s(\vv, \w_{22})
 \right),
\end{equation*}
and
\begin{equation*}
{G}_3 := \frac{1}{2} \left(Q_K^s(\vv, \w_{22}) - k_c^2 Q_{D_c}^s(\vv, \w_{22})\right),
\end{equation*}
where $Q_K^s({z}, {z}'):= Q_K({z}, {z}')+ Q_K({z}', {z})$ and $Q_{D_c}^s({z}, {z}'):= Q_{D_c}({z}, {z}')+ Q_{D_c}({z}', {z})$. Since ${G}$ contains the resonant term $\cos(k_c x)$, the solvability condition yields the following Stuart-Landau equation for the amplitude $A(\tau)$
\begin{equation}\label{SL}
\frac{{\rm d}A}{{\rm d}\tau}=\sigma A-L A^3, \qquad \sigma=-\frac{\langle {G}_1^{(1)}, {\Psi}\rangle}{\langle {v}, {\Psi}\rangle} \ {\rm and}\  L=-\frac{\langle {G}_1^{(3)}, {\Psi}\rangle}{\langle {v}, {\Psi}\rangle},
\end{equation}
where ${\Psi}$ denotes the adjoint eigenvector associated with the zero eigenvalue of the adjoint operator $(\mathscr{L}^c)^\dagger$, that is, ${\Psi}\in {\rm Ker}((\mathscr{L}^c)^\dagger)$. 

The Stuart–Landau equation~\eqref{SL} makes it possible to investigate the nature of the Turing-type bifurcation (super- or sub-critical) and the stability of the patterned state. In summary, in the super-critical case, which occurs when $L>0$, the nontrivial steady solution of the Stuart–Landau equation~\eqref{SL} is stable, and the amplitude of the emerging pattern can be estimated near the bifurcation point. Specifically, the amplitude of the steady pattern is 
\begin{equation}\label{eq:Ainfty}
A_\infty = \sqrt{\dfrac{\sigma}{L}}.
\end{equation}
Through~\eqref{expansion_sol}, \eqref{sol1}, and~\eqref{eq:Ainfty}, at first order in $\delta$, we obtain the following estimates for the amplitudes of $\rho_1$ and $\rho_2$ once that a steady pattern has formed
\begin{equation}\label{eq:winfty}
\delta \, \sqrt{\dfrac{\sigma}{L}} \quad \text{and} \quad \delta \, \sqrt{\dfrac{\sigma}{L}} \frac{J_{2,1}+k_c^2D_{c 2,1}}{-J_{2,2}+k_c^2D_{c 2,2}}.
\end{equation}
On the other hand, in the sub-critical case, corresponding to $L<0$, equation~\eqref{SL} does not admit a stable nontrivial solution and to determine the saturated pattern amplitude one would have to consider terms of order $O(\delta^5)$ in the asymptotic expansion~\eqref{expansion_sol}. This is beyond the scope \rr{of} the present study; yet, the third-order analysis presented here provides a useful classification, allowing one to identify subcritical regimes, commonly associated with hysteresis or bistability phenomena.
\rr{Finally, it is worth noting that the sign of $L$, computed at the corresponding Turing threshold, characterises the super- or sub-critical nature of the bifurcation. This classification is distinct from the quantitative prediction of the Stuart--Landau equation for the pattern amplitude, which is valid only in a sufficiently small neighbourhood of the bifurcation threshold.}

\section{Numerical simulations}\label{sec:numerics}
In this section, after summarising the set-up of numerical simulations (see Section~\ref{sec:setnum}), we present a sample of simulation results which confirm the results of pattern formation analyses carried out in Section~\ref{Sec:PatternFormation} as well as the formal asymptotics performed in the quasi-invariant regime of fast phenotype switching in Section~\ref{sec:mod1} (see Section~\ref{sec:numsimres}). 

Throughout this section, we let the phenotype switching kernels $M_i$ satisfy the additional assumptions~\eqref{eq:assumption_M_indipendent_of_y'} -- i.e. $M_i(y_i|y_i')\equiv M_i(y_i)$ -- to ensure that the results~\eqref{eq:psiunderassumption_M_indipendent_of_y'} hold so that we can explicitly compute the phenotype-averaged effective parameters \rr{$\la{D}_i\ra$, $\la{\alpha}_{ij}\ra$, $\la{r}_i\ra$, and $\la K_{ij} \ra$ defined via \eqref{eqn:def<Di><alphaij>} and \eqref{eqn:def<Kij>}} as
\begin{align}
\label{eqn:def<Di><alphaij>sim} 
    &\la{D}_i\ra := \int_{0}^1 D_i(y_i) M_i(y_i)  \dy_i, 
    \quad
    \la{\alpha}_{ij}\ra := \int_{0}^1 \alpha_{ij}(y_i) M_i(y_i)  \dy_i, \qquad i, j = 1,2.
\end{align}
\begin{align}
	\label{eqn:def<Kij>sim}
	&\rr{\la r_i \ra :=  \int_{0}^1
		r_{i}(y_i) M_i(y_i) \dy_i}, \quad 
	\la K_{ij} \ra := \int_{0}^1 \int_{0}^1
	K_{ij}(y_i, y_j') M_i(y_i) M_j(y_j') \dy_i \dy_j', \qquad i, j = 1,2,
\end{align}
\subsection{Set-up of numerical simulations}
\label{sec:setnum}
We solve numerically the coupled system of non-local reaction-cross-diffusion equations~\eqref{eqn:PDE_n1_n2_mutational_kernel_rescaled} posed on $\Omega := (-\pi,\pi)$ and $\Y_1 \equiv \Y_2 := [0,1]$, for $t \in (0,T]$ with $T>0$ being the final time of simulations, subject to the boundary conditions~\eqref{eq:bcs_rescaled} and complemented with initial data that form a small perturbation of uniform distributions, that is, for all $y_1, y_2 \in [0,1]$ we set
\begin{equation}\label{eqn:numerics_initial_condition}
    n^\varepsilon_1(0,x,y_1) = \overline{\rho}_1\left( 1 +
    0.005 \cos(3 x) \right), \qquad
    n^\varepsilon_2(0,x,y_2) = \overline{\rho}_2\left( 1 +
    0.005\sin(3 x)
    \right),
\end{equation}
with $\overline{\rho}_1$ and $\overline{\rho}_2$ defined via~\eqref{eq:steadystate_rho}. Note that, when the initial phenotype densities are defined via~\eqref{eqn:numerics_initial_condition}, the corresponding initial densities $\rho^\varepsilon_1(0,x)$ and $\rho^\varepsilon_2(0,x)$ form a small perturbation of the uniform steady state $(\overline{\rho}_1, \overline{\rho}_2)$ given by~\eqref{eq:steadystate_rho}. Numerical solutions are computed by combining finite differences for the differential terms and the Simpson $1/3$ rule for the integral terms, as described in detail in Appendix~\ref{app:numericalmethod}. 

The results we report on in this section, which are displayed in Figures~\ref{fig:EXP1_patterns_NO}-\ref{fig:EXP3}, are obtained by choosing sufficiently small values of $\varepsilon$, which are provided in the captions of Figures~\ref{fig:EXP1_patterns_NO}-\ref{fig:EXP3}. Furthermore, we choose model functions and parameters such that assumptions~\eqref{assMi}, \eqref{C_cond3}, and~\eqref{C_cond12} hold, as summarised in the next paragraphs and as detailed, for each of Figures~\ref{fig:EXP1_patterns_NO}-\ref{fig:EXP3}, in Table~\ref{table:numerics-PSPDE}. The corresponding values of the phenotype-averaged effective parameters \rr{$\la{D}_i\ra$, $\la{\alpha}_{ij}\ra$, $\la r_{i} \ra$, and $\la K_{ij} \ra$ defined via~\eqref{eqn:def<Di><alphaij>} and~\eqref{eqn:def<Kij>}}, along with the corresponding values of the bifurcation threshold, $\la\alpha_{21}\ra_c$, given by \eqref{TT_comp_}, and the critical spatial mode, $k_c$, given by~\eqref{kc_comp}, are provided in Table~\ref{table:numerics-SKT}. 
\rr{We remark that the definitions of the phenotype-dependent functions that are used in the numerical simulations are primarily chosen for illustrative purposes, since the focus here is on illustrating the results of pattern formation analyses and verifying the outputs of the formal asymptotics performed in the quasi-invariant regime of fast phenotype switching.}

\paragraph{Phenotype switching kernels.} As mentioned earlier, we let the phenotype switching kernels $M_i$ satisfy, in addition to assumptions~\eqref{assMi}, assumptions~\eqref{eq:assumption_M_indipendent_of_y'}. Specifically, introducing the Gaussian-like functions  
\begin{equation*}
g_{\mu_i, \sigma_i^2}(y_i) := \frac{1}{C_i} e^{-\frac{(y_i-\mu_i)^2}{2\sigma_i^2}}, \qquad C_i := \int_{0}^1e^{-\frac{(y_i-\mu_i)^2}{2\sigma_i^2}} \dy_i', \qquad i = 1,2,  
\end{equation*}
we consider both the case of unimodal phenotype switching kernels defined as
\begin{equation}\label{eqn:numerics_Mi_functions}
M_i(y_i) := g_{\mu_i, \sigma_i^2}(y_i),
\end{equation}
\rr{which correspond to scenarios where there is a single phenotype, $\mu_i$, that dominates the phenotypic shift}, and the case of bimodal phenotype switching kernels defined as
 \begin{equation}\label{eqn:numerics_Mi_functions_2peaks}
M_i(y_i) := (1-\lambda_i) g_{\mu_{i1}, \sigma_{i1}^2}(y_i) + \lambda_i g_{\mu_{i2}, \sigma_{i2}^2}(y_i), \qquad \lambda_i\in(0,1),
\end{equation} 
\rr{which correspond to scenarios where there are instead two phenotypes, $\mu_{i1}$ and $\mu_{i2}$, dominating the phenotypic shift.}

\paragraph{\rr{Proliferation rates and competition kernels.}} \rr{For simplicity, we consider constant proliferation rates, which correspond to the assumption that all members of population $i$ proliferate at the same rate, i.e.  $r_i(y_i) \equiv r_i \geq 0$, and, therefore, $\la r_i \ra\equiv r_i$ is also the intrinsic growth rate of the population. On the other hand,} we let the competition kernels $K_{ij}(y_i, y_j')$ be quadratic polynomials of the form
\begin{equation}\label{eqn:numerics_Kij_functions}
    K_{ij}(y_i, y_j') := A_{ij} + B_{ij} y_i + C_{ij} y_j' + D_{ij} y_i^2 + E_{ij}(y_j')^2 + F_{ij}y_iy_j',  \qquad A_{ij}, B_{ij}, C_{ij}, D_{ij}, E_{ij}, F_{ij}\in\R,
\end{equation}
and consider different values of the coefficients $A_{ij}$, $B_{ij}$, $C_{ij}$, $D_{ij}$, $E_{ij}$, and $F_{ij}$ corresponding to different scenarios, while all being such that $ K_{ij} \geq 0$ on $[0,1] \times [0,1]$. For instance, to consider the scenario where competition does not depend on the phenotypes of the interacting population members, we set all coefficients in~\eqref{eqn:numerics_Kij_functions} to zero except for $A_{ij}$. Furthermore, to consider the scenario where competition is weaker between members of populations $i$ and $j$ with similar phenotypes, we choose $D_{ij} = E_{ij} = 1$, $F_{ij} = -2$ and set the remaining coefficients to zero, so that the general form~\eqref{eqn:numerics_Kij_functions} reduces to 
\begin{equation}\label{eqn:K_ij=(y_i-y_j)^2}
K_{ij}(y_i, y_j') := (y_i-y_j')^2.
\end{equation}
On the other hand, choosing $A_{ij}=1$, $D_{ij} = E_{ij} = -1$, $F_{ij} = 2$, and setting the other coefficients in~\eqref{eqn:numerics_Kij_functions} to zero yields 
\begin{equation}\label{eqn:K_ij=1-(y_i-y_j)^2}
K_{ij}(y_i, y_j') := 1 - (y_i-y_j')^2,
\end{equation}
which corresponds to the opposite scenario where competition is stronger between members of populations $i$ and $j$ with similar phenotypes.

\paragraph{Diffusion coefficients.} We let the linear diffusion coefficients $D_i(y_i)$ be of the form
\begin{equation}\label{eqn:numerics_diff_functions}
D_i(y_i) := d_i  + f_{i}(y_i), \qquad d_i \in \R
\end{equation} 
with 
\begin{equation}\label{eqn:numerics_hi_functions0}
f_{i}(y_i) \equiv 0
\end{equation} 
or
\begin{equation}\label{eqn:numerics_hi_functions}
 f_{i}(y_i) := \frac{\gamma_{i}}{1 +\exp\left( -\eta_{i}\left(y_i - \frac12 \right) \right) }, \qquad \gamma_{i}, \eta_{i}\in \R,
\end{equation}
while the self-diffusion $(i=j)$ / cross-diffusion $(i\neq j)$ coefficients \rr{$\alpha_{ij}(y_i)$} are defined as 
\begin{equation}\label{eqn:numerics_csdiff_functions}
\alpha_{ij}(y_i) := a_{ij} +  f_{ij}(y_i), \qquad a_{ij} \in \R
\end{equation} 
with  
\begin{equation}\label{eqn:numerics_hij_functions0}
f_{ij}(y_i) \equiv 0
\end{equation} 
or
\begin{equation}\label{eqn:numerics_hij_functions}
 f_{ij}(y_i) := \frac{\gamma_{ij}}{1 +\exp\left( -\eta_{ij}\left(y_i - \frac12 \right) \right) }, \qquad \gamma_{ij}, \eta_{ij}\in \R,
\end{equation}
or 
\begin{equation}\label{eqn:numerics_pij_functions}
 f_{ij}(y_i) := \gamma_{ij} y_i + \eta_{ij}y_i^2, \qquad \gamma_{ij}, \eta_{ij}\in \R.
\end{equation}
Definitions~\eqref{eqn:numerics_hi_functions0} and~\eqref{eqn:numerics_hij_functions0} correspond to scenarios where linear diffusion and self-/cross-diffusion coefficients are not phenotype-dependent, while definitions~\eqref{eqn:numerics_hi_functions} and~\eqref{eqn:numerics_hij_functions} model scenarios where the values of the linear diffusion and self-/cross-diffusion coefficients undergo sharp monotone variation in phenotype, with steepness of the variation controlled by the parameters $\eta_i$ and $\eta_{ij}$. Note that, both in definition~\eqref{eqn:numerics_hi_functions} and in definition~\eqref{eqn:numerics_hij_functions}, the inflection point is in $y_i=\frac12$ for any $\gamma_{ij} \neq 0$ and $\eta_{ij} \neq 0$. Definition~\eqref{eqn:numerics_pij_functions} corresponds instead to the scenario where the values of the self-/cross-diffusion coefficients vary gradually with phenotype. 

\begin{table}\footnotesize
\centering
\renewcommand{\arraystretch}{1.2}
\arrayrulecolor{black}

\begin{tabular}{
|>{\columncolor{Mercury2}}c
!{\color{black}\vrule width 1.5pt}
c|c|c|}
\hline

\rowcolor{Mercury}
\multicolumn{1}{|c!{\color{black}\vrule width 1.5pt}}{} 
& \textbf{Figs. \ref{fig:EXP1_patterns_NO}, \ref{fig:EXP1_patterns_SI}, \ref{fig:EXP1-comparison-for-different-epsilon}} 
& \textbf{Fig. \ref{fig:EXP2}} 
& \textbf{Fig. \ref{fig:EXP3}} \\

\noalign{\color{black}\hrule height 1.5pt}

$D_1(y_1)$    
& \eqref{eqn:numerics_diff_functions}, \eqref{eqn:numerics_hi_functions0} with $d_1=0.001$ 
&  \eqref{eqn:numerics_diff_functions}, \eqref{eqn:numerics_hi_functions0} with $d_1=0.01$
&  \makecell{\eqref{eqn:numerics_diff_functions}, \eqref{eqn:numerics_hi_functions} with $d_1=0$ \\ $\gamma_1=0.04$, $\eta_1=20$}
\\ \hline

$D_2(y_2)$ 
&  \eqref{eqn:numerics_diff_functions}, \eqref{eqn:numerics_hi_functions0} with $d_2=0.01$ 
&  \eqref{eqn:numerics_diff_functions}, \eqref{eqn:numerics_hi_functions0} with \rr{$d_2=0.9$}  
&  \eqref{eqn:numerics_diff_functions}, \eqref{eqn:numerics_hi_functions0} with $d_2=0.02$ 
\\ \hline

$\alpha_{11}(y_1)$  
& \eqref{eqn:numerics_csdiff_functions}, \eqref{eqn:numerics_hij_functions0} with $a_{11}=0.0002$     
& \eqref{eqn:numerics_csdiff_functions}, \eqref{eqn:numerics_hij_functions0} with $a_{11}=0$ 
& \makecell{\eqref{eqn:numerics_csdiff_functions}, \eqref{eqn:numerics_hij_functions} with $a_{11}=0$ \\ $\gamma_{11}=0.06$, $\eta_{11}=20$}
\\ \hline

$\alpha_{12}(y_1)$ 
& \eqref{eqn:numerics_csdiff_functions}, \eqref{eqn:numerics_hij_functions0} with $a_{12}=0.02$   
& \makecell{\eqref{eqn:numerics_csdiff_functions}, \eqref{eqn:numerics_pij_functions} with $a_{12}=0$ \\ $\gamma_{12}=0.06$, $\eta_{12}=-0.03$}
& \makecell{\eqref{eqn:numerics_csdiff_functions}, \eqref{eqn:numerics_hij_functions} with $a_{12}=0.00002$ \\ $\gamma_{12}=-0.00002$, $\eta_{12}=20$}
\\ \hline

$\alpha_{21}(y_2)$  
& \makecell{\eqref{eqn:numerics_csdiff_functions}, \eqref{eqn:numerics_hij_functions} with $a_{21}=0$ \\ $\gamma_{21}=0.3$, $\eta_{21}=30$}
& \makecell{\eqref{eqn:numerics_csdiff_functions}, \eqref{eqn:numerics_pij_functions} with $a_{21}=0.5$ \\ $\gamma_{21}=0$, $\eta_{21}=-0.5$}
& \makecell{\eqref{eqn:numerics_csdiff_functions}, \eqref{eqn:numerics_hij_functions} with $a_{21}=8$ \\ $\gamma_{21}=-8$, $\eta_{21}=-20$}
\\ \hline

$\alpha_{22}(y_2)$ 
& \eqref{eqn:numerics_csdiff_functions}, \eqref{eqn:numerics_hij_functions0} with $a_{22}=0.0003$   
& \eqref{eqn:numerics_csdiff_functions}, \eqref{eqn:numerics_hij_functions0} with $a_{22}=0$ 
& \eqref{eqn:numerics_csdiff_functions}, \eqref{eqn:numerics_hij_functions0} with $a_{22}=0.00001$
\\ \hline

$r_1$          
& 0.5  
& 34 
& 10 \\
\hline

$r_2$          
& 0.3 
& 10 
& 8 \\
\hline

$\beta_{11}$       
& 6   
& 125.002  
& 75.721873 \\
\hline

$\beta_{12}$   
& 0.06 
& 6.923 
& 52.211524 \\
\hline

$\beta_{21}$    
& 0.002
& 0.027 
& 2.403204 \\
\hline

$\beta_{22}$        
& 0.05 
& 0.714 
& 8.016032 \\
\hline

$K_{11}(y_1, y_1')$  
& $(y_1 - y_1')^2$
& \makecell{$(y_1-y_1')^2$ in Fig.\ref{fig:EXP2}A\\
$1-(y_1-y_1')^2$ in Fig.\ref{fig:EXP2}B}
& $y_1y_1'$ 
\\ \hline

$K_{12}(y_1, y_2)$ 
& 1  
& $y_1^2$ 
& $y_1^2$ \\
\hline

$K_{21}(y_1, y_2)$   
& $-4y_1(y_1-1)$ 
& $y_2(2-y_2)$
& 1 
\\ \hline

$K_{22}(y_2, y_2')$    
& 1   
& $1-(y_2 - y_2')^2$
& 1      
\\ \hline

$M_1(y_1)$   
& \eqref{eqn:numerics_Mi_functions} with $\mu_1 = 0.5$, $\sigma_1^2=0.01$
& \eqref{eqn:numerics_Mi_functions} with $\mu_1 = 0.5$, $\sigma_1^2=0.02$
& \makecell{\eqref{eqn:numerics_Mi_functions_2peaks} with 
$\mu_{11}=0.25$, $\mu_{12}=0.5$ \\ $\sigma_{11}^2=\sigma_{12}^2=0.002$
\\
$\lambda_1 = 0.50$ in Fig.\ref{fig:EXP3}A\\
$\lambda_1 = 0.35$ in Fig.\ref{fig:EXP3}B}
\\ \hline

$M_2(y_2)$  
& \makecell{\eqref{eqn:numerics_Mi_functions} with $\sigma_2^2=0.01$ \\
$\mu_2 = 0.45$  in Fig.\ref{fig:EXP1_patterns_NO}\\ 
$\mu_2 = 0.50$ in Fig.\ref{fig:EXP1_patterns_SI}}
& \eqref{eqn:numerics_Mi_functions} with $\mu_2 = 0.5$, $\sigma_2^2=0.02$
& \makecell{\eqref{eqn:numerics_Mi_functions} with $\sigma_2^2=0.002$ \\
$\mu_2 = 0.35$ in Fig.\ref{fig:EXP3}A\\
$\mu_2 = 0.50$ in Fig.\ref{fig:EXP3}B} 
\\ \hline

\end{tabular}

\caption{Functions and parameters used to carry out numerical simulations of the rescaled phenotype-structured SKT model~\eqref{eqn:PDE_n1_n2_mutational_kernel_rescaled} complemented with definitions~\eqref{eqn:K1}-\eqref{eq:mathcal_M_definition}.}
\label{table:numerics-PSPDE}
\end{table}

\begin{table}\footnotesize
\centering
\renewcommand{\arraystretch}{1.2}
\arrayrulecolor{black}

\begin{tabular}{
|>{\columncolor{Mercury2}}c
!{\color{black}\vrule width 1.5pt}
c|c|c|}
\hline

\rowcolor{Mercury}
\multicolumn{1}{|c!{\color{black}\vrule width 1.5pt}}{} 
& \textbf{Figs. \ref{fig:EXP1_patterns_NO}, \ref{fig:EXP1_patterns_SI}, \ref{fig:EXP1-comparison-for-different-epsilon}} 
& \textbf{Fig. \ref{fig:EXP2}} 
& \textbf{Fig. \ref{fig:EXP3}} \\

\noalign{\color{black}\hrule height 1.5pt}

\rr{$\la r_1\ra$}
& 0.5
& 34
& 10
\\ \hline

\rr{$\la r_2\ra$}
& 0.3
& 10
& 8
\\ \hline

$\la K_{11}\ra$  
& 0.01
& \makecell{0.02 in Fig.\ref{fig:EXP2}A\\ 0.98 in Fig.\ref{fig:EXP2}B}
& \makecell{0.140625 in Fig.\ref{fig:EXP3}A \\ 0.113906 in Fig.\ref{fig:EXP3}B}
\\ \hline

$\la K_{12}\ra$ 
& 1  
& 0.26
& \makecell{0.157250 in Fig.\ref{fig:EXP3}A \\ 0.129125 in Fig.\ref{fig:EXP3}B}
\\ \hline

$\la K_{21}\ra$   
& 0.98
& 0.74 
& 1 
\\ \hline

$\la K_{22}\ra$    
& 1   
& 0.98      
& 1 \\
\hline

$\la D_1 \ra$    
& 0.001     
& 0.01
& \makecell{0.010163 in Fig.\ref{fig:EXP3}A \\ 0.007211 in Fig.\ref{fig:EXP3}B}
\\ \hline

$\la D_2 \ra$ 
& 0.01 
& 0.9  
& 0.02 \\
\hline

$\la \alpha_{11} \ra$  
& 0.0002     
& 0
& \makecell{0.015244 in Fig.\ref{fig:EXP3}A \\ 0.010817 in Fig.\ref{fig:EXP3}B}
\\ \hline

$\la\alpha_{12}\ra$ 
& 0.02 
& 0.0222
& \makecell{0.000015 in Fig.\ref{fig:EXP3}A \\ 0.000016 in Fig.\ref{fig:EXP3}B}
\\ \hline

$\la\alpha_{21}\ra$  
& \makecell{0.15 in Fig.\ref{fig:EXP1_patterns_SI}\\
0.0875 in Fig.\ref{fig:EXP1_patterns_NO}}
& 0.37
& \makecell{0.447375 in Fig.\ref{fig:EXP3}A \\ 4 in Fig.\ref{fig:EXP3}B}
\\ \hline

$\la\alpha_{22}\ra$ 
& 0.0003
& 0 
& 0.00001 \\
\hline

$\la\alpha_{21}\ra_c$  
& 0.110014
& \makecell{0.36028 in Fig.\ref{fig:EXP2}A \\ 17.64 in Fig.\ref{fig:EXP2}B}
& \makecell{0.422951 in Fig.\ref{fig:EXP3}A \\ 2.091835 in Fig.\ref{fig:EXP3}B}
\\ \hline

$\,k_c$  
& 2.093276
& \makecell{4.052429 in Fig.\ref{fig:EXP2}A \\ 4.053737 in Fig.\ref{fig:EXP2}B}
& \makecell{9.162583 in Fig.\ref{fig:EXP3}A \\ 5.856378 in Fig.\ref{fig:EXP3}B}
\\ \hline

\end{tabular}

\caption{Values of the phenotype-averaged effective parameters \rr{$\la r_{i} \ra$,} $\la K_{ij} \ra$, $\la{D}_i\ra$, and $\la{\alpha}_{ij}\ra$ defined via~\eqref{eqn:def<Kij>sim} and \eqref{eqn:def<Di><alphaij>sim}, the bifurcation threshold $\la\alpha_{21}\ra_c$ given by~\eqref{TT_comp_}, and the critical spatial mode $k_c$ given by~\eqref{kc_comp} corresponding to the model functions and parameters listed in Table~\ref{table:numerics-PSPDE}.}
\label{table:numerics-SKT}
\end{table}

\subsection{Main results of numerical simulations}\label{sec:numsimres}

\begin{figure}[!ht]
    \centering
    \includegraphics[width=0.45\linewidth]{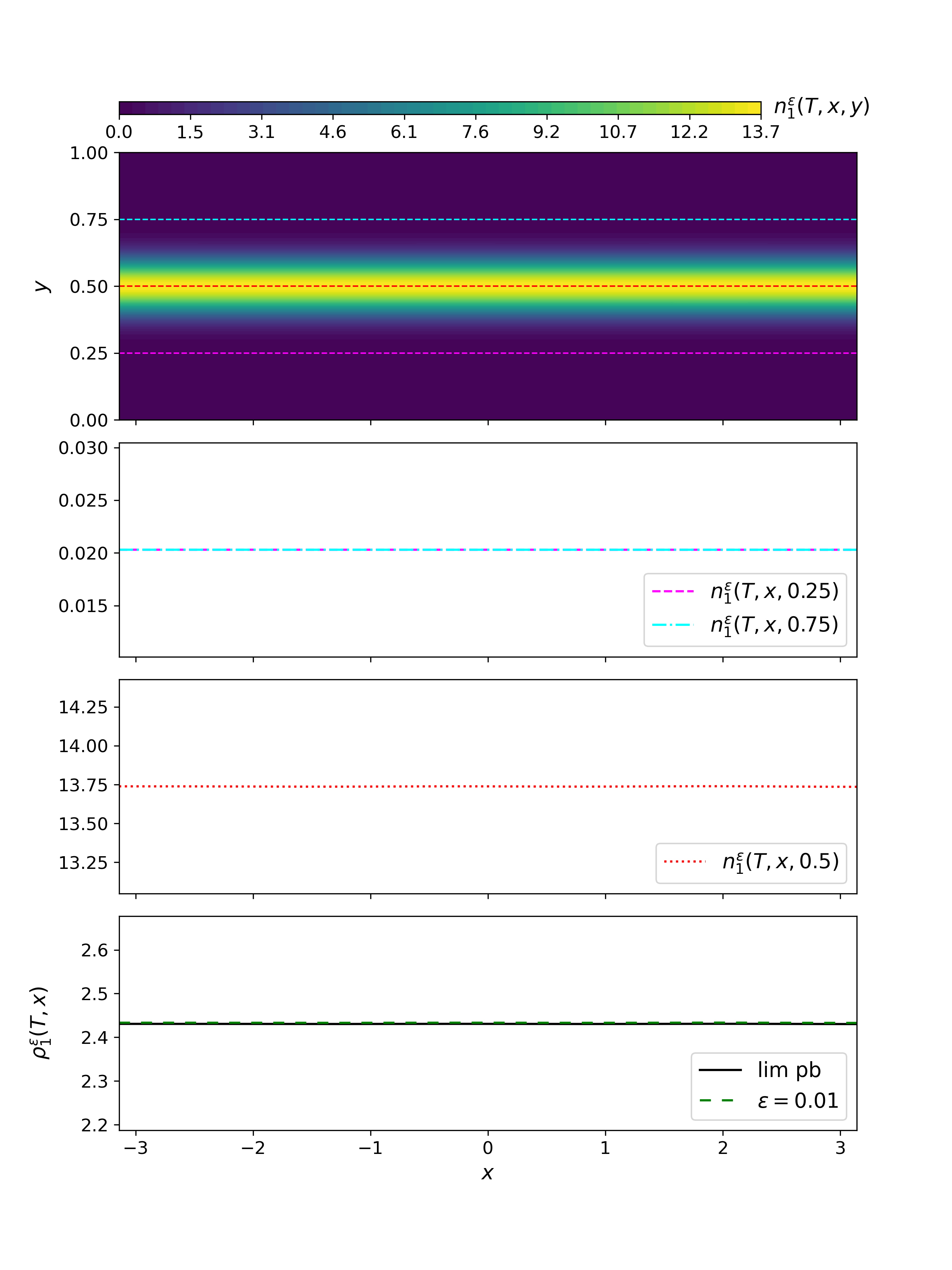}
    \includegraphics[width=0.45\linewidth]{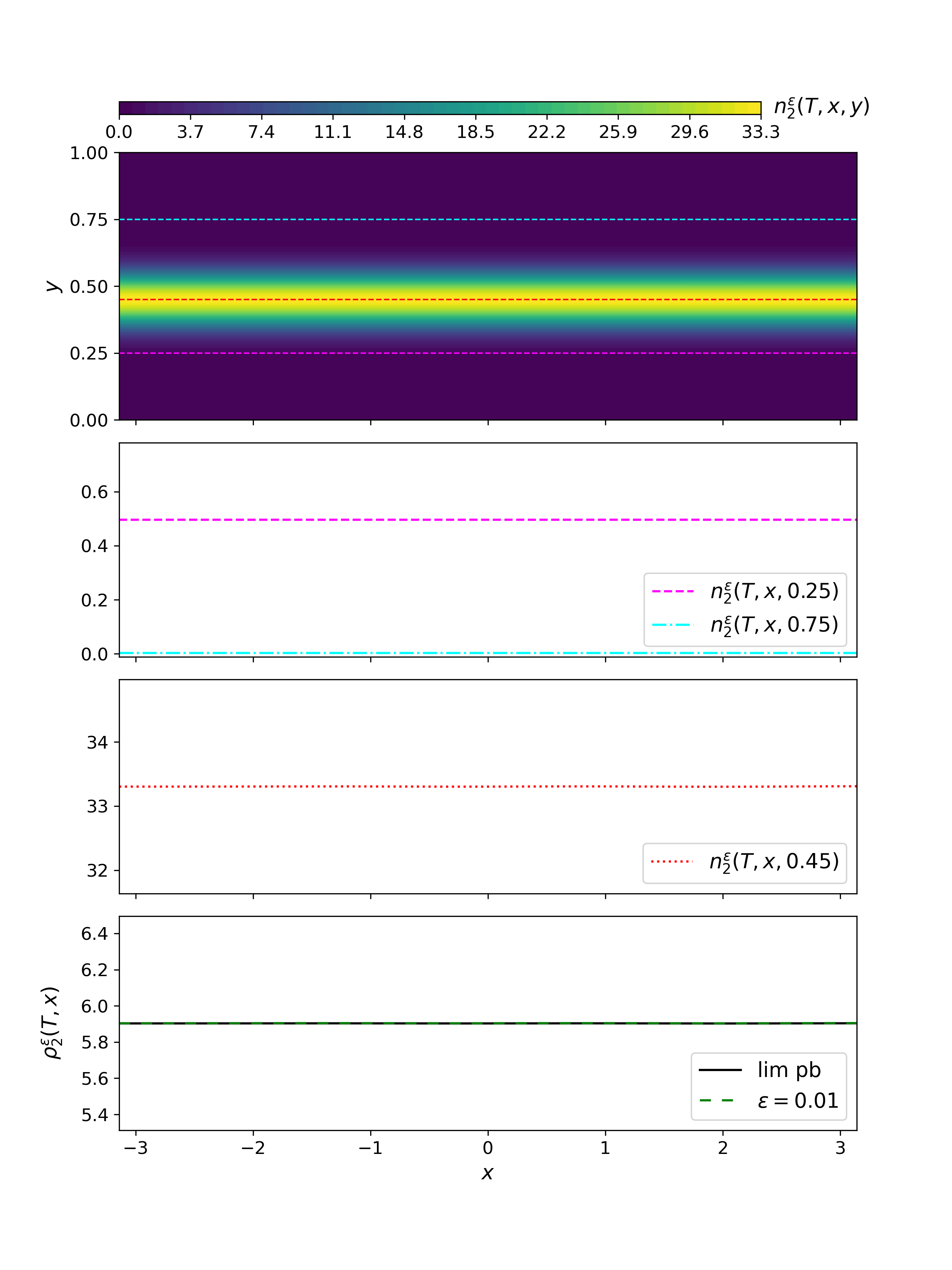}
    \caption{{\bf Pattern formation does not occur for $\la \alpha_{21} \ra < \la{\alpha}_{21}\ra_c$.} Plots of $n^\varepsilon_i(T,x,y_i)$ (top panel), $n^\varepsilon_i(T,x,0.25)$ and $n^\varepsilon_i(T,x,0.75)$ (middle-upper panel), $n^\varepsilon_i(T,x,0.5)$ (middle-lower panel), and $\rho^\varepsilon_i(T,x)$  (bottom panel, coloured lines) for $i=1$ (left column) and $i=2$ (right column), obtained by solving numerically the phenotype-structured SKT model~\eqref{eqn:PDE_n1_n2_mutational_kernel_rescaled}, subject to the boundary conditions~\eqref{eq:bcs_rescaled} and the initial data~\eqref{eqn:numerics_initial_condition}. The set-up of numerical simulations is described in Section~\ref{sec:setnum}, with the scaling parameter $\varepsilon= 0.01$ and the final time of simulations $T=1000$. The model functions and parameters employed, which are listed in Table~\ref{table:numerics-PSPDE}, are such that $\la \alpha_{21} \ra < \la{\alpha}_{21}\ra_c$. The choices of the model parameters and functions are the same as those of Figure~\ref{fig:EXP1_patterns_SI} except for the value of the parameter $\mu_2$ in the definition~\eqref{eqn:numerics_Mi_functions} of the phenotype switching kernel $M_2$. The black lines in the bottom panels highlight $\rho_1$ and $\rho_2$ obtained by solving numerically the limiting problem~\eqref{eqn:SKT_phenotype}-\eqref{eq:bcs_rescaledrho_limit}, subject to initial data analogous to~\eqref{eqn:numerics_initial_condition}, under the effective parameter values in Table~\ref{table:numerics-SKT}.  
    }
    \label{fig:EXP1_patterns_NO}
\end{figure}

\begin{figure}[!ht]
    \centering
    \includegraphics[width=0.45\linewidth]{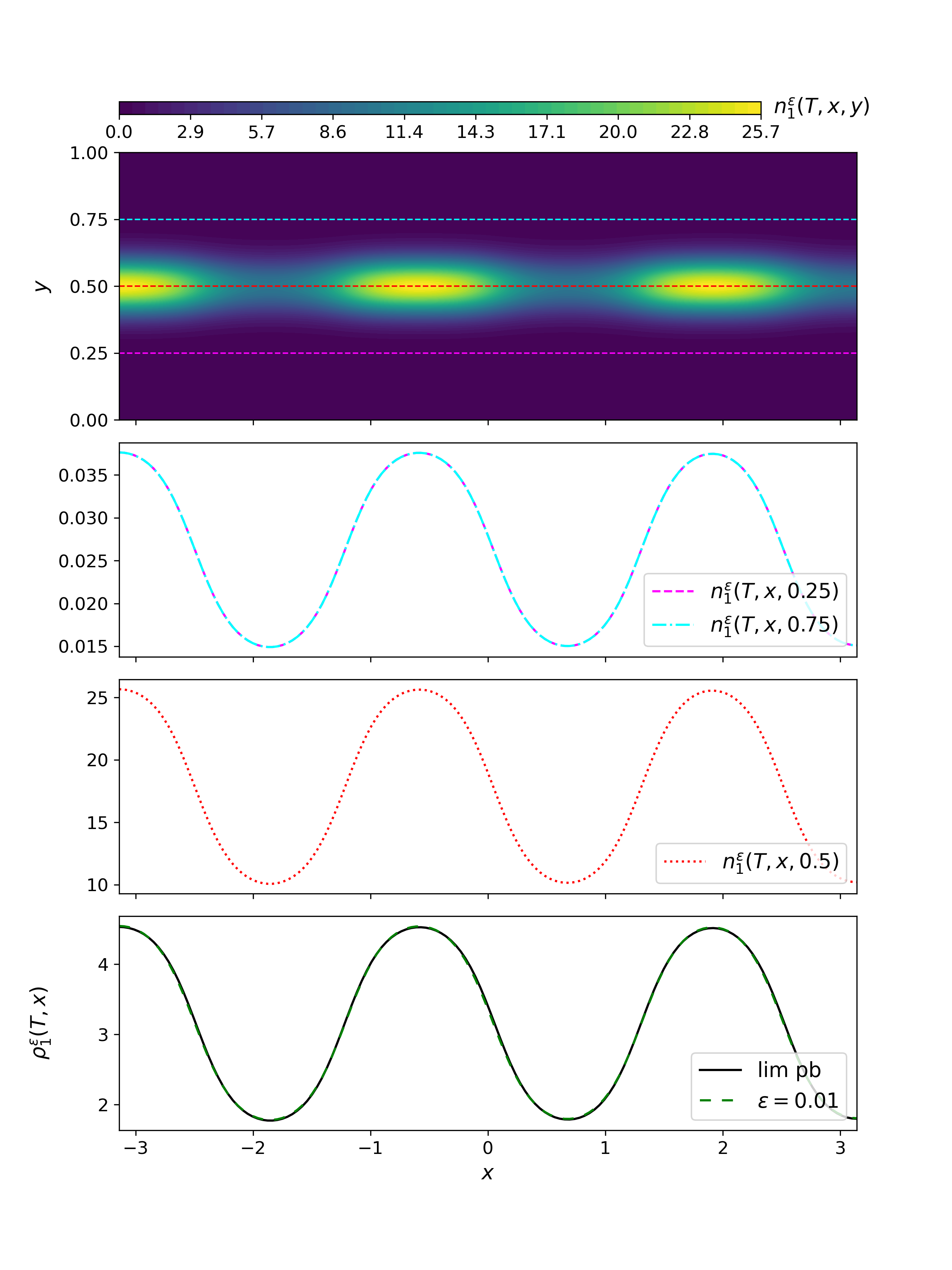}
    \includegraphics[width=0.45\linewidth]{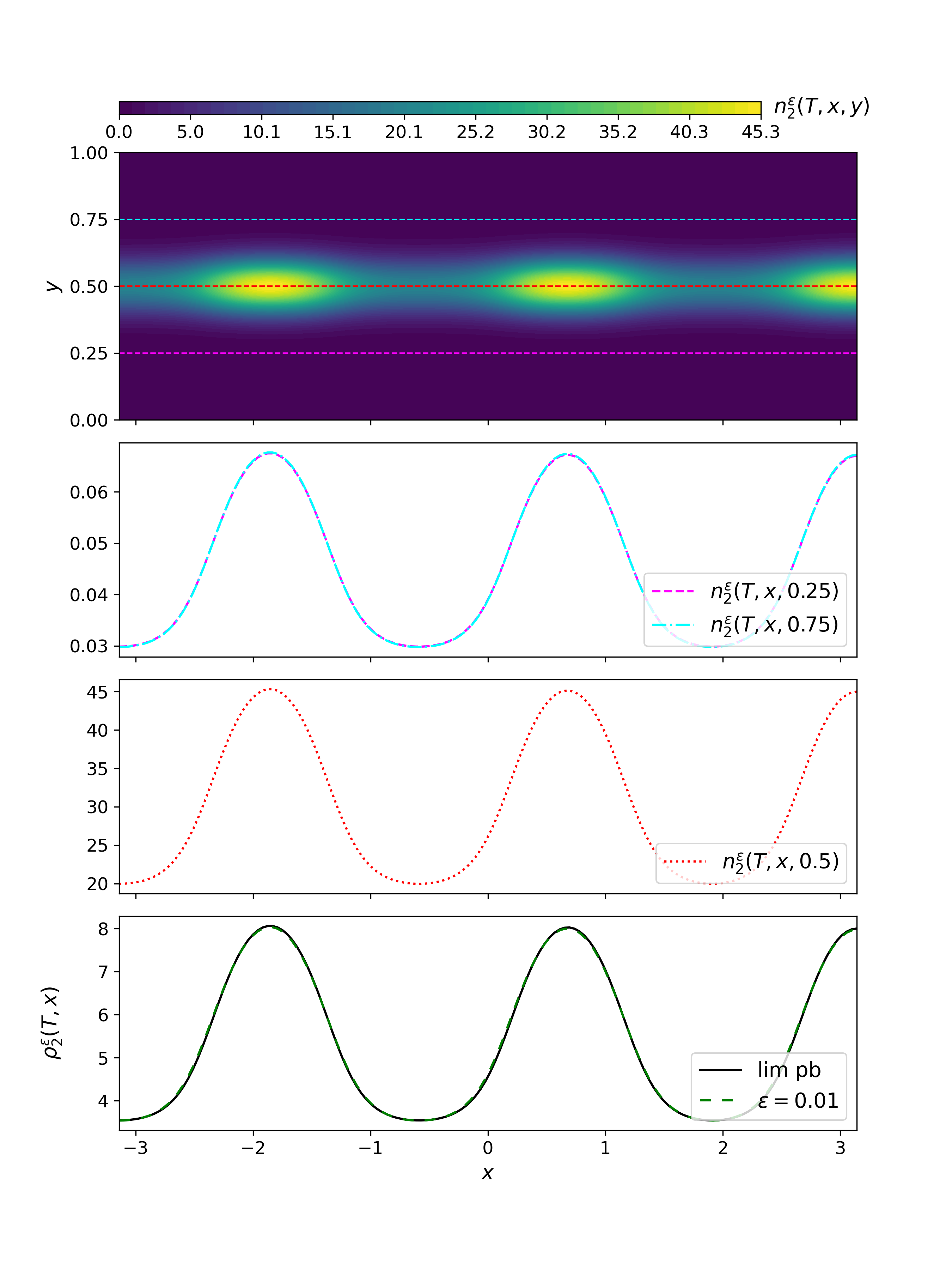}
    \caption{{\bf Spatial patterns emerge for $\la \alpha_{21} \ra > \la{\alpha}_{21}\ra_c$.} Plots of $n^\varepsilon_i(T,x,y_i)$ (top panel), $n^\varepsilon_i(T,x,0.25)$ and $n^\varepsilon_i(T,x,0.75)$ (middle-upper panel), $n^\varepsilon_i(T,x,0.5)$ (middle-lower panel), and $\rho^\varepsilon_i(T,x)$ (bottom panel, coloured lines) for $i=1$ (left column) and $i=2$ (right column), obtained by solving numerically the phenotype-structured SKT model~\eqref{eqn:PDE_n1_n2_mutational_kernel_rescaled}, subject to the boundary conditions~\eqref{eq:bcs_rescaled} and the initial data~\eqref{eqn:numerics_initial_condition}. The set-up of numerical simulations is described in Section~\ref{sec:setnum}, with the scaling parameter $\varepsilon= 0.01$ and the final time of simulations $T=1000$. The model functions and parameters employed, which are listed in Table~\ref{table:numerics-PSPDE}, are such that $\la \alpha_{21} \ra > \la{\alpha}_{21}\ra_c$. The choices of the model parameters and functions are the same as those of Figure~\ref{fig:EXP1_patterns_NO} except for the value of the parameter $\mu_2$ in the definition~\eqref{eqn:numerics_Mi_functions} of the phenotype switching kernel $M_2$. The black lines in the bottom panels highlight $\rho_1$ and $\rho_2$ obtained by solving numerically the limiting problem~\eqref{eqn:SKT_phenotype}-\eqref{eq:bcs_rescaledrho_limit}, subject to initial data analogous to~\eqref{eqn:numerics_initial_condition}, under the effective parameter values in Table~\ref{table:numerics-SKT}.
    }
    \label{fig:EXP1_patterns_SI}
\end{figure}

\begin{figure}[!ht]
  \centering
  \begin{tabular}{@{}>{\centering\arraybackslash}m{1cm}@{} m{1.1\linewidth}}
    \textbf{A} & \includegraphics[width=0.42\linewidth]{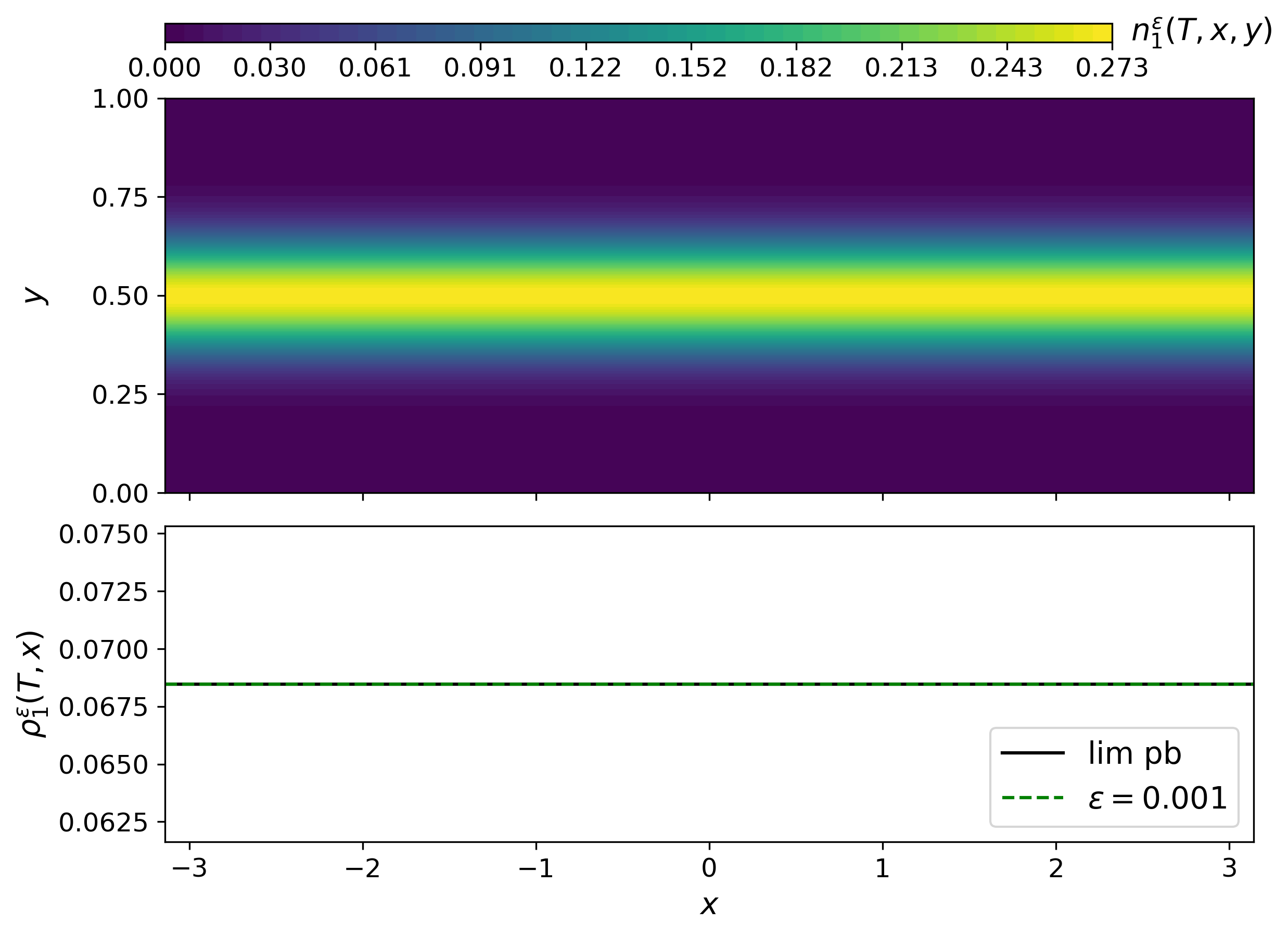}
    \includegraphics[width=0.42\linewidth]{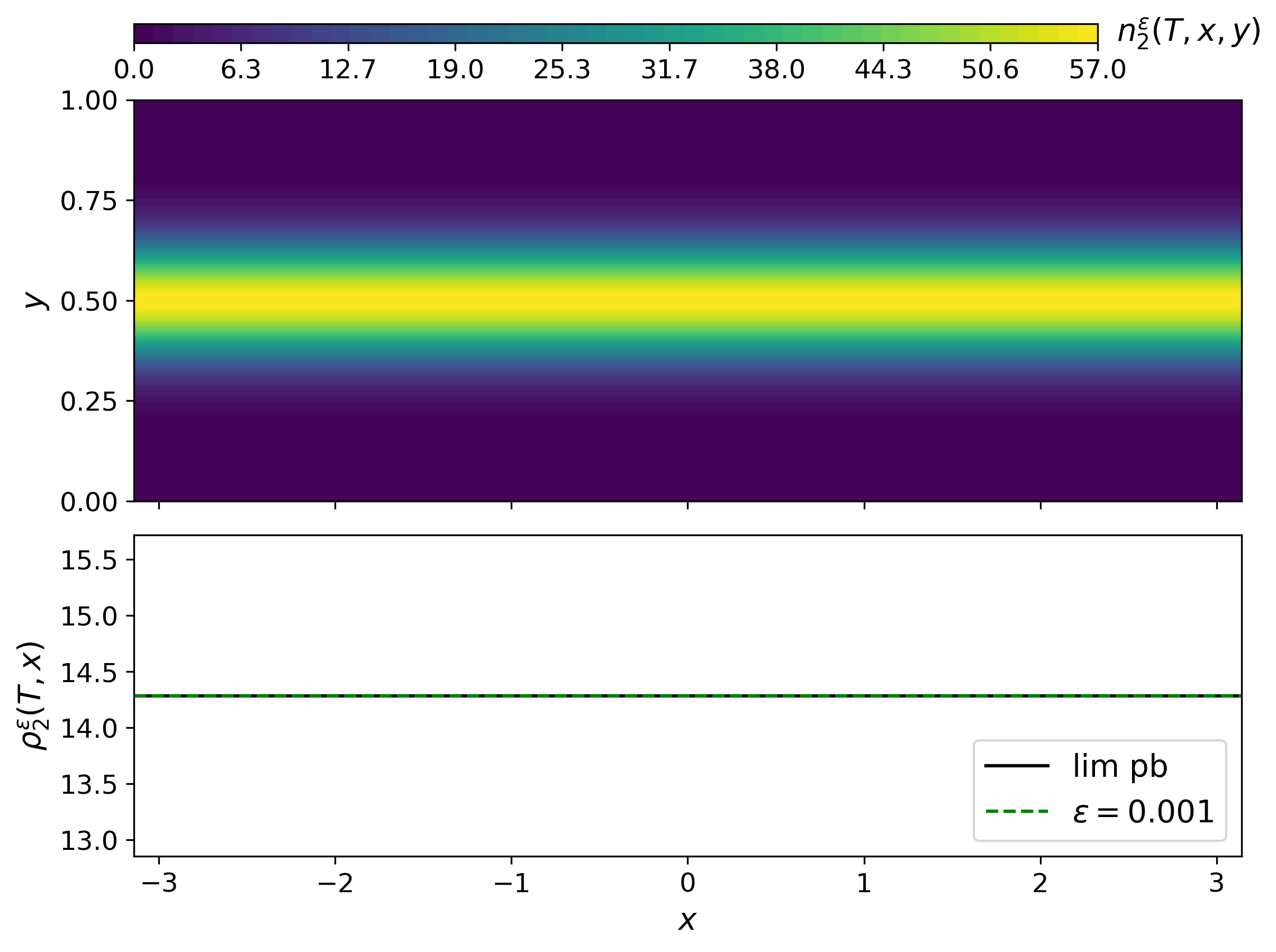} \\
    \textbf{B} & \includegraphics[width=0.42\linewidth]{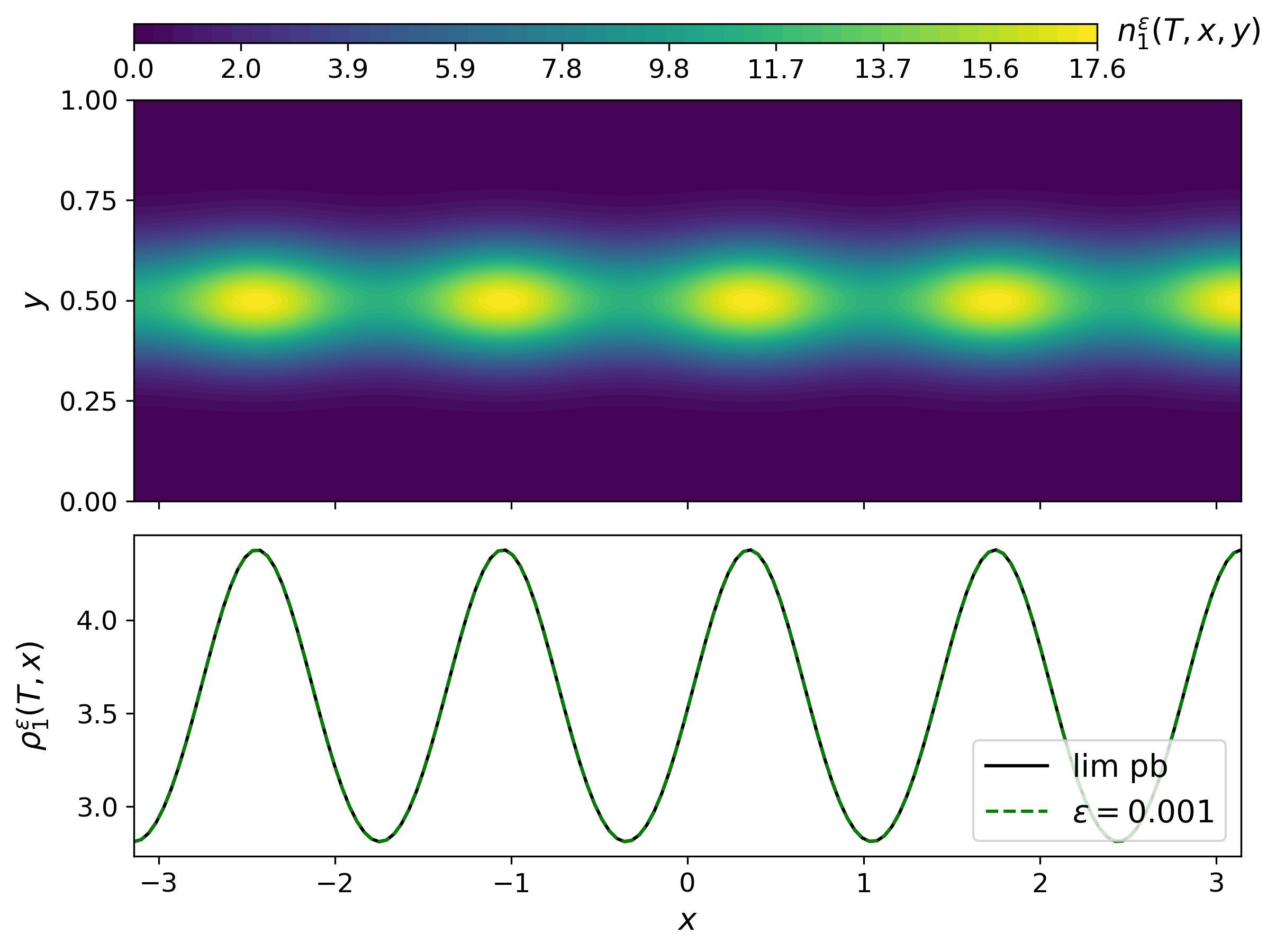}
    \includegraphics[width=0.42\linewidth]{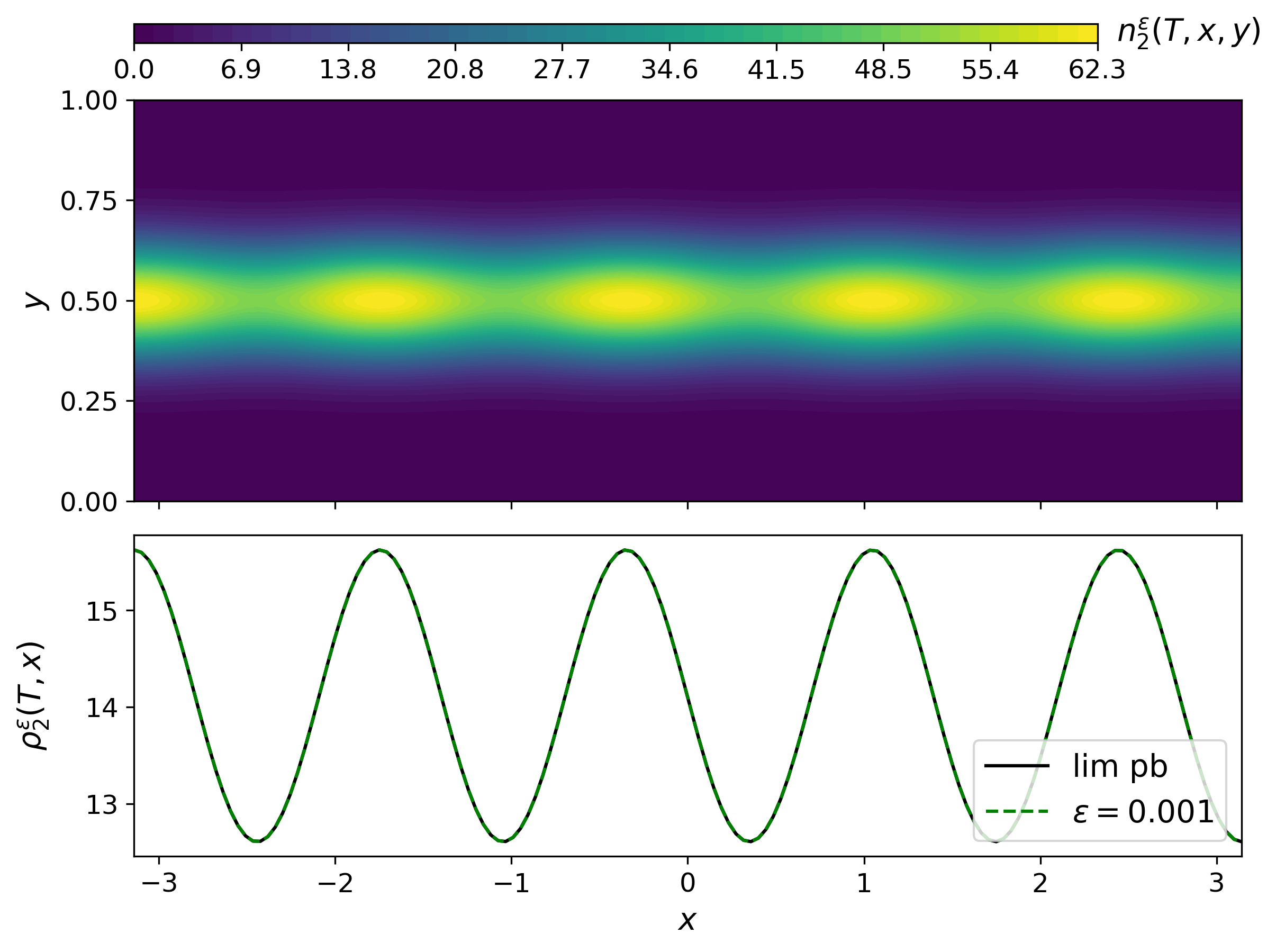}
  \end{tabular}
    \caption{{\bf Pattern formation does not occur for $\la \alpha_{21} \ra < \la{\alpha}_{21}\ra_c$, while spatial patterns emerge for $\la \alpha_{21} \ra > \la{\alpha}_{21}\ra_c$.} Plots of $n^\varepsilon_i(T,x,y_i)$ (top sub-panel) and $\rho^\varepsilon_i(T,x)$  (bottom sub-panel, coloured lines) for $i=1$ (left column) and $i=2$ (right column), obtained by solving numerically the phenotype-structured SKT model~\eqref{eqn:PDE_n1_n2_mutational_kernel_rescaled}, subject to the boundary conditions~\eqref{eq:bcs_rescaled} and the initial data~\eqref{eqn:numerics_initial_condition}. The set-up of numerical simulations is described in Section~\ref{sec:setnum}, with the scaling parameter $\varepsilon= 0.001$ and the final time of simulations $T=1000$. The model functions and parameters employed, which are listed in Table~\ref{table:numerics-PSPDE}, are such that $\la \alpha_{21} \ra < \la{\alpha}_{21}\ra_c$ (panels A) and $\la \alpha_{21} \ra > \la{\alpha}_{21}\ra_c$ (panels B). For panels A and B, the choices of the model parameters and functions are the same, except for the definition of the competition kernel $K_{11}(y_1,y_1')$, which is defined via~\eqref{eqn:K_ij=(y_i-y_j)^2} for the simulations of panels A and via~\eqref{eqn:K_ij=1-(y_i-y_j)^2} for the simulations of panels B. The black lines in the bottom sub-panels highlight $\rho_1$ and $\rho_2$ obtained by solving numerically the limiting problem~\eqref{eqn:SKT_phenotype}-\eqref{eq:bcs_rescaledrho_limit}, subject to initial data analogous to~\eqref{eqn:numerics_initial_condition}, under the effective parameter values in Table~\ref{table:numerics-SKT}.}
    \label{fig:EXP2}
\end{figure}

\paragraph{Emergence of spatial patterns above the bifurcation threshold $\la{\alpha}_{21}\ra_c$.} The results of numerical simulations displayed in the bottom panels of Figures~\ref{fig:EXP1_patterns_NO} and~\ref{fig:EXP1_patterns_SI} and in the bottom panels of Figures~\ref{fig:EXP2}A and~\ref{fig:EXP2}B show that, consistently with the analytical results presented in Section~\ref{Sec:Turing_instability}, a Turing-type bifurcation occurs when the bifurcation parameter $\la \alpha_{21}\ra$, defined via~\eqref{eqn:def<Di><alphaij>}, crosses the critical value $\la{\alpha}_{21}\ra_c$, with $\la{\alpha}_{21}\ra_c$ given by~\eqref{TT_comp_}, whereby: spatial patterns emerge for $\la \alpha_{21} \ra > \la{\alpha}_{21}\ra_c$ (see bottom panels of Figure~\ref{fig:EXP1_patterns_SI} and Figure~\ref{fig:EXP2}B); pattern formation does not occur for $\la \alpha_{21} \ra < \la{\alpha}_{21}\ra_c$ (see bottom panels of Figure~\ref{fig:EXP1_patterns_NO} and Figure~\ref{fig:EXP2}A), since small perturbations present in initial data decay and, after a short initial transient time interval, the densities $\rho^\varepsilon_1(t,x)$ and $\rho^\varepsilon_2(t,x)$ relax onto the components of the steady state $(\overline{\rho}_1, \overline{\rho}_2)$ given by~\eqref{eq:steadystate_rho}. 

Note that, as detailed in Table~\ref{table:numerics-PSPDE}, for the numerical simulations of Figures~\ref{fig:EXP1_patterns_NO} and~\ref{fig:EXP1_patterns_SI}: the cross-diffusion coefficient $\alpha_{12}$ is defined via~\eqref{eqn:numerics_csdiff_functions} and~\eqref{eqn:numerics_hij_functions0}, while the phenotype-dependent cross-diffusion coefficient $\alpha_{21}(y_2)$ is defined via~\eqref{eqn:numerics_csdiff_functions} and~\eqref{eqn:numerics_hij_functions}, with $\eta_{21}=30$, thus exhibiting a sharp transition at $y_2=0.5$;  the choices of the model parameters and functions are the same, except for the value of the parameter $\mu_2$ in the definition~\eqref{eqn:numerics_Mi_functions} of the phenotype switching kernel $M_2$, that is, $\mu_2 = 0.45$ for the simulations of Figure~\ref{fig:EXP1_patterns_NO} and $\mu_2 = 0.5$ for the simulations of Figure~\ref{fig:EXP1_patterns_SI}. Because of the sharp profile of $\alpha_{21}$, such a relatively small difference in the value of $\mu_2$ results in the bifurcation parameter $\la \alpha_{21}\ra$ being below or above the threshold $\la \alpha_{21}\ra_c$. This shows how even relatively small differences in phenotype switching kernels can lead to substantial differences in spatiotemporal dynamics of the phenotype-structured SKT model~\eqref{eqn:PDE_n1_n2_mutational_kernel_rescaled}, in scenarios where the phenotype-dependent cross-diffusion coefficient, the average of which acts as bifurcation parameter, undergoes sharp variations in phenotype. 

Note also that, as detailed again in Table~\ref{table:numerics-PSPDE}, for Figures~\ref{fig:EXP2}A and~\ref{fig:EXP2}B: the phenotype-dependent cross-diffusion coefficients $\alpha_{12}(y_1)$ and $\alpha_{21}(y_2)$, defined via~\eqref{eqn:numerics_csdiff_functions} and~\eqref{eqn:numerics_pij_functions}, vary gradually with phenotype; the choices of the model parameters and functions are the same, except for the definition of the competition kernel $K_{11}(y_1,y_1')$, which is defined via~\eqref{eqn:K_ij=(y_i-y_j)^2} for the simulations of Figure~\ref{fig:EXP2}A and via~\eqref{eqn:K_ij=1-(y_i-y_j)^2} for the simulations of Figure~\ref{fig:EXP2}B. These results demonstrate that also differences in the phenotype dependence of competition kernels can result in substantial differences in spatiotemporal dynamics of the phenotype-structured SKT model~\eqref{eqn:PDE_n1_n2_mutational_kernel_rescaled}, even in scenarios where the values of the phenotype-dependent cross-diffusion coefficients undergo gradual variations in phenotype. Note that, in contrast to Figures~\ref{fig:EXP1_patterns_NO} and~\ref{fig:EXP1_patterns_SI}, here the bifurcation threshold is crossed due to a difference in the value of $\la \alpha_{21} \ra_c$ between Figures~\ref{fig:EXP2}A and~\ref{fig:EXP2}B, while the value of $\la\alpha_{21}\ra$ does not change, as reported in Table \ref{table:numerics-SKT}.

\begin{figure}[!ht]
    \centering
    \includegraphics[width=0.45\linewidth]{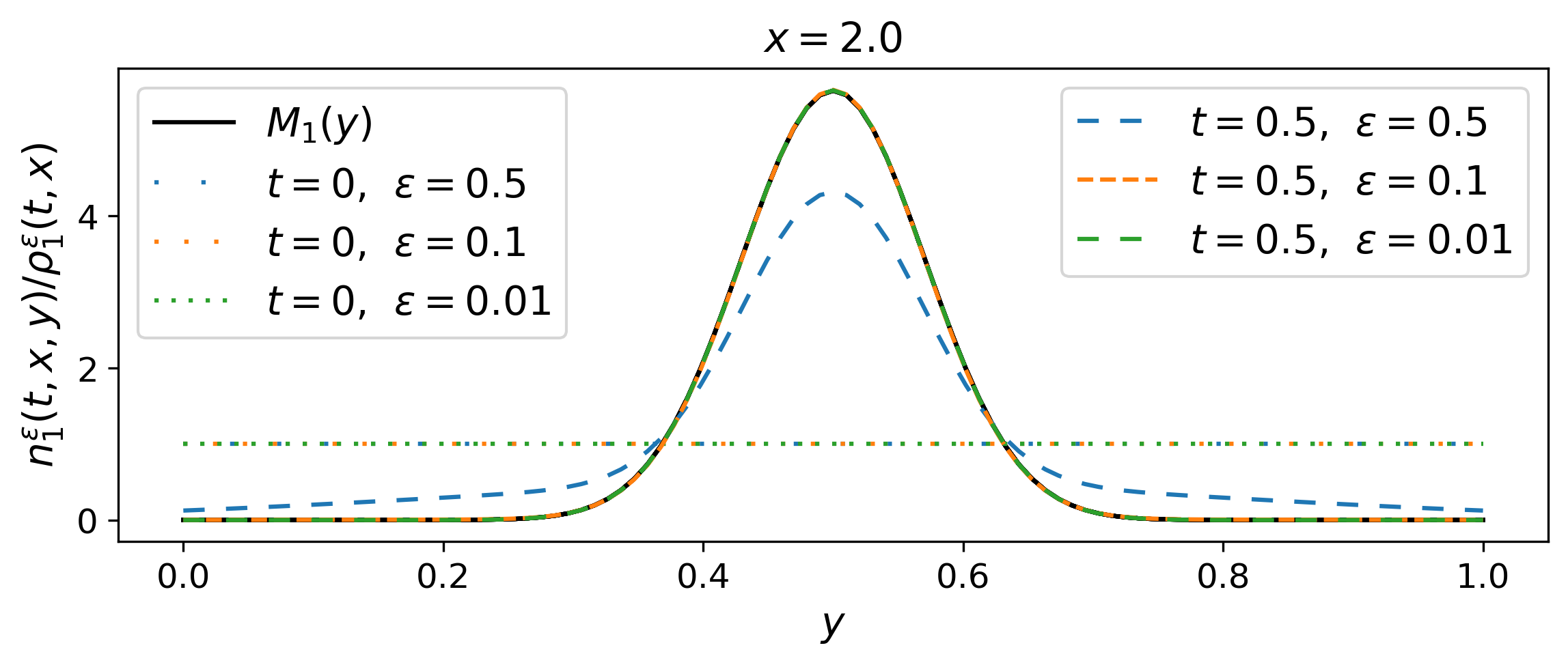}
    \includegraphics[width=0.45\linewidth]{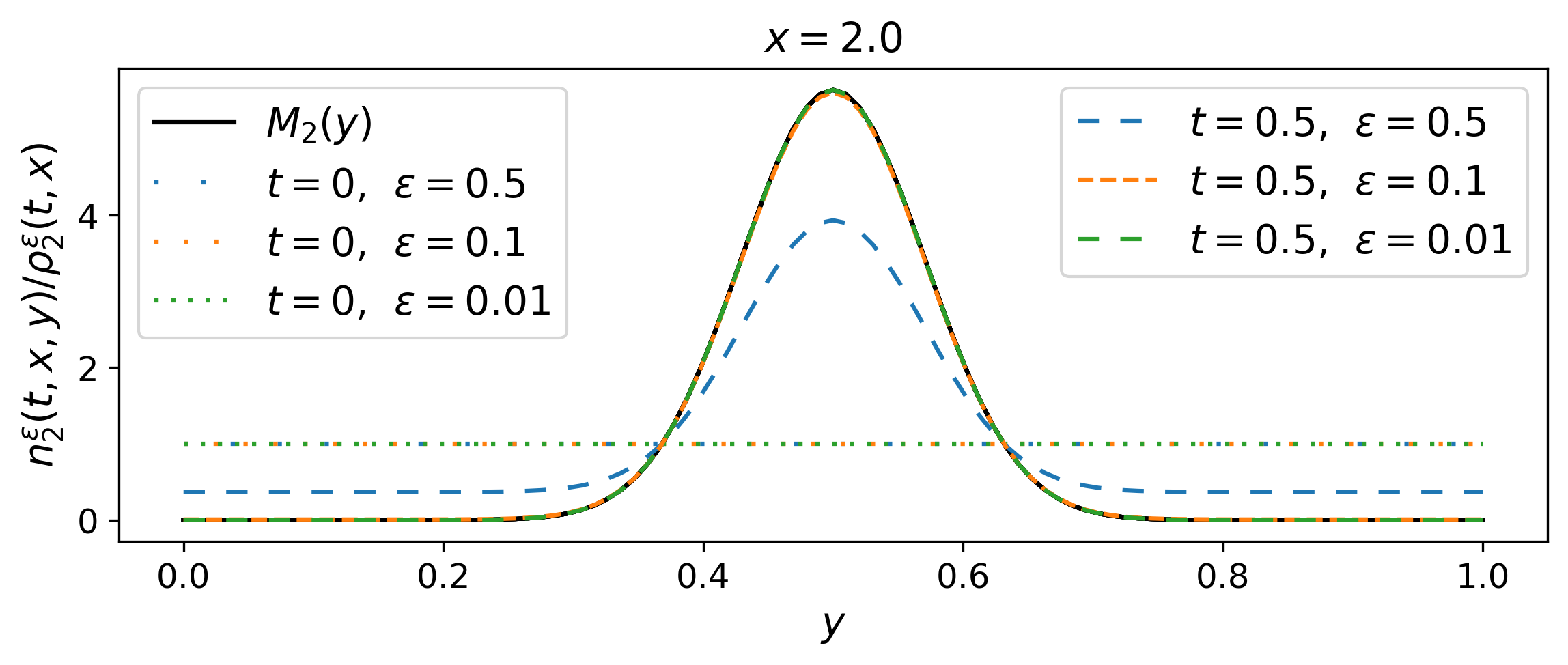}
    \includegraphics[width=0.45\linewidth]{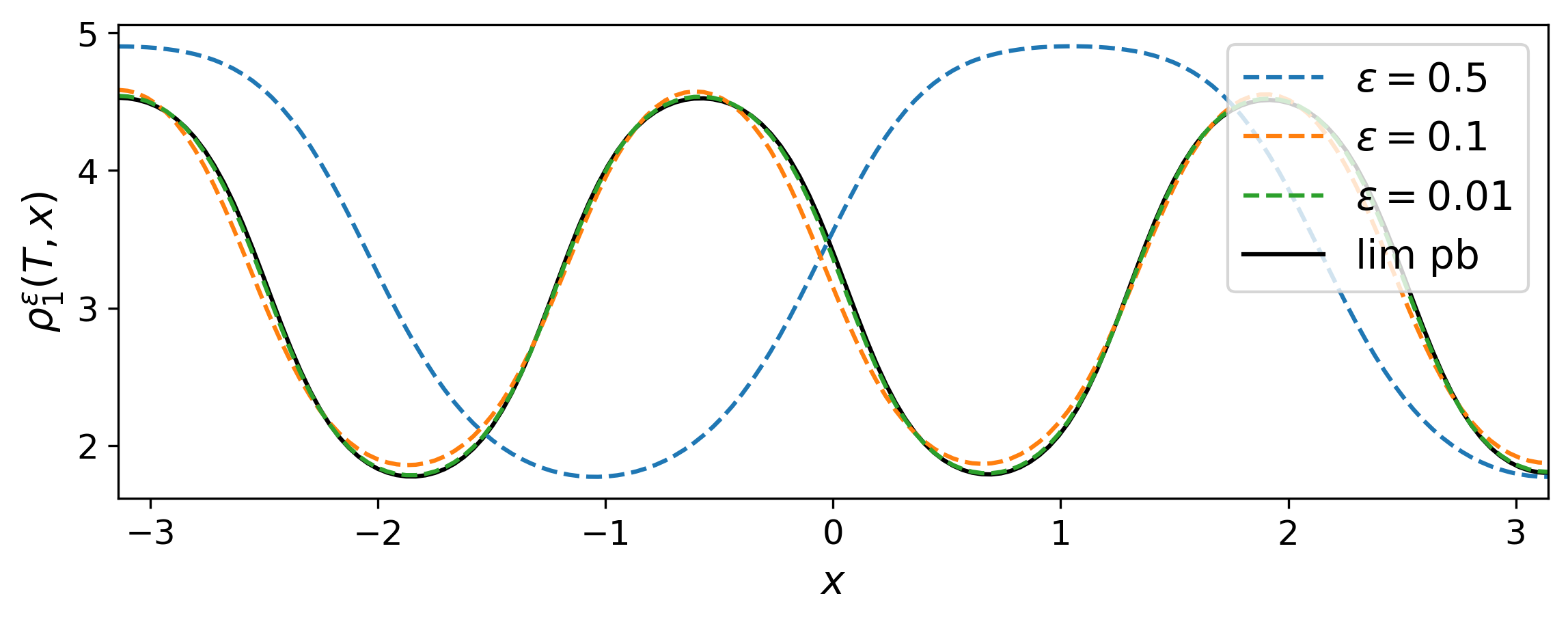}
    \includegraphics[width=0.45\linewidth]{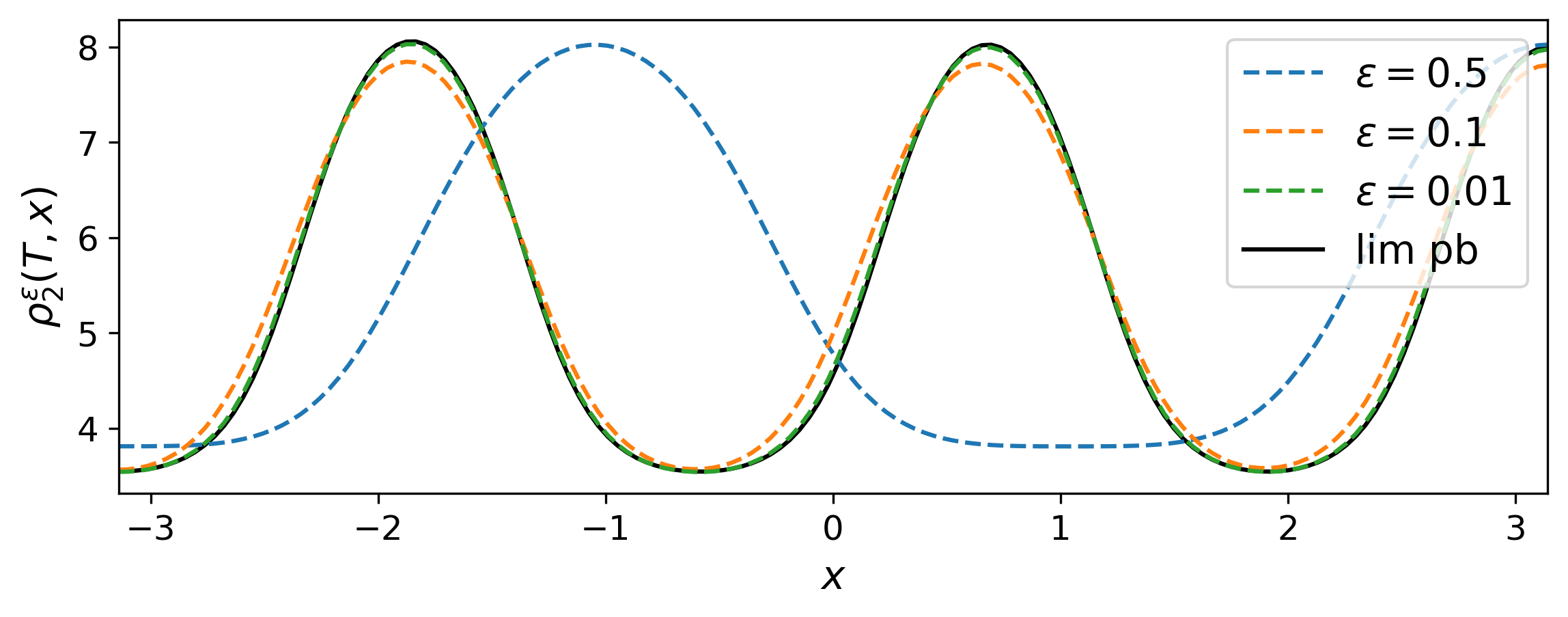}
    \caption{{\bf Agreement between the non-local problem~\eqref{eqn:PDE_n1_n2_mutational_kernel_rescaled}-\eqref{eq:bcs_rescaled} as $\varepsilon \to 0$ and the limiting problem~\eqref{eqn:SKT_phenotype}-\eqref{eq:bcs_rescaledrho_limit}.} Plots of $n^\varepsilon_i(t,x,y_i)/\rho^\varepsilon_i(t,x)$ at $x=2$ for $t =0$ and $t=0.5$ (top panel, coloured lines) and $\rho^\varepsilon_i(T,x)$  (bottom panel, coloured lines) for $i=1$ (left column) and $i=2$ (right column), obtained by solving numerically the phenotype-structured SKT model~\eqref{eqn:PDE_n1_n2_mutational_kernel_rescaled}, subject to the boundary conditions~\eqref{eq:bcs_rescaled} and the initial data~\eqref{eqn:numerics_initial_condition}. The set-up of numerical simulations is described in Section~\ref{sec:setnum}, with the scaling parameter $\varepsilon \in \{0.5,0.1,0.01\}$ and the final time of simulations $T=1000$. The model functions and parameters employed are the same as those of Figure~\ref{fig:EXP1_patterns_SI}. The black lines in the top panels highlight $M_1(y_1)$ and $M_2(y_2)$, while the black lines in the bottom panels highlight $\rho_1$ and $\rho_2$ obtained by solving numerically the limiting problem~\eqref{eqn:SKT_phenotype}-\eqref{eq:bcs_rescaledrho_limit}, subject to initial data analogous to~\eqref{eqn:numerics_initial_condition}, under the effective parameter values in Table~\ref{table:numerics-SKT} for Figure~\ref{fig:EXP1_patterns_SI}. }  
    \label{fig:EXP1-comparison-for-different-epsilon}
\end{figure}

\paragraph{Agreement between the non-local problem~\eqref{eqn:PDE_n1_n2_mutational_kernel_rescaled}-\eqref{eq:bcs_rescaled} as $\varepsilon \to 0$ and the limiting problem~\eqref{eqn:SKT_phenotype}-\eqref{eq:bcs_rescaledrho_limit}.} 
As shown by the plots in the middle and bottom panels of Figures~\ref{fig:EXP1_patterns_NO} and~\ref{fig:EXP1_patterns_SI}, the shape of $n_i^\varepsilon(T,x,y_i)$ for different values of $y_i$ is the same as that of $\rho_i^\varepsilon(T,x)$ up to a multiplicative factor, both for $i=1$ and $i=2$. This is due to the fact that, consistently with the formal calculations of Section~\ref{sec:mod1}, for $\varepsilon$ sufficiently small we have $n_i^\varepsilon(T,x,y_i) \approx \rho_i^\varepsilon(T,x) \, M_i(y_i)$ for almost every $t > 0$. In this regard, the results of Figures~\ref{fig:EXP1_patterns_NO} and~\ref{fig:EXP1_patterns_SI} are complemented by those of the bottom panels of Figure~\ref{fig:EXP1-comparison-for-different-epsilon}, which demonstrate that reducing the value of $\varepsilon$ leads to an excellent quantitative agreement between $\rho^\varepsilon_1$ and $\rho^\varepsilon_2$ computed as the integrals of the components of the numerical solution to the non-local problem~\eqref{eqn:PDE_n1_n2_mutational_kernel_rescaled}-\eqref{eq:bcs_rescaled}, subject to the initial data~\eqref{eqn:numerics_initial_condition}, and the components $\rho_1$ and $\rho_2$ of the numerical solution to the limiting problem~\eqref{eqn:SKT_phenotype}-\eqref{eq:bcs_rescaledrho_limit}, subject to initial data analogous to~\eqref{eqn:numerics_initial_condition}. Moreover, as shown by the plots in the top panels of Figure~\ref{fig:EXP1-comparison-for-different-epsilon}, for $\varepsilon$ sufficiently small the components $n_1^\varepsilon$ and $n_2^\varepsilon$ of the numerical solution to the non-local problem~\eqref{eqn:PDE_n1_n2_mutational_kernel_rescaled}-\eqref{eq:bcs_rescaled} are such that $n_1^\varepsilon(t,x,y_1)/\rho^\varepsilon_1(t,x) \approx M_1(y_1)$ and $n_2^\varepsilon(t,x,y_2)/\rho^\varepsilon_2(t,x) \approx M_2(y_2)$ for almost every $t > 0$. Taken together, these results validate the formal procedure employed in Section~\ref{sec:mod1} to obtain the limiting problem~\eqref{eqn:SKT_phenotype}-\eqref{eq:bcs_rescaledrho_limit} from the non-local problem~\eqref{eqn:PDE_n1_n2_mutational_kernel_rescaled}-\eqref{eq:bcs_rescaled} in the asymptotic regime $\varepsilon \to 0$. \rr{Finally, it is worth noting that for larger values of $\varepsilon$ (e.g. $\varepsilon = 0.5$ in Figure~\ref{fig:EXP1-comparison-for-different-epsilon}), the non-local problem~\eqref{eqn:PDE_n1_n2_mutational_kernel_rescaled}-\eqref{eq:bcs_rescaled} can still produce spatial patterns, which may however substantially differ from those obtained in the limiting case $\varepsilon \to 0$.}

\begin{figure}[!ht]
  \centering
  \begin{tabular}{@{}>{\centering\arraybackslash}m{1cm}@{} m{1.1\linewidth}}
    \textbf{A \squaremarker} & \includegraphics[width=0.42\linewidth]{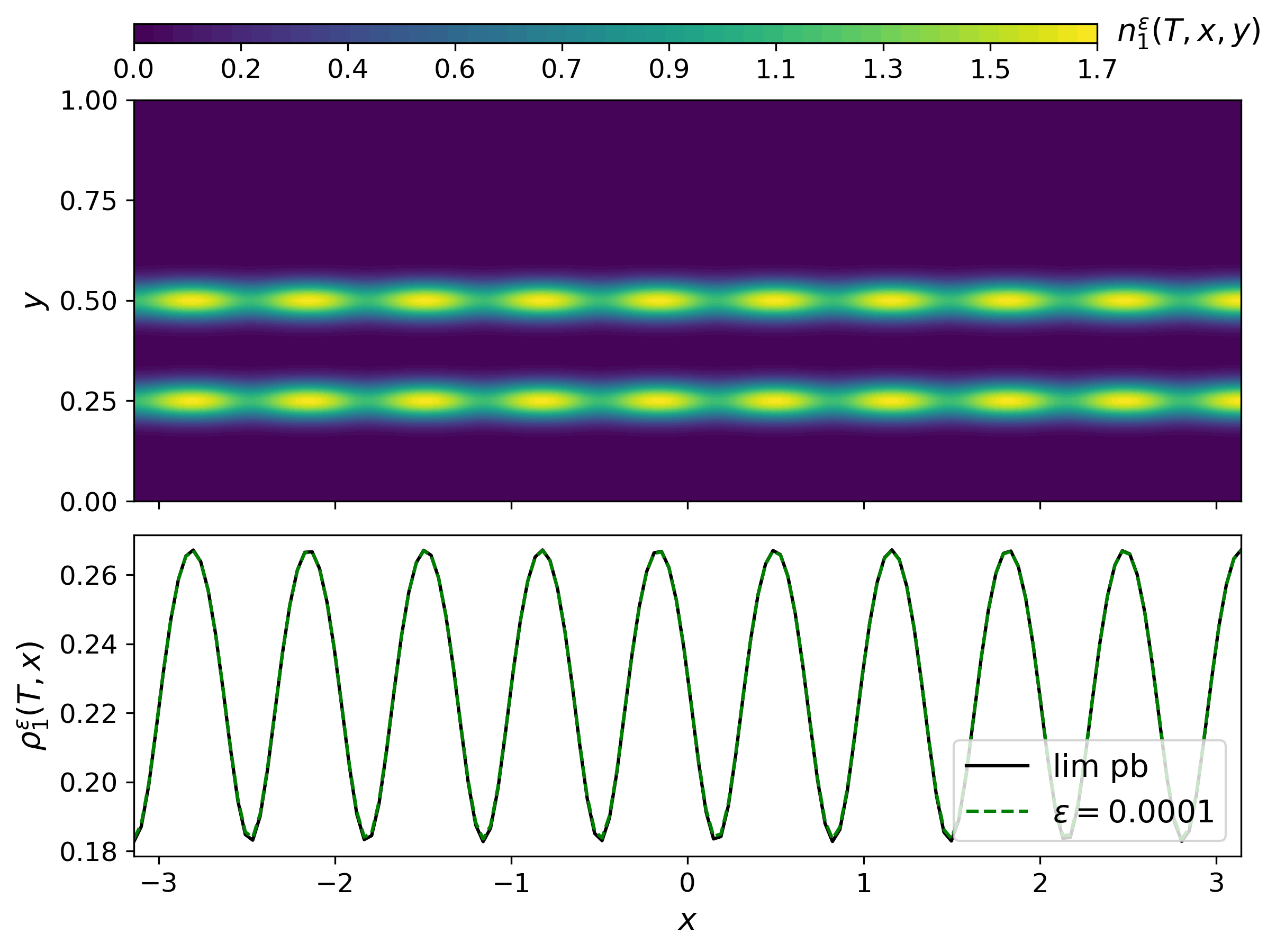}
    \includegraphics[width=0.42\linewidth]{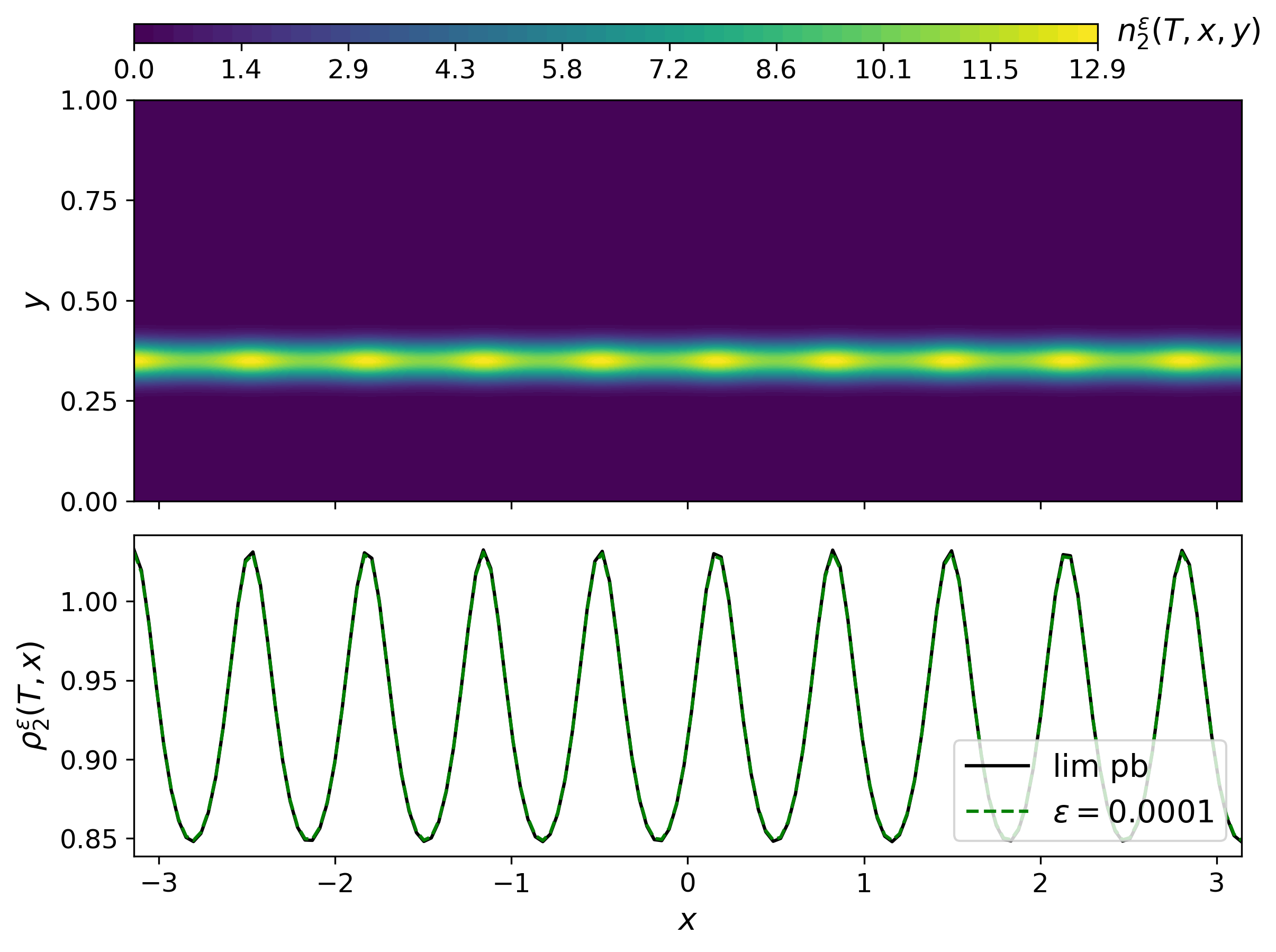} \\
    \textbf{B \circlemarker} & \includegraphics[width=0.42\linewidth]{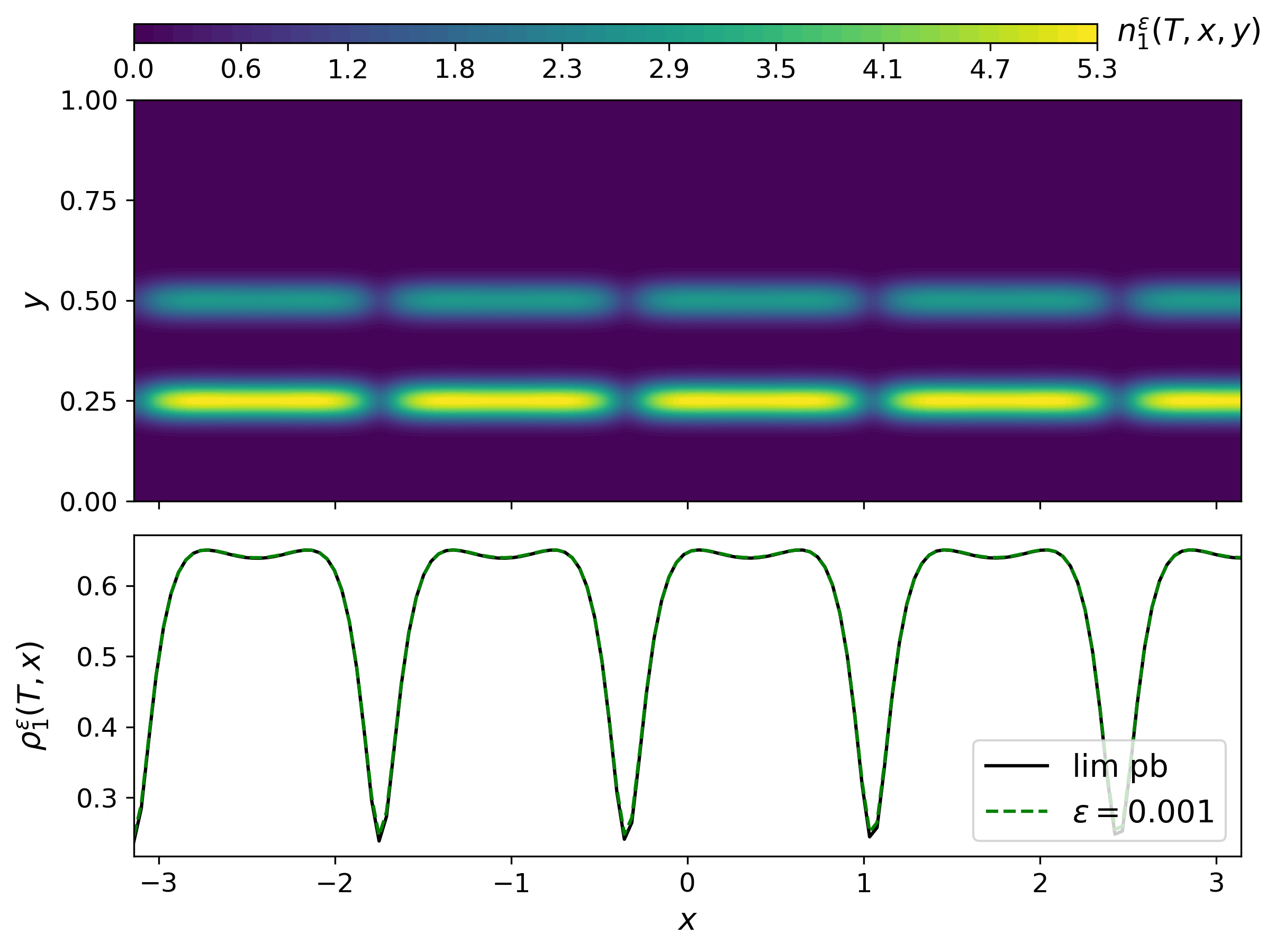}
    \includegraphics[width=0.42\linewidth]{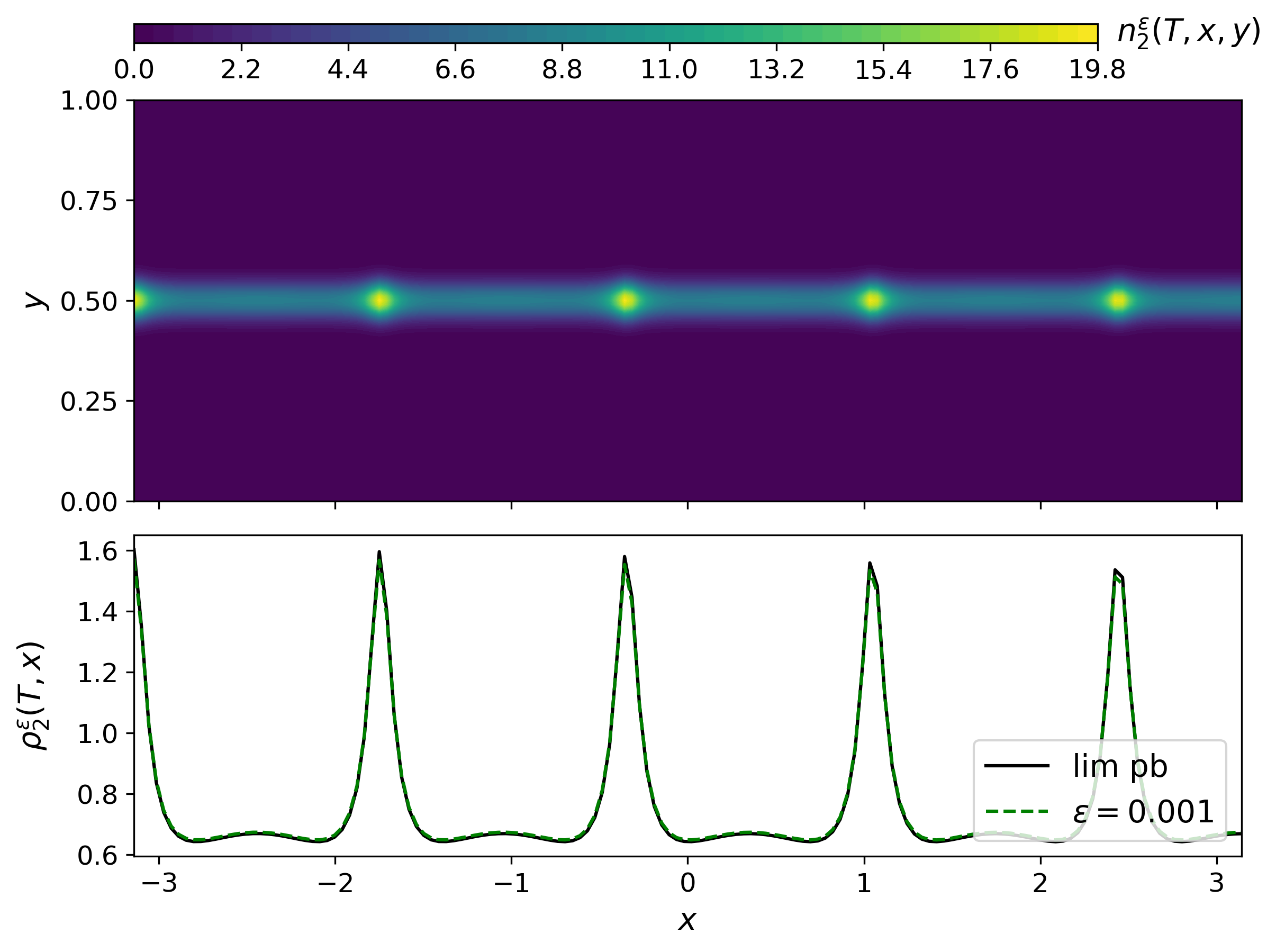}
  \end{tabular}
  \caption{{\bf Dependence of the form of spatial patterns on the critical mode $k_c$.} Plots of $n^\varepsilon_i(T,x,y_i)$ (top sub-panel) and $\rho^\varepsilon_i(T,x)$  (bottom sub-panel, coloured lines) for $i=1$ (left column) and $i=2$ (right column), obtained by solving numerically the phenotype-structured SKT model~\eqref{eqn:PDE_n1_n2_mutational_kernel_rescaled}, subject to the boundary conditions~\eqref{eq:bcs_rescaled} and the initial data~\eqref{eqn:numerics_initial_condition}. The set-up of numerical simulations is described in Section~\ref{sec:setnum}, with the scaling parameter $\varepsilon= 0.0001$ and the final time of simulations $T=1000$. The model functions and parameters employed, which are listed in Table~\ref{table:numerics-PSPDE}, are such that $\la \alpha_{21} \ra > \la{\alpha}_{21}\ra_c$ (for both panels A and panels B) and $k_c=9.16$ (panels A) and $k_c=5.86$ (panels B). The Turing-type bifurcation driving pattern formation is sub-critical in the case of panels B (i.e. $L = -2.455$) and super-critical in the case of panels A (i.e. $L = 41.837$). \rr{Panels A and B correspond to the values of $(\lambda_1, {\mu_{11}})$ highlighted by the orange circle and the blue square, respectively, in Figures~\ref{fig:Turing-diagrams} and~\ref{fig:alpha21-kc-as-functions-of-mu2-lambda1}}. The black lines in the bottom sub-panels highlight $\rho_1$ and $\rho_2$ obtained by solving numerically the limiting problem~\eqref{eqn:SKT_phenotype}-\eqref{eq:bcs_rescaledrho_limit}, subject to initial data analogous to~\eqref{eqn:numerics_initial_condition}, under the effective parameter values in Table~\ref{table:numerics-SKT}.    }\label{fig:EXP3}
\end{figure}

\paragraph{Dependence of the form of spatial patterns on the critical mode $k_c$.} The results of numerical simulations displayed in Figure~\ref{fig:EXP3} show that, consistently with the analytical results presented in Section~\ref{Sec:Turing_instability}, the shape of the spatial patterns that are formed \rr{depends} on the critical mode $k_c$ given by~\eqref{kc_comp}. In fact, the patterns of Figure~\ref{fig:EXP3}A, which correspond to $k_c=9.16$, differ from those of Figure~\ref{fig:EXP3}B, which correspond to $k_c=5.86$. The Turing-type bifurcation driving pattern formation is sub-critical in the case of Figure~\ref{fig:EXP3}B, since the quantity $L$ given by~\eqref{SL} is negative (i.e. $L = -2.455$), whereas it is super-critical in the case of Figure~\ref{fig:EXP3}A, as $L$ is positive (i.e. $L = 41.837$). In the latter case the amplitudes of $\rho_1^\varepsilon(T,x)$ and $\rho_2^\varepsilon(T,x)$, computed as
$$
\dfrac{1}{2} \left(\max_{x \in \Omega} \rho_1^\varepsilon(T,x) - \min_{x \in \Omega} \rho_1^\varepsilon(T,x)\right) \approx 0.04177
$$
and
$$
\dfrac{1}{2} \left(\max_{x \in \Omega} \rho_2^\varepsilon(T,x) - \min_{x \in \Omega} \rho_2^\varepsilon(T,x)\right) \approx 0.09140,
$$
are in agreement with the corresponding estimates obtained through weakly nonlinear analysis in Section~\ref{sec:wnla}, which are given by~\eqref{eq:winfty} and are $0.04497$ and $0.09603$, respectively.

Note that, as detailed in Table~\ref{table:numerics-PSPDE}, for Figures~\ref{fig:EXP3}A and~\ref{fig:EXP3}B: the phenotype-dependent cross-diffusion coefficients $\alpha_{12}(y_1)$ and $\alpha_{21}(y_2)$, defined via~\eqref{eqn:numerics_csdiff_functions} and~\eqref{eqn:numerics_hij_functions}, exhibit a sharp transition at $y_1=0.5$ and $y_2=0.5$, respectively;  the choices of the model parameters and functions are the same, except for the value of the parameter $\lambda_1$ in the definition~\eqref{eqn:numerics_Mi_functions_2peaks} of the phenotype switching kernel $M_1$ and the value of the parameter $\mu_2$ in the definition~\eqref{eqn:numerics_Mi_functions} of the phenotype switching kernel $M_2$, that is, $\lambda_1 = 0.5$ and $\mu_2=0.35$ for the simulations of Figure~\ref{fig:EXP3}A and $\lambda_1=0.35$ and $\mu_2=0.5$ for the simulations of Figure~\ref{fig:EXP3}B. Note also that, while the value of the bifurcation threshold $\la{\alpha}_{21}\ra_c$ is different between the simulations of Figure~\ref{fig:EXP3}A and Figure~\ref{fig:EXP3}B, the condition $\la \alpha_{21} \ra > \la{\alpha}_{21}\ra_c$ is met for the simulations of both Figure~\ref{fig:EXP3}A and Figure~\ref{fig:EXP3}B. Hence, the results of Figure~\ref{fig:EXP3} further corroborate the idea that relatively small differences in phenotype switching kernels can lead to substantial differences in the spatial patterns produced by the phenotype-structured SKT model~\eqref{eqn:PDE_n1_n2_mutational_kernel_rescaled} in scenarios where the phenotype-dependent cross-diffusion coefficient, the average of which acts as bifurcation parameter, undergoes sharp variations in phenotype. 

\begin{figure}[!ht]
	\centering
	\includegraphics[width=0.45\linewidth]{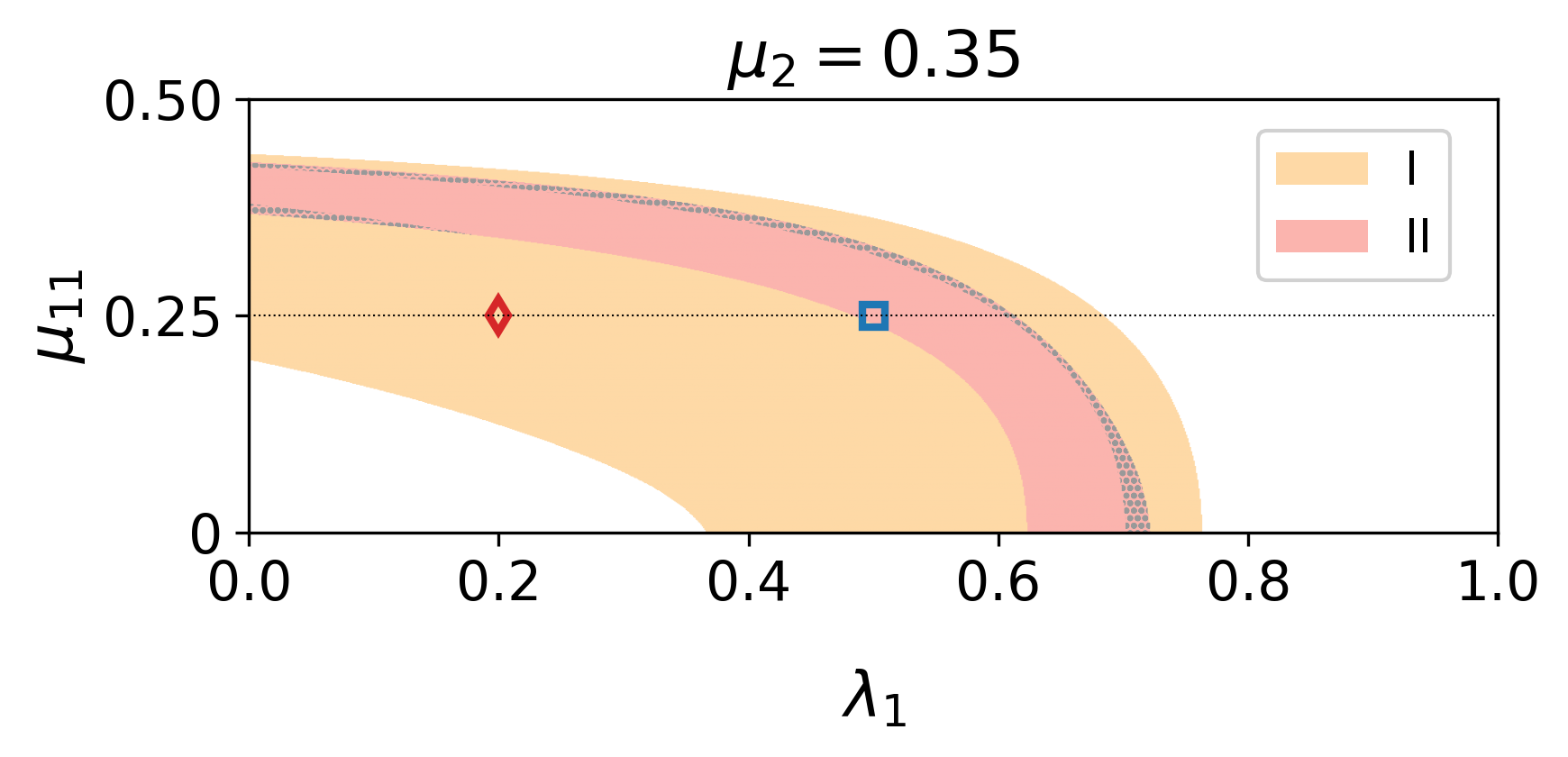}	
	\includegraphics[width=0.45\linewidth]{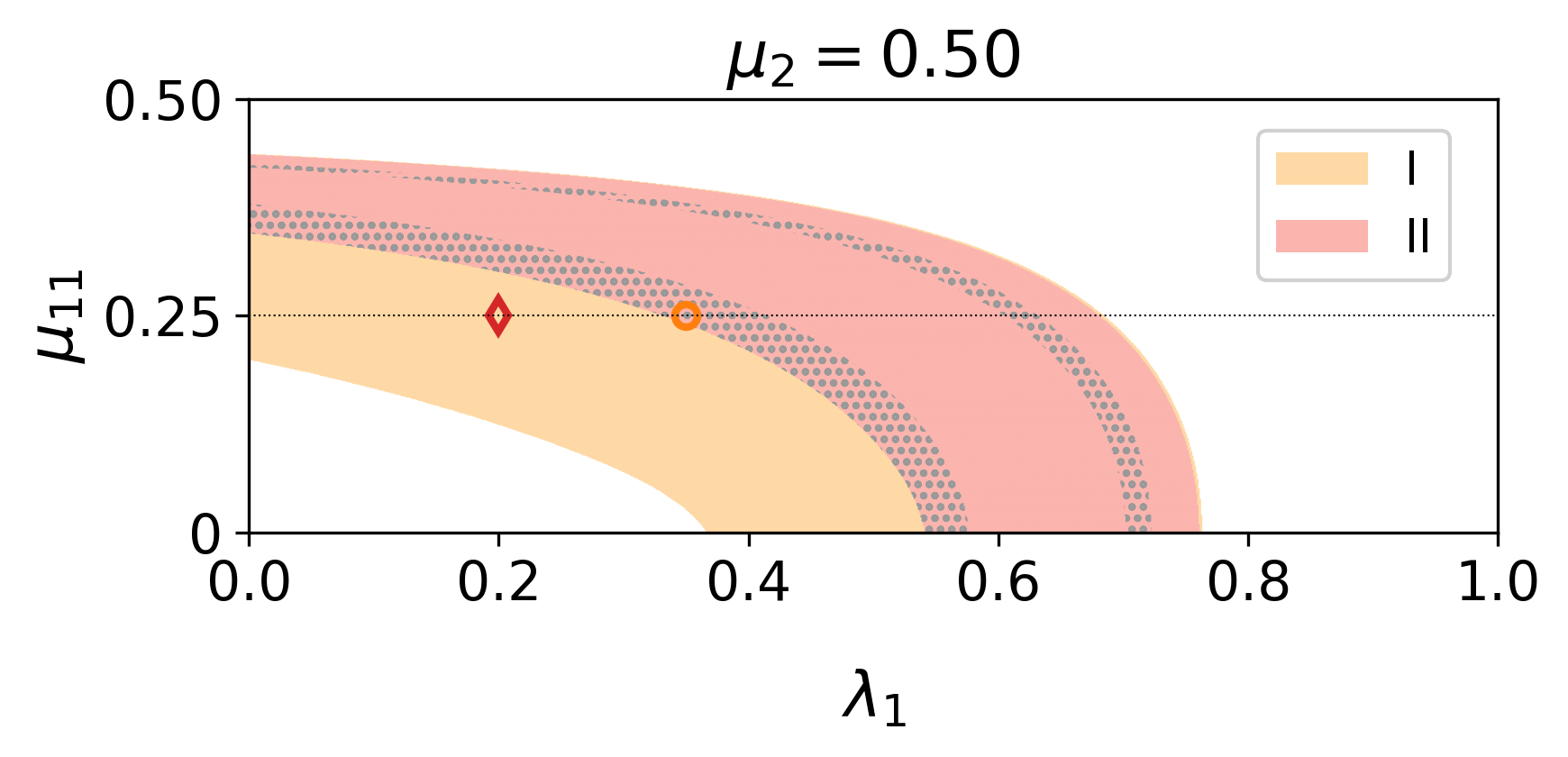}
	\caption{
	\rr{
		{\bf Phenotype distributions act as eﬀective control parameters for pattern formation: impact of phenotype structuring on the coexistence uniform-in-space steady state and Turing instability.} Plots of the Turing instability diagrams in the parameter space $(\lambda_1, \mu_{11})$ (i.e. the phenotype switching parameters in the definition~\eqref{eqn:numerics_Mi_functions_2peaks} of $M_1(y_1)$) for the limiting problem \eqref{eqn:SKT_phenotype}-\eqref{eq:bcs_rescaledrho_limit}. The plots in the left and right panels correspond to a different choice of the value of the parameter $\mu_2$ in the definition~\eqref{eqn:numerics_Mi_functions} of $M_2(y_2)$.
		The region II is the Turing instability region, while region I is the existence and stability region of the coexistence uniform-in-space steady state. In the rest of the parameter space, such a steady state does not exist (blank areas). The dotted subregions within region II correspond to the sub-critical regime ($L<0$), while the rest of region II is super-critical ($L>0$).
		All the other effective parameters in \eqref{eqn:SKT_phenotype}-\eqref{eq:bcs_rescaledrho_limit} correspond to the model functions underlying the simulation results of Figure~\ref{fig:EXP3}. In particular, the blue square in the left panel diagram and the orange circle in the right panel diagram highlight the values of $(\lambda_1, \mu_{11})$, that is, (0.50, 0.25) and (0.35, 0.25), respectively, corresponding to the simulation results of Figure~\ref{fig:EXP3}A and Figure~\ref{fig:EXP3}B. Moreover, the red diamonds highlight an exemplar value of $(\lambda_1, \mu_{11})$, i.e. (0.20, 0.25), for which numerical simulations were carried out to verify that the coexistence uniform-in-space steady state is stable for parameter values corresponding to region I (results not shown).}
}
	\label{fig:Turing-diagrams}
\end{figure}

\rr{These results correspond to the values of $(\lambda_1, {\mu_{11}})$ highlighted by the blue square and the orange circle within the Turing instability regions in the diagrams of Figure~\ref{fig:Turing-diagrams}. Simulations were also carried out for the value of $(\lambda_1, {\mu_{11}})$ highlighted by the red diamonds in the diagrams of Figure~\ref{fig:Turing-diagrams}, i.e. $(\lambda_1, {\mu_{11}}) = (0.20, 0.25)$, which was chosen as an exemplar corresponding to stability of the coexistence uniform-in-space steady state. As expected, in this case, spatial heterogeneity is suppressed and the numerical solution tends to a uniform-in-space steady state, which may however depend on the chosen value of $\varepsilon$ (results not shown).
}

\paragraph{Phenotype distributions act as eﬀective control parameters for pattern formation.} The results of \rr{the} numerical simulations displayed in Figures~\ref{fig:EXP1_patterns_NO} and~\ref{fig:EXP1_patterns_SI} and in Figure~\ref{fig:EXP3} show, respectively, that phenotype structuring can impact both the formation/suppression and the shape of spatial patterns. 
\rr{
To provide a broader perspective on this, Figure~\ref{fig:Turing-diagrams} displays the Turing instability diagrams in the $(\mu_{11},\lambda_1)$ parameter space for two different choices of $M_2(y_2)$. These diagrams highlight how changes in $M_1(y_1)$ resulting from changes in the values of the phenotype switching parameters $\mu_{11}$ and $\lambda_1$ can modify the existence regions of the coexistence uniform-in-space steady state and occurrence of Turing instability. Moreover, comparing the two diagrams in Figure~\ref{fig:Turing-diagrams} one sees that changing $M_2(y_2)$ can alter the extent of the Turing instability region, even when the other model functions and parameters are kept fixed. In particular, the transition between super- and sub-critical regimes, identified by the sign of the cubic coefficient $L$ in the Stuart--Landau equation, demonstrates that phenotype structuring controls not only the onset of Turing instability, but also the nonlinear character of the corresponding bifurcation.}

\begin{figure}[!ht]
	\centering
	\makebox[\textwidth][c]{%
		\includegraphics[width=0.36\textwidth]{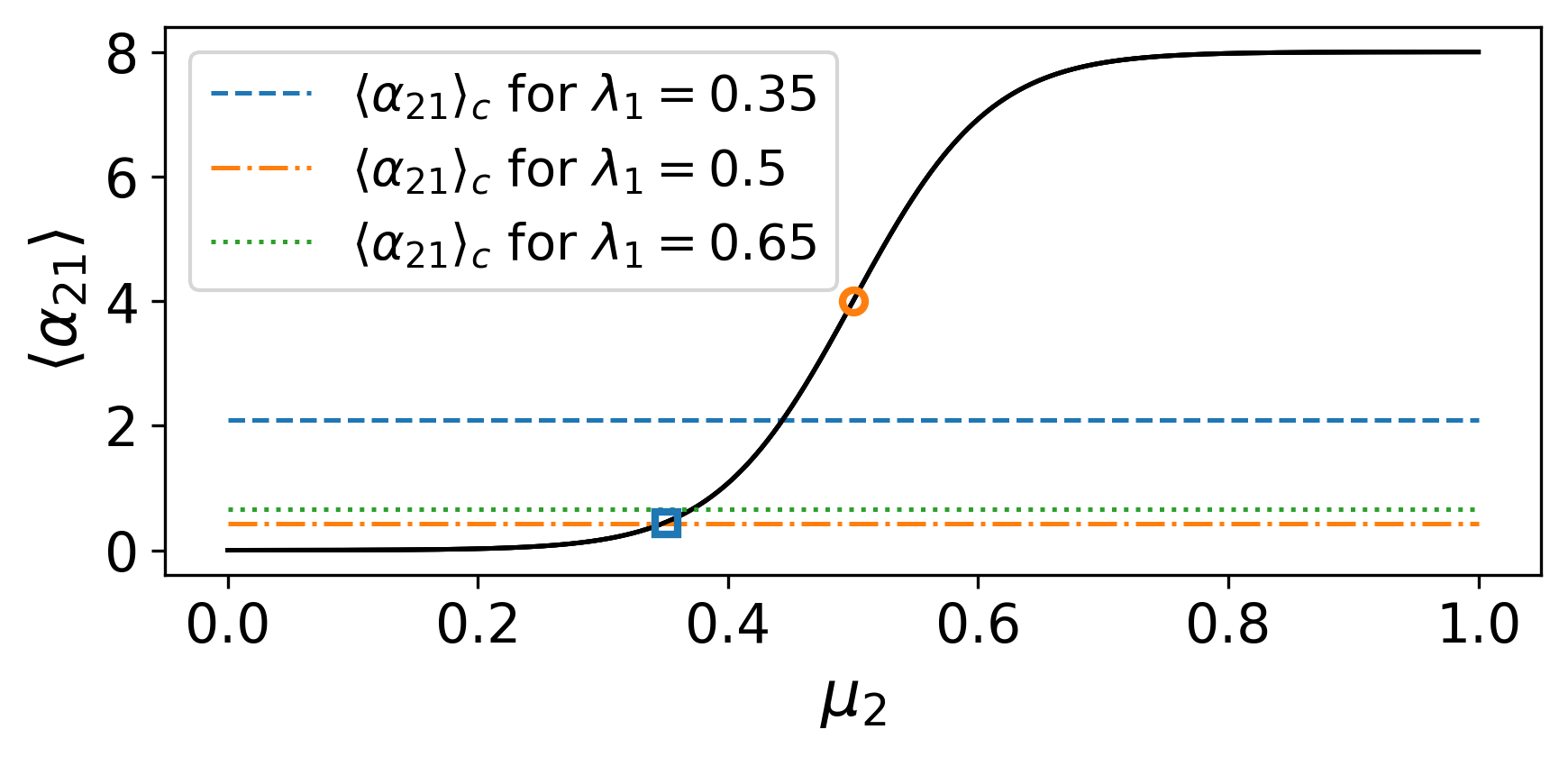}
		\includegraphics[width=0.36\textwidth]{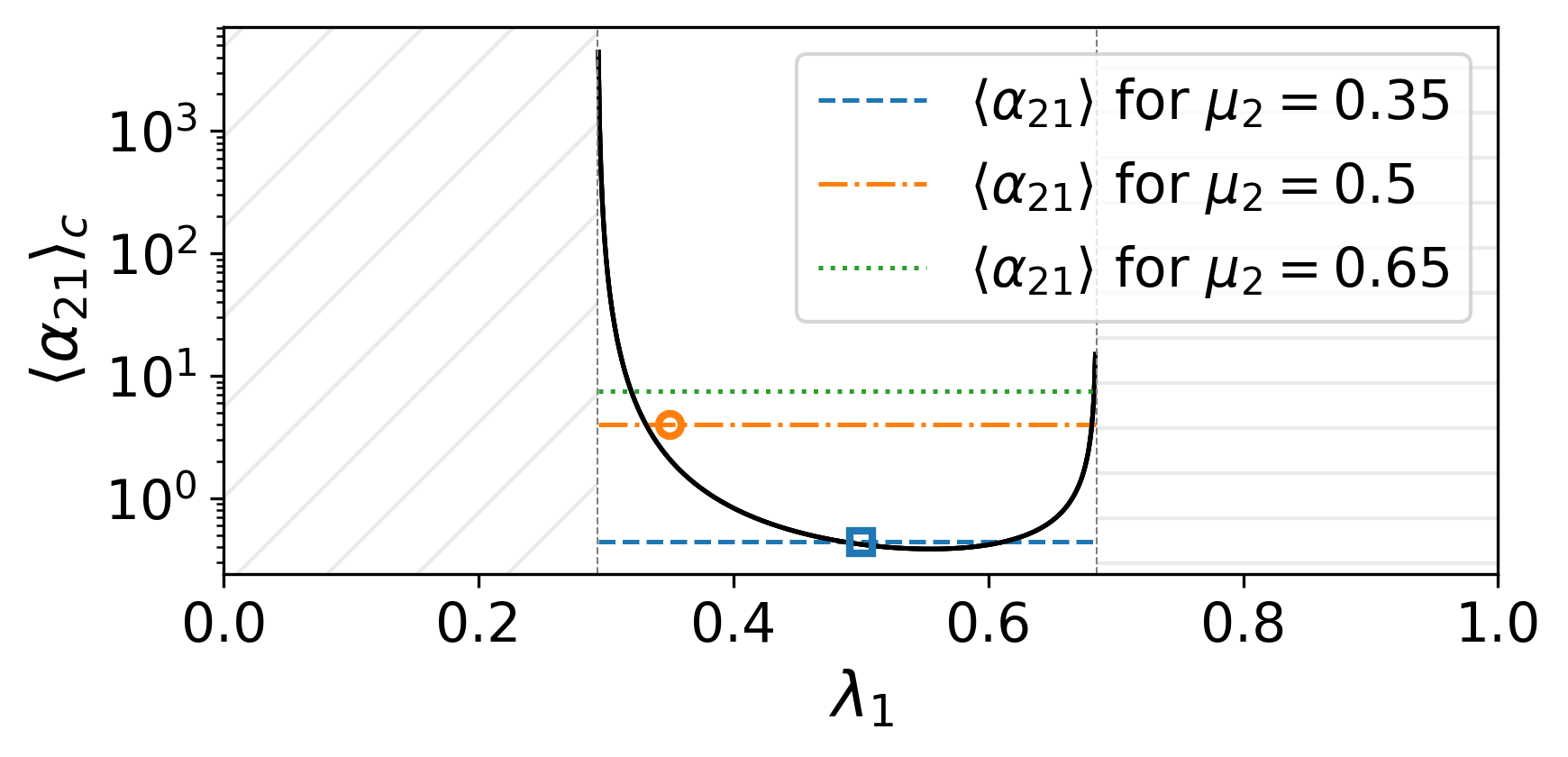}
		\includegraphics[width=0.36\textwidth]{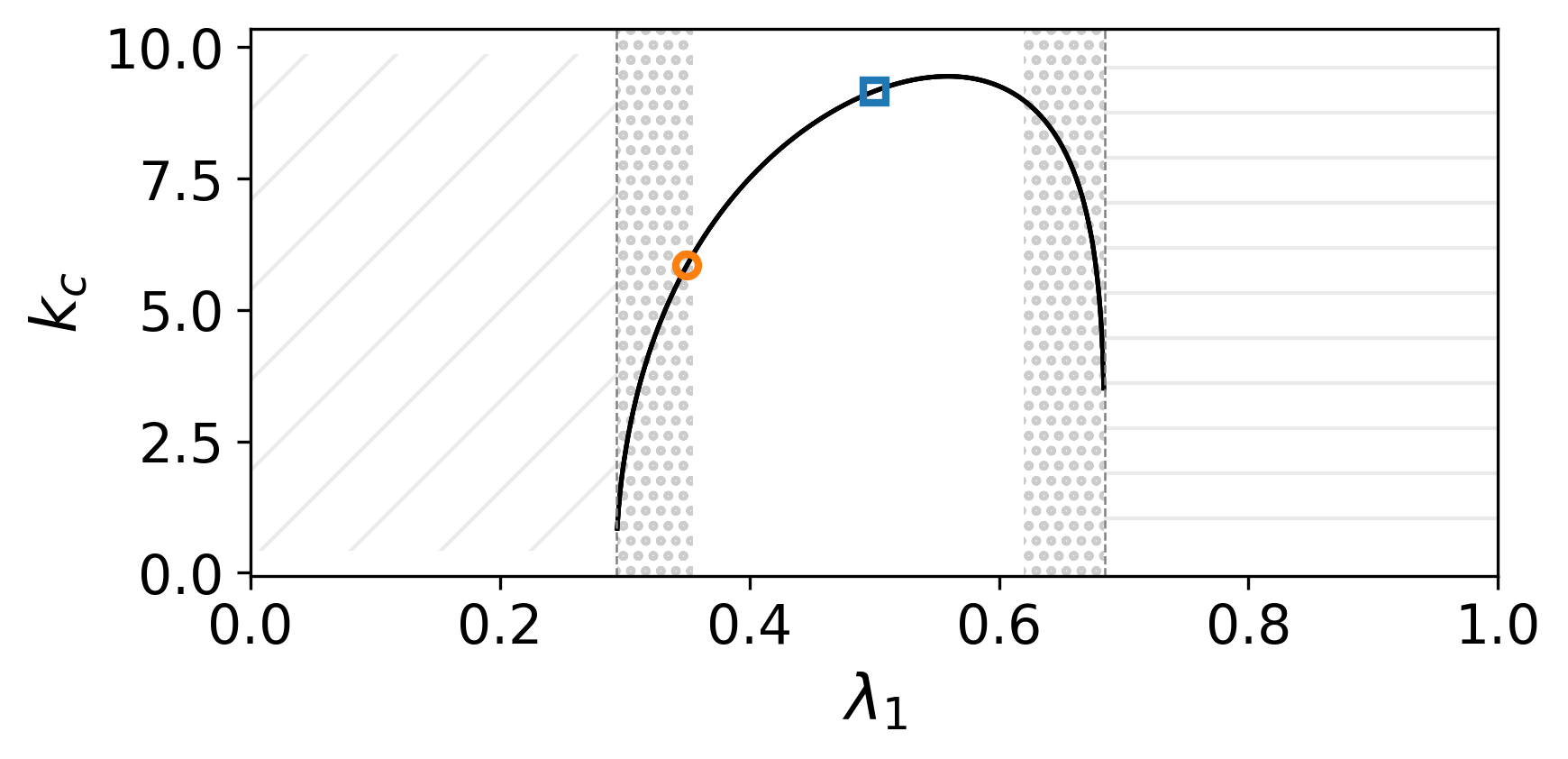}
	}
	\caption{
		\rr{
		{\bf Phenotype distributions act as eﬀective control parameters for pattern formation: impact of phenotype structuring on the Turing threshold and the critical spatial mode.} Plots of the bifurcation parameter $\la\alpha_{21}\ra$, defined via~\eqref{eqn:def<Di><alphaij>sim}, as a function of the parameter $\mu_2$ in the definition~\eqref{eqn:numerics_Mi_functions} of $M_2(y_2)$ (left panel, black line), and the bifurcation threshold $\la\alpha_{21}\ra_c$, given by \eqref{TT_comp_}, and the critical mode $k_c$, given by~\eqref{kc_comp}, as functions of the parameter $\lambda_1$ in the definition~\eqref{eqn:numerics_Mi_functions_2peaks} of $M_1(y_1)$ (central and right panels, black lines), for the choices of the model functions underlying the simulation results of Figure~\ref{fig:EXP3}. The coloured lines in the left panel highlight $\la \alpha_{21}\ra_c$, while the coloured lines in the central panel highlight $\la\alpha_{21}\ra$, with the blue and orange lines highlighting the values of $\la \alpha_{21}\ra_c$ (left panel) and $\la\alpha_{21}\ra$ (central panel) corresponding to the simulation results of Figures~\ref{fig:EXP3}A and~\ref{fig:EXP3}B, respectively. Moreover, the blue square and the orange circle highlight the values of  $\la\alpha_{21}\ra$ (left panel), $\la\alpha_{21}\ra_c$ (central panel), and $k_c$ (right panel) corresponding to the simulation results of Figures~\ref{fig:EXP3}A and~\ref{fig:EXP3}B, respectively. In the central and right panels, the diagonally hatched region, where $\lambda_1\lesssim0.29$, corresponds to the range of values of $\lambda_1$ for which the coexistence uniform-in-space steady state is stable to space-dependent perturbations, and thus pattern formation does not occur, while the horizontally hatched region, where $\lambda_1\gtrsim0.68$, corresponds to the range of values of $\lambda_1$ for which the coexistence uniform-in-space steady state ceases to exist. In the right panel, the grey dot-patterned regions correspond to the range of values of $\lambda_1$ for which $L<0$ (i.e. the Turing-type bifurcation is sub-critical), while the blank region corresponds to the range of values of $\lambda_1$ (i.e. $0.36 \lesssim \lambda_1  \lesssim 0.62$) for which $L>0$ (i.e. the Turing-type bifurcation is super-critical). These plots correspond to the profiles for $\mu_{11}=0.25$ marked by the black-dotted horizontal lines in the diagrams of Figure~\ref{fig:Turing-diagrams}.
	}}
	\label{fig:alpha21-kc-as-functions-of-mu2-lambda1}
\end{figure}

\rr{To further elucidate the impact of phenotype structuring on pattern formation, the plots in Figure~\ref{fig:alpha21-kc-as-functions-of-mu2-lambda1} 
show how the bifurcation parameter $\la\alpha_{21}\ra$, the critical threshold $\la\alpha_{21}\ra_c$, and the critical spatial mode $k_c$ depend on the shape of $M_1(y_1)$ and $M_2(y_2)$.} In more detail, the plots in the left panel of Figure~\ref{fig:alpha21-kc-as-functions-of-mu2-lambda1} show that, increasing the value of the parameter $\mu_2$ in the definition~\eqref{eqn:numerics_Mi_functions} of $M_2(y_2)$ leaves the value of the bifurcation threshold $\la \alpha_{21}\ra_c$ unchanged while causing an increase in the value of the bifurcation parameter $\la \alpha_{21}\ra$. This is to be expected since, for the choices of the model functions underlying the simulation results of Figure~\ref{fig:EXP3}, amongst the competition kernels and the linear diffusion and the self-diffusion/cross-diffusion coefficients, the only $y_2$-dependent function is the cross-diffusion coefficient $\alpha_{21}$. Increasing the value of the parameter $\mu_2$ can then drive the system across the Turing-type bifurcation threshold, thereby activating pattern formation. \rr{Since for the small values of $\varepsilon$ considered here $n_2^\varepsilon(t,x,y_2) \approx \rho^\varepsilon_2(t,x) M_2(y_2)$ for almost every $t > 0$, and thus the phenotype distribution of population 2 is determined by $M_2(y_2)$, these findings demonstrate how the phenotype distribution acts as an effective control parameter for Turing instability, activating or suppressing the onset of the instability by determining the value of the phenotype-averaged effective cross-diffusion coefficient $\la \alpha_{21}\ra$.}

Moreover, as shown by the plots in the central and right panels of Figure~\ref{fig:alpha21-kc-as-functions-of-mu2-lambda1}, changing the value of the parameter $\lambda_1$ in the definition~\eqref{eqn:numerics_Mi_functions_2peaks} of $M_1(y_1)$ leaves the value of the bifurcation parameter $\la \alpha_{21}\ra$ unchanged while leading to changes in the value of the bifurcation threshold $\la \alpha_{21}\ra_c$ and, consequently, of the critical modes $k_c$, as a result of changes simultaneously induced in the values of several of the other phenotype-averaged effective parameters defined via~\eqref{eqn:def<Kij>sim} and \eqref{eqn:def<Di><alphaij>sim}. This is again to be expected since, for the choices of the model functions underlying the simulation results of Figure~\ref{fig:EXP3}, amongst the competition kernels and the linear diffusion and the self-diffusion/cross-diffusion coefficients, there is a number of $y_1$-dependent functions. \rr{Since for the small values of $\varepsilon$ considered here $n_1^\varepsilon(t,x,y_1) \approx \rho^\varepsilon_1(t,x) M_1(y_1)$ for almost every $t > 0$ and, therefore, the phenotype distribution of population 1 is determined by $M_1(y_1)$, these findings indicate that the phenotype distribution acts as an effective control parameter not only for Turing instability but also for pattern selection.} In more detail, a phenotype distribution corresponding to larger values of the bifurcation threshold $\la \alpha_{21}\ra_c$ is more likely to suppress pattern formation, since larger values of the bifurcation parameter $\la \alpha_{21}\ra$ are required to make the coexistence uniform-in-space steady state unstable to space-dependent perturbations. Furthermore, changes in the phenotype distribution lead to changes in the critical mode $k_c$, which in turn lead to changes in the spatial scale of the emerging patterns. In fact, a phenotype distribution corresponding to larger values of $k_c$ leads to the emergence of spatially finer segregation structures, whereas a phenotype distribution corresponding to smaller values of $k_c$ brings about spatially coarser segregation structures. Note that, when approaching the boundary of the instability region, the bifurcation threshold increases while $k_c$ decreases, indicating that pattern formation becomes less likely to occur and the corresponding spatial structures get coarser. 

Finally, the plots in the central and right panels of Figure~\ref{fig:alpha21-kc-as-functions-of-mu2-lambda1} also show that, for certain values of $\lambda_1$, and thus for certain shapes of $M_1(y_1)$ (i.e. the phenotype distribution of population 1), the system can be driven out \rr{of} the Turing instability region (for $\lambda_1\lesssim0.29$) or the coexistence uniform-in-space steady state can \rr{cease} to exist (for $\lambda_1\gtrsim 0.68$), and the Turing-type bifurcation causing pattern formation can either be super-critical (for $0.36 \lesssim \lambda_1  \lesssim 0.62$) or sub-critical.  

In summary, for choices of the model functions considered here: changing the value of the parameter $\mu_2$  in the definition~\eqref{eqn:numerics_Mi_functions} of $M_2(y_2)$ leads to changes in the phenotype distribution of population 2 which affect the value of the bifurcation parameter $\la\alpha_{21}\ra$, thereby controlling the distance from the Turing-type bifurcation threshold without modifying the structure of the instability region; by contrast, changing the value of $\lambda_1$ in the definition~\eqref{eqn:numerics_Mi_functions_2peaks} of $M_1(y_1)$ leads to changes in the phenotype distribution of population 1 which modify both the value of the bifurcation threshold $\la \alpha_{21}\ra_c$ and the value of the critical spatial mode $k_c$, thus reshaping the instability region itself.

\section{Conclusions}
\label{sec:discussion}
In this work, we proposed a generalisation of the well-known SKT model to account for continuous phenotypic structuring across linear diﬀusion, self/cross-diﬀusion, and intra-/inter-population competition. First we have shown how a form of the classical SKT model, wherein parameters are written in terms of continuous weighted averages of the phenotype-dependent functions of the generalised structured model, with weights given by the phenotype distributions of the two populations, can be obtained in the quasi-invariant regime of fast phenotype switching. Then, still assuming fast phenotype switching and extending classical Turing-like linear and weakly nonlinear analyses, we have investigated the conditions for the emergence of spatial patterns, identified a Turing-type bifurcation threshold leading to pattern formation, and determined the nature of such a bifurcation (super- or sub-critical) as well as the stability of the patterned state.

\rr{An important consequence of considering the quasi-invariant fast phenotype-switching regime is that, in this regime, detailed information on the mechanisms driving pattern-forming dynamics of the phenotype-structured model can be extracted through analytical techniques for the corresponding phenotype-averaged limiting model. Of note is the fact that the effective coefficients of the limiting model retain dependence on the phenotype distributions of the two populations, thereby making it possible to connect phenotype structuring to Turing instability onset, critical mode selection, and the bifurcation criticality. In more detail,} the results obtained make it possible to draw connections between phenotype-dependent model functions and the emergence of population-scale aggregate spatial dynamics, highlighting how phenotype distributions can act as effective control parameters for diffusion-driven instability by affecting the onset of pattern formation, the instability region, the critical spatial mode, and the nonlinear character of the Turing-type bifurcation. These findings are complemented by numerical simulations, which validate the formal asymptotics and confirm the predictions of the pattern formation analyses.

While Turing-type pattern formation analyses have been previously carried out for non-local PDE models of evolutionary dynamics~\citep{genieys2006adaptive,genieys2007dynamics}, phenotypically structured model of chemotaxis~\citep{LorenziPainter2025}, and chemically structured model of signalling bacteria~\citep{ridgway2025weakly}, as far as we are aware, the one presented here is the first such analysis for a phenotype-structured SKT model. In this regard, it would be important to carry out rigorous linear and weakly nonlinear analyses in the asymptotic regime of fast phenotype switching, which in this paper we only studied on the formal level. From the analytical point of view, it would also be relevant to investigate global existence for the coupled system of non-local reaction-cross-diffusion equations of the proposed phenotype-structured SKT model and address whether an appropriately rescaled version of this model admits travelling wave solutions that exhibit phenotype structuring, a point that has been of significant interest in related models of spatial-evolutionary dynamics -- see the recent review~\citep{LorenziPainterVilla2025} and references therein.

\rr{From an application point of view, while the current study is not tied to a specific ecological system, spatially structured microbial communities provide one possible setting in which competition-dependent movement and reversible phenotype switching associated with different motility states may synergistically drive the dynamics of the system.}
Future investigations may therefore benefit from a focused ecological application, thereby refining the model assumptions and allowing certain key questions to be addressed. 

From a modelling standpoint, it would be interesting to extend the phenotype-structured model presented here by making the phenotype switching kernels space-time dependent, in order to take into account the fact that phenotype switching processes may be shaped by spatiotemporal environmental changes. In the light of the formal asymptotic results presented here, in the quasi-invariant regime of fast phenotype switching, we expect the phenotype distributions of the two populations in such an extended model to mirror spatiotemporal environmental changes, bringing about richer dynamics, including the formation of more complex spatial structures with composite shapes resulting from the fact that the critical spatial mode, and thus the form of the spatial patterns, may vary throughout the spatial domain depending on the local environmental conditions.
     
\rr{Although the present study focuses on the SKT model as a prototypical framework for competition-driven cross-diffusion and spatial segregation, the phenotype structuring approach and the fast-switching asymptotic employed here are not intrinsically restricted to such a particular model and may be adapted to other spatial population models with different forms of phenotype-dependent movement and interactions.}
Taken together, the results of this initial study point the way towards novel compelling research directions on phenotype-structured models of cross-diffusion competitive systems. 


\section*{Acknowledgements} 
The authors gratefully acknowledge support from the Italian Ministry of University and Research (MUR) through the grant PRIN2022-PNRR project (No. P2022Z7ZAJ) ``A Unitary Mathematical Framework for Modelling Muscular Dystrophies'' (CUP: E53D23018070001) funded by the European Union NextGenerationEU. GG and TL are members of INdAM-GNFM.

\appendix
\section{Numerical method}
\label{app:numericalmethod}
In this appendix we present the scheme employed to solve numerically the rescaled system of non-local reaction-cross-diffusion equations~\eqref{eqn:PDE_n1_n2_mutational_kernel_rescaled} subject to the homogeneous Neumann boundary conditions~\eqref{eq:bcs_rescaled}, which is based on finite differences for the differential terms and the Simpson $1/3$ rule for the integral terms, under assumptions~\eqref{eq:assumption_M_indipendent_of_y'}. For compactness of notation, the scheme is here presented for the original system of non-local reaction-cross-diffusion equations~ \eqref{eqn:PDE_n1_n2_mutational_kernel} subject to the homogeneous Neumann boundary conditions~\eqref{eq:bcs}. Moreover, for simplicity we focus on a one-dimensional spatial domain $\Omega$, but the scheme can easily be generalised to two- and three-dimensional spatial domains.

We introduce the temporal, spatial, and phenotypic steps $\Delta t, \Delta x, \Delta y > 0$, and consider the discrete points $t_n = n \Delta t$,  \rr{$x_{i} = x_0 + i \Delta x $, $y_{j} = j \Delta y $, for, respectively, $n = 0, \ldots, N_t$, $i = 0, \ldots, N_x-1$, and $j = 0, \ldots, N_y-1$, with $N_t$, $N_x$, $N_y\in\mathbb{N}$  and $N_y$ odd. In the case of $\Omega=(-\pi,\pi)$, we define $x_0 := -\pi$.} Moreover, for a function $u(t,x,y)$ we define $u_{i,j}^{n} := u(t_n, x_i, y_j)$ and, for convenience, by using the map \rr{$k = i + j N_x$}, we order, at any time point, the values over the spatial and phenotypic discretised domains into a single vector $\textbf{U}^n$ of length $N_x N_y$, with elements $U^n_k = u^n_{i,j}$.

We discretise the derivatives in time and space, respectively, via
\begin{equation}
	\label{eqn:numerics_time_derivative}
	\partial_t u(t_n, x_{i}, y_j) = \frac{u_{i,j}^{n+1} - u_{i,j}^{n} }{\Delta t} + \rr{O(\Delta t)}
\end{equation}
and 
\begin{equation}
	\label{eqn:numerics_laplacian}
	\Delta_x u(t_n, x_i, y_j)
	= 
	\frac{ u_{i+1,j}^{n} - 2u_{i,j}^{n} + u_{i-1,j}^{n}}{\Delta x^2}
	+ \rr{O(\Delta x^2)},
\end{equation}
and we calculate integrals in phenotype by using the following formula 
\begin{equation}
	\label{eqn:numerics_trapezoidal}
	\int_0^1 u(t_n,x_i,y) \dy 
	=
	\frac{\Delta y}{3} 
	\left(
	u_{i,0}^{n} 
	+ 
	4 \sum_{\substack{j' = 1 \\ j' \text{ odd}}}^{\rr{N_y-2}} u_{i,j'}^{n} 
	+ 
	2 \sum_{\substack{j' = 1 \\ j' \text{ even}}}^{\rr{N_y-3}} u_{i,j'}^{n} 
	+ 
	u_{i,N_y-1}^{n}
	\right)
	+ \rr{O(\Delta y^4)}.
\end{equation}
Homogeneous Neumann boundary conditions are directly included in the approximation of the Laplacian. In fact, by using the central difference scheme corresponding to
\begin{equation}
	\label{eqn:numerics_boundarycondition}
	\partial_x u(t_n, x_i, y_j)
	= 
	\frac{ u_{i+1,j}^{n} - u_{i-1,j}^{n}}{2\Delta x}
	+ \rr{O(\Delta x^2)} = 0,\qquad i \in\{0, \rr{N_x-1}\},
\end{equation}
we obtain the values of $u$ at the ghost points $x_{-1}$ and \rr{$x_{N_x}$}, which are applied in \eqref{eqn:numerics_laplacian} at $x_0$ and \rr{$x_{N_x-1}$}.\\

Under assumptions~\eqref{eq:assumption_M_indipendent_of_y'}, rearranging terms, we write the system \eqref{eqn:PDE_n1_n2_mutational_kernel} as 
\begin{equation*}
\begin{cases}
\displaystyle
	\partial_t n_1 
	- D_1(y_1) \Delta_x n_1 
	- r_1 n_1 
	- \theta_1 \left(M_1(y_1) \rho_1 - n_1 \right) - \Delta_x \left[ ( \alpha_{11}(y_1)\rho_1 + \alpha_{12}(y_1)\rho_2 ) n_1 \right]
	+  \mathcal{K}_1[n_1,n_2] n_1
	= 0,
		\\[0.9em]
\displaystyle
	\partial_t n_2 
	- D_2(y_2) \Delta_x n_2 
	- r_2 n_2 
	- \theta_2 \left(M_2(y_2) \rho_2 - n_2 \right)
	- \Delta_x \left[ ( \alpha_{21}(y_2)\rho_1 + \alpha_{22}(y_2)\rho_2 ) n_2 \right]
	+  \mathcal{K}_2[n_1,n_2] n_2
	= 0.
\end{cases}
\end{equation*}
In what follows, the vectors $\textbf{U}^n$ and $\textbf{V}^n$ represent, respectively, the values of $u^n_{i,j}$ and $v^n_{i,j}$. Hence, using the discretisation formulas \eqref{eqn:numerics_time_derivative}--\eqref{eqn:numerics_boundarycondition}, we approximate the linear terms as
\begin{equation*}
\begin{cases}
\displaystyle
	\partial_t n_1 
	- D_1(y_1) \Delta_x n_1 
	- r_1 n_1 
	- \theta_1 \left(M_1(y_1) \rho_1 - n_1 \right)
	\approx \text{A}_1 \textbf{U}^{n+1} - \frac{1}{\Delta t} \textbf{U}^{n},
	\\[0.9em]
\displaystyle
	\partial_t n_2 
	- D_2(y_2) \Delta_x n_2 
	- r_2 n_2 
	- \theta_2 \left(M_2(y_2) \rho_2 - n_2 \right)
	\approx \text{A}_2 \textbf{V}^{n+1} - \frac{1}{\Delta t} \textbf{V}^{n},
\end{cases}
\end{equation*}
where $\text{A}_1$ and $\text{A}_2$ are the matrices of the coefficients of the elements of $\textbf{U}^{n+1}$ and $\textbf{V}^{n+1}$. Moreover, since the remaining terms in the system are nonlinear, we use the fixed-point iteration method to linearise them.
When solving for $\left(\textbf{U}^{n+1}, \textbf{V}^{n+1}\right)$, we set a tolerance $\varepsilon_p > 0$, initialise the vectors
\begin{align*}
	\textbf{U}^{n+1;\,0} = \textbf{U}^{n} 
	, \qquad \textbf{V}^{n+1;\,0} = \textbf{V}^{n}, 
\end{align*} and iteratively solve, for $p\geq 0$, the two linear systems 
\begin{equation*}
\begin{cases}
\displaystyle
	\left(\text{A}_1 + \text{P}_1\left[\textbf{U}^{n+1;\, p}, \textbf{V}^{n+1;\, p}\right]\right)\textbf{U}^{n+1;\, p+1} = \frac{1}{\Delta t} \textbf{U}^{n}, 
	\\[0.9em]
\displaystyle
	\left(\text{A}_2 + \text{P}_2\left[\textbf{U}^{n+1;\, p}, \textbf{V}^{n+1;\, p}\right]\right)\textbf{V}^{n+1;\, p+1} = \frac{1}{\Delta t}  \textbf{V}^{n},
\end{cases}
\end{equation*}
where $P_1$ and $P_2$ are the discretisation formulas for the remaining terms evaluated at the known vectors $\textbf{U}^{n+1;\, p}$ and $\textbf{V}^{n+1;\, p}$, 
until the following condition is met
$$
\max \left\{ \left\|\textbf{U}^{n+1;\, p+1} - \textbf{U}^{n+1;\, p}\right\|_2,  \left\| \textbf{V}^{n+1;\, p+1} - \textbf{V}^{n+1;\, p}\right\|_2  \right\} < \varepsilon_p.
$$
Then we set 
$
\textbf{U}^{n+1} = \textbf{U}^{n+1;\, p+1}$ and $
\textbf{V}^{n+1} = \textbf{V}^{n+1;\, p+1},
$
and proceed to the next time point. Otherwise, if the fixed-point iteration algorithm does not converge to a solution $\left(\textbf{U}^{n+1}, \textbf{V}^{n+1}\right)$ within a preset maximal number of iterations, we run an adaptive step, whereby we refine the discretisation on the spatial, temporal, and phenotypic domains, apply a piecewise linear interpolation of the solution at the previous step, $(\textbf{U}^{n}, \textbf{V}^{n})$, on the refined mesh, calculate the refined matrices $A_1$ and $A_2$, and resume the solver to find $(\textbf{U}^{n+1}, \textbf{V}^{n+1})$.\\

\noindent The limiting problem~\eqref{eqn:SKT_phenotype}-\eqref{eq:bcs_rescaledrho_limit} is consistently solved by employing a similar approach: the same finite difference schemes are employed to discretise the differential terms and the same fixed-point iteration algorithm is used to linearise the nonlinearities. \\

\noindent These numerical schemes are coded in Python and the scripts are available at the link \url{https://github.com/davidecusseddu/PS-SKT}.

\bibliographystyle{unsrtnat}
\bibliography{references}  

\end{document}